\documentclass[%
 reprint,longbibliography,preprintnumbers,
nofootinbib,
 amsmath,amssymb,
 aps,
pre,
]{revtex4-2}
\pdfoutput=1
\usepackage{overpic}
\usepackage{booktabs}
\usepackage{graphicx}
\usepackage[utf8]{inputenc}
\usepackage{flushend}
\usepackage{dcolumn}
\usepackage{bm}
\usepackage{balance}

\usepackage[normalem]{ulem}

\usepackage[colorlinks = true,
            linkcolor = blue,
            urlcolor  = blue,
            citecolor =green,
            anchorcolor = blue]{hyperref}
\usepackage{verbatim}
\usepackage{color,ulem}
\usepackage{xcolor}
\usepackage[english]{babel}

\usepackage[utf8]{inputenc}
\input Starburst.fd
\newcommand*\initfamily{\usefont{U}{Starburst}{xl}{n}}\initfamily

\makeatletter 
    
\renewcommand\onecolumngrid{
\do@columngrid{one}{\@ne}%
\def\set@footnotewidth{\onecolumngrid}
\def\footnoterule{\kern-6pt\hrule width 1.5in\kern6pt}%
}

\renewcommand\twocolumngrid{
        \def\footnoterule{
        \dimen@\skip\footins\divide\dimen@\thr@@
        \kern-\dimen@\hrule width.5in\kern\dimen@}
        \do@columngrid{mlt}{\tw@}
}%

\makeatother

\newcommand{\beq}{\begin{eqnarray}}
\newcommand{\eeq}{\end{eqnarray}}
\usepackage{amsmath}
\usepackage{tikz}
\usetikzlibrary{decorations.pathmorphing}
\usetikzlibrary{shapes.misc}
\tikzset{cross/.style={cross out, draw=black, minimum size=8*(#1-\pgflinewidth), inner sep=0pt, outer sep=0pt},
cross/.default={1pt}}
\usetikzlibrary{patterns,math}
\begin{document}

\title{Pion dynamics in a soft-wall AdS-QCD model}

 \author{Xuanmin Cao$^{1}$}
\email{caoxm@jnu.edu.cn}
\author{Matteo Baggioli$^{2,3}$}%
 \email{b.matteo@sjtu.edu.cn}
 \author{Hui Liu$^{1}$}
 \email{tliuhui@jnu.edu.cn}
 \author{Danning Li$^{1}$}
 \email{lidanning@jnu.edu.cn}
 \vspace{1cm}
 
 \affiliation{$^{1}$Department of physics and Siyuan Laboratory, Jinan University, Guangzhou 510632, China.}
\affiliation{$^{2}$Wilczek Quantum Center, School of Physics and Astronomy, Shanghai Jiao Tong University, Shanghai 200240, China}
\affiliation{$^{3}$Shanghai Research Center for Quantum Sciences, Shanghai 201315, China}

\begin{abstract}
Pseudo-Goldstone modes appear in many physical systems and display robust universal features. First, their mass $m$ obeys the so-called Gell-Mann-Oakes-Renner (GMOR) relation $f^2\,m^2=H\,\bar{\sigma}$, with $f$ the Goldstone stiffness, $H$ the explicit breaking scale and $\bar{\sigma}$ the spontaneous condensate. More recently, it has been shown that their damping $\Omega$ is constrained to follow the relation $\Omega=m^2 D_\varphi$, where $D_\varphi$ is the Goldstone diffusivity in the purely spontaneous phase. Pions are the most paradigmatic example of pseudo-Goldstone modes and they are related to chiral symmetry breaking in QCD. In this work, we consider a bottom-up soft-wall AdS-QCD model with broken ${\rm{SU}}(2)_L \times {\rm{SU}}(2)_R$ symmetry and we study the nature of the associated pseudo-Goldstone modes -- the pions. In particular, we perform a detailed investigation of their dispersion relation in presence of dissipation, of the role of the explicit breaking induced by the quark masses and of the dynamics near the critical point. Taking advantage of the microscopic information provided by the holographic model, we give quantitative predictions for all the coefficients appearing in the effective description. In particular, we estimate the finite temperature behavior of the kinetic parameter $\mathfrak{r^2}$ defined as the ration between the Goldstone diffusivity $D_\varphi$ and the pion attenuation constant $D_A$. Interestingly, we observe important deviations from the value $\mathfrak{r^2}=3/4$ computed in chiral perturbation theory in the limit of zero temperature.
\end{abstract}

\maketitle
\tableofcontents
\section{Introduction}
Nambu-Goldstone modes (NGMs) are massless excitations related to the spontaneous symmetry breaking (SSB) of global symmetries \cite{Beekman:2019pmi}. Their appearance and their properties follow from the Goldstone theorem and its generalizations \cite{PhysRev.117.648,PhysRev.127.965,Watanabe:2011ec,Low:2001bw}. Because of their protected massless nature, NGMs have a strong impact on the infrared (IR) low energy dynamics and they constitute the fundamental building blocks for low-energy effective field theory and hydrodynamics \cite{Nicolis:2015sra,Baggioli:2022pyb,Son:2002zn,Pich:2018ltt}. 

To provide a more concrete manifestation of these statements, let us consider two paradigmatic examples. The first one is that of superfluids, phases of matter which spontaneously break a global ${\rm{U}}(1)$ symmetry \cite{Schmitt:2014eka}. The low-energy properties of superfluids are described in terms of the ${\rm{U}}(1)$ Goldstone mode \cite{Son:2002zn} which is ultimately responsible for the propagation of second sound \cite{doi:10.1063/1.3248499} and their infinite DC conductivity as well.\footnote{See \cite{PhysRevB.104.205132} for a modern derivation of this statement using higher-form mixed anomalies.} A second, and perhaps even more familiar, example is that of solids. Solids are physical systems which spontaneously break translational invariance with their atoms sit at preferred positions, a.k.a. ``long-range order". Solids exhibit Goldstone modes usually called phonons which correspond to what you hear knocking your desk.\footnote{\label{f2}This identification is slightly imprecise. For example, transverse sound in solids does not correspond directly to transverse phonons but rather to a collective mixture of transverse phonons and transverse momentum fluctuations \cite{Baggioli:2022pyb}.} As a matter of fact, Goldstones and the SSB of translations are the ultimate reason for the rigidity of your desk.\\
Despite symmetries and NGMs are an old subject, there are still several open questions in particular when dissipative/open systems or out-of equilibrium setups are considered, e.g. \cite{Minami:2015uzo,Hongo:2019qhi,PhysRevLett.126.141601}, or non ``standard'' symmetries are spontaneously broken, e.g. \cite{Hofman:2018lfz}.

In the past decades, Holography, or gauge-gravity duality, has emerged as a useful complementary tool to study physical systems where standard perturbative methods fail, with many applications to condensed matter physics \cite{zaanen2015holographic,Hartnoll:2016apf}, QCD \cite{Casalderrey-Solana:2011dxg}, etc. SSB has historically been one of the first applications of this duality in the form of the so-called holographic superconductor model \cite{Hartnoll:2008vx}.\footnote{Because of historical reasons, we keep using this jargon well aware that the model does not describe the physics of superconductors since the boundary ${\rm{U}}(1)$ symmetry remains global, at least using standard boundary conditions.} The holographic model mimics the condensation of a charged scalar operator which induces the spontaneous symmetry breaking of a global ${\rm{U}}(1)$ symmetry -- a superfluid phase. The low-energy dynamics of the dual field theory is in perfect agreement with relativistic superfluid hydrodynamics \cite{Herzog:2011ec} and displays, as expected, the presence of a well-defined Goldstone mode in the spectrum. Similarly, in more recent years, large effort has been devoted to holographic phases which break spontaneously translations \cite{Baggioli:2022pyb}. In a large class of models, the presence of propagating Goldstone bosons (phonons)\footnote{The same argument of footnote \ref{f2} applies here.} has been verified, e.g. \cite{Alberte:2017oqx,Amoretti:2019kuf}, and their dynamics matched to relativistic viscoelasticity theory \cite{Armas:2019sbe,Ammon:2020xyv}. Interestingly, even more exotic type of Goldstone modes have been discussed in holography such as type II/B NGMs \cite{Amado:2013xya,Baggioli:2020edn}, out-of-equilibrium NGMs \cite{Ishigaki:2020vtr} and NGMs of spontaneously broken higher-form symmetries \cite{Hofman:2017vwr,Iqbal:2021tui}. The graviton itself has been recently identified as a Goldstone boson of an exotic biform symmetry \cite{Hinterbichler:2022agn}.

Interestingly, the dynamics of pNGMs displays a certain degree of universality which was first realized in the context of chiral perturbation theory for pions \cite{PhysRev.175.2195}. In particular, it was early recognized that their mass $m_\pi$ satisfies the so-called Gell-Mann-Oakes-Renner (GMOR) relation which, at zero temperature, is given by:
\begin{equation}\label{gmor}
f_\pi^2 m_\pi^2\,=\,\left(m_u+m_d\right)\,\bar{\sigma}
\end{equation}
in terms of the up and down quark masses $m_u,m_d$ which explicitly break chiral symmetry. $\bar{\sigma}\equiv \langle \bar{\psi}\psi\rangle$ is the chiral condensate which signals the spontaneous breaking of chiral symmetry and $f_\pi$ is the pion decay constant which, as we will see in detail later, plays the role of the Goldstone susceptibility. The mass $m_\pi$ is usually denoted as \textit{screening mass} in contrapposition to the \textit{pole mass}.\footnote{As we will explicitly see, at finite temperature, the two definitions of mass do not coincide.} From a physical perspective, the GMOR relation, Eq.~\eqref{gmor}, is tantamount to recognizing that the mass squared of the pNGMs is linearly proportional to the strength of explicit breaking (the quark masses) and the amount of SSB (the chiral condensate). More in general, we do expect the mass of pNGMs to vanish by setting either the explicit breaking scale or the spontaneous one to zero. As we will see, this property is a generic feature of pseudo-Goldstone modes which goes beyond the specific case of pions. 
The validity of the GMOR relation, Eq.\eqref{gmor}, has been verified directly in holographic QCD models both at zero temperature \cite{PhysRevLett.95.261602} and finite temperature \cite{Cao:2021tcr}. Moreover, a similar expression has been discussed at length in the context of holographic phonons \cite{Ammon:2019wci,Amoretti:2018tzw,Andrade:2018gqk} and also in holographic models with pseudo-spontaneous breaking of a global ${\rm{U}}(1)$ symmetry \cite{Donos:2021pkk,Ammon:2021pyz}. More formal analyses can be found in \cite{Argurio:2015wgr,Amoretti:2016bxs}.

By direct computation in a large class of holographic models with broken translations \cite{Amoretti:2018tzw,Baggioli:2019abx,Ammon:2019wci,Andrade:2018gqk,Donos:2019txg,Amoretti:2021fch,Donos:2019hpp}, it was only recently realized that in finite temperature dissipative systems also the damping, e.g. relaxation rate or linewidth, of pNGMs follow a universal relation in terms of other physical parameters of the system. In particular, at small frequency (energy), the dispersion relation of the pNGM is expected to have a universal form of the type:
\begin{equation}
\omega=\pm \,\omega_0 -\frac{i}{2}\,\Omega\,+\,\dots
\end{equation}
where $\omega_0=v \,k_0$ is the pole mass and $\Omega$ is the already mentioned relaxation rate (at zero wave-vector).\footnote{In order to avoid confusion, let us clarify our jargon. Given a dispersion relation $\omega(k)$, its imaginary part, $-\mathrm{Im}[\omega(k)]$, represents the inverse of its relaxation time, $\tau^{-1}$. The latter is also indicated as ``relaxation rate", ``damping'' or ``linewidth".} $v$ is the speed of the propagating sound mode in the purely spontaneous phase\footnote{For simplicity, we consider only type I Goldstone modes. Same arguments can be made for type II NMGs \cite{Delacretaz:2021qqu}.} and $k_0$ is the screening mass, which was indicated as $m_\pi$ for the pions above. As we will describe in detail in the next Section, the damping  $\Omega$ in the limit of soft-explicit breaking is universally given by:
\begin{equation}\label{uni}
\Omega= D_\varphi k_0^2\,,
\end{equation}
where $D_\varphi$ is the Goldstone diffusivity (see more details later).
The universal relation, Eq.~\eqref{uni}, has been verified numerically and via perturbative semi-analytical methods in a large class of holographic models for translations \cite{Amoretti:2018tzw,Baggioli:2019abx,Ammon:2019wci,Andrade:2018gqk,Donos:2019txg,Amoretti:2021fch,Donos:2019hpp} and also in holographic models with global $U(1)$ symmetry \cite{Donos:2021pkk,Ammon:2021pyz}. Moreover, it has been derived in effective field theories for quasicrystals and phason modes in \cite{Baggioli:2020haa,Baggioli:2020nay}, it has been proven within pions hydrodynamics using positivity of entropy production in \cite{Grossi:2020ezz}, and more generally in \cite{Armas:2021vku}, and it has finally been demonstrated using the locality requirements of the hydrodynamics framework in \cite{Delacretaz:2021qqu}.

This universal relation might have important physical consequences for several phenomena in nature, such as the $T$-linear resistivity of strange metals \cite{Delacretaz:2021qqu,Baggioli:2022pyb}. An even more natural framework where this relation might be at play is QCD, in particular around the chiral critical point. In the limit of two flavors of massless quarks, the chiral phase transition in QCD is in the same universality class as the classical four components Heisenberg antiferromagnet \cite{Rajagopal:1992qz} and it can be effectively described by a non-abelian version of superfluid hydrodynamics reflecting the SU$(2)_L \times$ SU$(2)_R \simeq O(4)$ symmetry \cite{Son:1999pa}. The pseudo-Goldstone mode nature of the pions might produce strong effects and their mass $m_\pi$ can modify the hydrodynamics description in the chiral limit with potentially measurable consequences. In particular, in the presence of a finite pion mass, the hydrodynamic theory is ordinary hydrodynamics at long distances, and superfluid-like at short distances \cite{Grossi:2020ezz}. Recently, a more systematic study of this problem has been initiated in \cite{Grossi:2021gqi} and an initial exploration using numerical simulations has appeared in \cite{Florio:2021jlx}. In this series of works, the universal relation in Eq.~\eqref{uni} has been derived using positivity of entropy production but its validity has not been confirmed yet in QCD simulations or experiments.\\

Holography has been playing a complementary role also in this direction. Chiral symmetry breaking and the effects of finite quark masses have been studied in many instances \cite{Rodrigues:2018pep,Colangelo:2011sr,Evans:2016jzo,Bartz:2016ufc,Ballon-Bayona:2021ibm,Sui:2009xe,Gherghetta:2009ac,Li:2012ay,Li:2016smq,Chelabi:2015gpc,Chelabi:2015cwn,Chen:2018msc,Fang:2015ytf}. More recently, the properties of pions and their dynamics in finite temperature holographic QCD models have been discussed \cite{MartinContreras:2021yfz,Cao:2021tcr,Cao:2020ryx,Lv:2018wfq}, as well as in the NJL model~\cite{Sheng:2020hge}. Nevertheless, a complete description is still lacking. In this work, we provide a step forward in this program by continuing the study of pions in finite temperature holographic QCD models. Among the various tasks, we will test directly the validity of the universal relation in Eq.~\eqref{uni} in a more realistic QCD model, generalizing the results obtained in the ${\rm{U}}(1)$ holographic toy models \cite{Ammon:2021pyz}.\\

The manuscript is organized as follows. In Section \ref{sec:damping} we review the basic properties of pseudo-Goldstone modes at finite temperature; in Section \ref{sec:pions} we focus on the case of pions and summarize the main concepts therein; in Section \ref{sec:model} we present the holographic QCD model; in Section \ref{sec:results} we describe in detail all our results and finally in Section \ref{sec:out} we provide an outlook and some open questions for the future. Appendix \ref{appendix} provides further details about the methods and the numerical techniques.
\section{The damping of pseudo-Goldstone modes}\label{sec:damping}
In this Section, we provide more details about the low-energy description of pNGMs and a more formal definition of the universal relation in Eq.\eqref{uni}. For a simple treatment in the case of a global ${\rm{U}}(1)$ symmetry we refer to \cite{Ammon:2021pyz}, for a more complete discussion and formal derivation see \cite{Delacretaz:2021qqu,Armas:2021vku}.\\
Let us start by considering a system with a spontaneously broken symmetry and a type I Goldstone mode with dispersion relation
\begin{equation}\label{didi}
    \omega=\pm v\, k -\frac{i}{2}\,D_A\,k^2 +\dots
\end{equation}
with $v$ the velocity and $D_A$ the leading quadratic coefficient of attenuation constant. The latter has not to be confused with the wave-vector dependent damping $\Gamma_k$ used later which is defined as:
\begin{equation}
    \Gamma_k\equiv -\mathrm{Im}\left[\omega(k)\right]\,.
\end{equation}
Eq.\eqref{didi} can be for example identified with the dispersion relation of second sound in a superfluid or that of shear sound in a standard solid. $D_A$ arises because of the dissipative finite temperature effects and it is usually not discussed in the zero temperature field theory treatments. 

At zero wave-vector $k=0$, and finite frequency $\omega$, the Goldstone\footnote{The Goldstone mode $\varphi$ has always to be identified with the phase of the complex order parameter $\psi= A \exp(i \varphi)$ where $A$ is the amplitude. The fluctuations of the amplitude mode $\delta A$ are known to be gapped and they are therefore usually not included in the low-energy effective description. For simplicity, they will not be considered in this work either.} retarded Green's function is given by:

\begin{equation}\label{green1}
    G_{\varphi\varphi}(\omega,k=0)=\frac{1}{\chi_Q\,\omega^2}-\frac{i}{\omega}\,\Xi+\dots
\end{equation}
where $\chi_Q$ is the susceptibility of the conserved quantity $Q$ associated to the spontaneously broken symmetry. The ellipsis indicates higher-order corrections in the frequency $\omega$ which are neglected. For example, in the case of a spontaneously broken $U(1)$ (superfluid), we have $\chi_Q\equiv \chi_{\rho\rho}$ which is the charge density susceptibility; in the case of spontaneously broken translations (solid), $\chi_Q\equiv \chi_{PP}$ is the momentum susceptibility. The $\Xi$ parameter appearing in Eq.\eqref{green1} is formally defined via the following Kubo's formula
\begin{equation}
    \Xi= \lim_{\omega \rightarrow 0}\,\omega\,\mathrm{Im}G_{\varphi\varphi}(\omega,k=0)
\end{equation}
and it is related to the Goldstone diffusivity $D_\varphi$ as:
\begin{equation}\label{eq:Goldstonediffusivity}
    D_\varphi=\frac{\Xi}{\chi_{\varphi\varphi}}
\end{equation}
with $\chi_{\varphi\varphi}$ the Goldstone susceptibility. Let us give once more some concrete examples. For a superfluid state, $\Xi$ is usually denoted at $\zeta_3$ \cite{Herzog:2011ec} and $\chi_{\varphi\varphi}=\mu/\rho_s$ with $\mu$ the chemical potential and $\rho_s$ the superfluid density. In a solid, $\chi_{\varphi\varphi}=1/G$ for transverse phonons and $1/(K+G)$ for longitudinal ones, with $G,K$ respectively the bulk and shear moduli \cite{Baggioli:2022pyb}. Importantly, the Goldstone diffusivity is not the attenuation constant $D_A$ in Eq.\eqref{didi} but only part of it. More precisely $D_A=D_Q+D_\varphi$ with $D_Q$ the diffusion constant of the conserved charge associated to the spontaneously broken symmetry. Also, $D_Q=\sigma_Q/\chi_Q$ with $\sigma_Q$ the ``conductivity'' associated to the conserved current $J_Q$ and given by:
\begin{equation}\label{eq:sigmaQ}
        \sigma_Q\equiv - \lim_{\omega \rightarrow 0}\frac{1}{\omega} \mathrm{Im}G_{J^QJ^Q}(\omega,k=0)\,.
    \end{equation}
For a superfluid $\sigma_Q$ is the electric conductivity and $J_Q$ the electric current; for translations $\sigma_Q$ is the viscosity and $J_Q\equiv T_{ij}$ the shear stress. Finally, let us remind the Reader that the velocity in Eq.\eqref{didi} satisfies the following identity:
\begin{equation}\label{eq:soundvelocity1}
    v^2=\frac{1}{\chi_Q\,\chi_{\varphi\varphi}}\,.
\end{equation}

Let us now introduce a small external source breaking explicitly the already spontaneously broken symmetry. In the regime of small explicit breaking, the associated current $J_Q$ is an ``almost'' conserved quantity and the corresponding conserved charge $Q$ slowly relaxes in time with a rate proportional to the amount of explicit breaking itself. In this limit, the dispersion in Eq.\eqref{didi} gets modified into:
\begin{equation}\label{didi2}
    \omega=\pm\,\omega_0 -\frac{i}{2}\,\Omega\,+\dots
\end{equation}
where, for simplicity, the wave-vector dependent terms are not displayed (see \cite{Ammon:2021pyz,Delacretaz:2021qqu,Armas:2021vku} for a treatment of those corrections).\\
This implies that the original Goldstone mode acquires both a finite mass $\omega_0$ and a finite relaxation rate $\Omega$ which is now independent of the wave-vector $k$ (cfr. Eq.\eqref{didi}). In the language of hydrodynamics, the damping  $\Omega$ is known as ``phase-relaxation rate" and it is appears in the Josephson constraint as a relaxation term
\begin{equation}
    \left(\partial_t+\Omega\right) \varphi= \dots
\end{equation}
where the ellipsis indicates terms which are not relevant to the present discussion and therefore omitted. 
Before continuing, let us pause for a moment on the meaning of the parameter $\Omega$. Even in absence of explicit breaking, a phase-relaxation mechanism ($\Omega 
\neq 0$) is known to appear in presence of topological defects, e.g. vortices in superfluids \cite{Davison:2016hno} or dislocations/disclinations in solids \cite{Delacretaz:2017zxd}. In our scenario, the origin of the damping term $\Omega$ is different. It is not connected to the existence of any defects and it disappears in absence of explicit breaking. In other words, this damping term is an effect of the pseudo-spontaneous breaking of the global symmetry.

Back to our discussion, the pNGM mass $\omega_0$ is proportional to both the explicit and spontaneous breaking scales and obeys a GMOR-type relation. It appears in the static Goldstone correlator:
\begin{equation}\label{stat}
    G_{\varphi\varphi}(\omega=0,k)\,=\,-\frac{\chi_{\varphi\varphi}}{k^2+k_0^2}
\end{equation}
where
\begin{equation}
    k_0\equiv \frac{\omega_0}{v}
\end{equation}
is the screening mass. The latter can be therefore obtained independently from the static correlator in Eq.\eqref{stat} and from the pNGM dispersion at zero wave-vector, Eq.\eqref{didi2}.\\
Finally, as already discussed above, the damping  $\Omega$ in the limit of soft explicit breaking satisfies the following relation:
\begin{equation}\label{eq:diffusionrelation}
    \Omega=k_0^2 D_\varphi
\end{equation}
which can be thought as the dissipative counterpart of the more known GMOR relation. Notice that, following Eq.\eqref{didi2}, the damping  can be directly extracted from the pNGM dispersion using:
\begin{equation}\label{eq:dampingwidth}
    \Omega \equiv -2 \lim_{k\rightarrow 0}\,\mathrm{Im}\left[\omega(k)\right]\,.
\end{equation}

\subsection{The case of Pions}
\label{sec:pions}
In order to connect more explicitly the language of \cite{Delacretaz:2021qqu,Armas:2021vku} with the QCD framework recently discussed in \cite{Grossi:2020ezz,Grossi:2021gqi,Florio:2021jlx}, in this short section, we provide more details about the case of pions. \\
Neglecting the dynamics of the heavier quarks, QCD can be described by a ${\rm{U}}(1)\times{\rm{SU}}(2)_L \times {\rm{SU}}(2)_R$ symmetry which is spontaneously broken below the chiral critical point to ${\rm{U}}(1) \times {\rm{SU}}(2)_V$ by the presence of a finite chiral condensate $\bar{\sigma}\equiv \langle \bar{\psi}\psi\rangle$. In the limit of zero quark masses $m_q=0$, i.e., the chiral limit, the low-energy description needs to take into account the presence of massless Goldstone degrees of freedom and it takes the form of a non-abelian superfluid hydrodynamics introduced in \cite{Son:1999pa} and then formally generalized in \cite{Jain:2016rlz}. In presence of finite quark masses $m_q\neq0$, then chiral symmetry is not an exact symmetry from the start. This effect must be taken into account in the hydrodynamics description as done in \cite{Pujol:2002na,Son:2001ff,Son:2002ci} and more recently in \cite{Grossi:2020ezz,Grossi:2021gqi}. \\

For simplicity, we will focus on QCD in the limit of two flavors. In this approximation, the low-energy description is based on a four-component vector field:
\begin{equation}
    \phi_a=\left(\sigma, \Vec{\pi}\right)
\end{equation}
where $\sigma$ is the meson and $\Vec{\pi}$ the pions degrees of freedom. The vacuum expectation value of this vector is given by:
\begin{equation}
    \langle \phi_a \rangle= \left(\bar{\sigma},0\right)
\end{equation}
where $\bar{\sigma}$ is the chiral condensate breaking chiral symmetry. The pions degrees of freedom can be described using the phase fluctuations $\varphi=\pi/\bar{\sigma}$ which will be indicated in the rest of this work as the Goldstone degrees of freedom.\\
Under precise assumptions, the hydrodynamic equations describing this system are then given by (see for example \cite{Soloviev:2021syx}):
\begin{align}
    & \partial_t \varphi =\,- \mu_5 + D_\varphi\left(\nabla^2-m_{scr}^2\right)\varphi\,,\\
    &\partial_t \mu_5= v^2 \left(-\nabla^2+m_{scr}^2\right)\varphi+D_5 \nabla^2 \mu_5\,,\\
    & \partial_t \delta\sigma=D_\varphi\left(\nabla^2-m_\sigma^2\right)\delta \sigma\,.
\end{align}
Here $\varphi$ indicates the Goldstone degrees of freedom, $\mu_5$ the axial chemical potential and $\delta \sigma$ the meson fluctuations. Additionally, $m_{scr}$ is the pion screening mass and $m_\sigma$ the sigma meson mass. The Goldstone degrees of freedom couple to the axial charge fluctuations where $\mu_5,D_5$ are the axial chemical potential and diffusion constant. On the contrary, at least in this approximation, the $\sigma$ mode, which plays the role of the Higgs/amplitude mode, appears decoupled. This decoupling approximation might not always be valid and the effects of the amplitude mode could become important close to the critical point (see for example \cite{Donos:2022xfd,Donos:2022qao}). We will return on this point in the outlook; for the moment, we will ignore the effects of the amplitude mode in our description.\\
In Fourier space, the dispersion relation of the pNGMs is the solution of the following equation:
\begin{equation}\label{eq:dispersion}
   - \omega^2+\omega_k^2-i \,\omega \,\Gamma_k\,=\,0
\end{equation}
with:
\begin{equation}\label{eq:dispersionreal}
    \omega_k^2=v^2\left(k^2+m_{scr}^2\right)+\dots
\end{equation}
where $v^2=f_s^2/f_t^2$ give the pion velocity and $m_{scr}$ its screening mass. $f_s$ is usually called pion decay constant while $f_t^2$ corresponds to the axial charge susceptibility. The two coincide at zero temperature. Finally, one can define the pole mass:
\begin{equation}\label{eq:soundvelocity}
    m_p^2=v^2 m_{scr}^2
\end{equation}
which was indicated in the previous section as $\omega_0$. The pion decay constant corresponds to what was indicated as Goldstone susceptibility in the previous section, $f_s^2=1/\chi_{\varphi\varphi}$, and the axial susceptibility with $f_t^2=\chi_Q$. This treatment is in complete analogy with spin wave fluctuations in Heisenberg antiferromagnets where $f_s$ corresponds to the so-called stiffness and $f_t^2$ to the magnetic susceptibility in a direction perpendicular to magnetization \cite{PhysRev.188.898}.\\

Additionally, the damping is given, up to quadratic order, by\footnote{Originally \cite{Son:2002ci}, the damping  was assumed to be:
\begin{equation}\label{nene}
    \Gamma_k= \kappa_1 m_{scr}^2 + \left(D_5+\kappa_2\right) k^2\,
\end{equation}
with $\kappa_1\neq \kappa_2$.
In Eq.\eqref{dam}, following the more recent results in \cite{Grossi:2021gqi}, we have fixed $\kappa_1=\kappa_2=D_\varphi$. As discussed in the main text, this is equivalent to assume the universal relation \eqref{uni} for the pseudo-Goldstone modes damping .}:
\begin{equation}\label{dam}
    \Gamma_k= D_\varphi m_{scr}^2 + \left(D_5+D_\varphi\right) k^2\,.
\end{equation}
Here, $D_5$ is the axial charge diffusion constant corresponding to what was indicates as $D_Q$ in the previous general section. Moreover, the product $D_\varphi m_{scr}^2$ coincides with the damping  $\Omega$ implying the universal damping relation discussed in the previous section. Likewise, from Eq.\eqref{nene}, we can identify the coefficient $D_A$ used above as $D_A=D_5+D_\varphi$.\\
Finally, one can define a useful phenomenological parameter:
\begin{equation}\label{parameterr}
    \mathfrak{r}^2\equiv \frac{D_\varphi}{D_\varphi+D_5}=\frac{D_\varphi}{D_A}
\end{equation}
which parametrizes the ratio between the Goldstone diffusivity and the pion attenuation constant.
In chiral perturbation theory, the latter is found to be equal to $3/4$ in the limit of zero temperature \cite{Torres-Rincon:2022ssx}.

\section{The holographic model}
\label{sec:model}

In the holographic framework, the hard-wall~\cite{erlich_qcd_2005} and soft-wall~\cite{karch_linear_2006}  AdS/QCD models provide an ideal starting point to study the spontaneous symmetry breaking of chiral symmetry, by promoting the 4D global symmetry ${\rm{SU}}(N_f)_L\times {\rm{SU}}(N_f)_R$ of QCD to a gauge symmetry in 5D. The chiral symmetry is spontaneously broken to ${\rm{SU}}(N_f)_V$ when the bulk scalar field in the model gets a non-vanishing expectation value, which could be self-consistently derived from its bulk equation of motion. As a result, the pions appear as the Goldstone modes of the spontaneous chiral symmetry breaking. Another advantage of these two models is that the explicit breaking could also be easily considered, when the quark masses are introduced as non-vanishing sources for the corresponding operators. Thus, in such holographic QCD models, one has all the ingredients to investigate the nature of the pseudo-Goldstone modes. 

Compared to the hard-wall model, the soft-wall model introduces an IR scale through the dilaton field profile
\begin{equation}\label{eq-dilaton}
    \Phi(z)=\mu_g^2 z^2,
\end{equation} 
which breaks the conformal symmetry of the dual field theory. Here $\mu_g$ is a parameter with mass dimension one related to the dynamics of confinement and to the $\Lambda_{QCD}$ scale. This parameter provides the possibility to describe the Regge behavior, i.e. the observation that the mass square of the highly excited states $m_n^2$ satisfies $m_n^2\propto n$, with $n$ the excitation quantum number. As a phenomenological model, we will keep the dilaton field as a background field without dynamics, just like the original soft-wall model. Also, we will consider only the lightest two flavors, i.e. u, d quarks. This implies $N_f=2$. ~\footnote{If you are interested in more flavors case, you can read the $N_f=4$ case in ~\cite{Chen:2021wzj}}

The action of the soft-wall model is easily constructed from symmetries and it reads  \begin{eqnarray}\label{bulkaction}
S=\int d^5 x\sqrt{g}\, e^{-\Phi}\, {\rm{Tr}}\left[|D X|^2-V(|X|)-\frac{1}{g_5^2}\left(F_L^2+F_R^2\right)\right],\nonumber\\
\end{eqnarray}
with $g$ the determinant of the metric $g_{MN}$, $g_5$ the gauge coupling. $X$ is a bulk matrix-valued scalar field dual to the 4D operator $\bar{\psi}^\alpha \psi^\beta$ (with $\alpha$ and $\beta$ flavor indices). $F_{MN,L/R}$ is the left/right hand gauge field strengths. The latter are defined via the gauge potentials $A_{L,R}=A_{L,R}^a \tau^a$ ($\tau^a$ the generator of ${\rm{SU}}(N_f)$ group) using:
\begin{equation}
    F_{MN,L/R}=\partial_M A_{N,L/R}-\partial_N A_{M,L/R}-i[A_{M,L/R},A_{N,L/R}]\nonumber
\end{equation}
where $D_MX=\partial_MX-iA_{LM}X+iXA_{RM}$ is the covariant derivative of scalar field with $M ,N =\{0,1,2,3,4\}$. Finally, $V(|X|)$ is the scalar potential which will be specified in the following. By mapping the holographic result for the two-point function of the vector current to the 4D field theory calculation, one could fix $g_5^2=12\pi^2/N_c$ \cite{Cherman:2008eh}. In this work, we will focus on the realistic case with $N_c=3$. 

In the original soft-wall model, the background metric is taken to simply be that of five-dimensional Anti-de Sitter spacetime, 
\begin{equation}
    ds^2=\frac{1}{z^2}\left (dt^2-dx^i dx^i-dz^2\right ),
\end{equation}
with $i=1,2,3$ representing the spatial dimensions and z the fifth radial dimension. Moreover, only the mass term is kept in the scalar potential which reads $ V(|X|)=m_5^2\,|X|^2$. In order to have an operator with conformal dimension $\Delta=3$, we fix $m_5^2=-3$. For convenience, the AdS radius is also fixed to be $L=1$. It is interesting to notice that this simple construction accurately describes the spectrum of light vector mesons, including the highly excited states~\cite{karch_linear_2006}.

To study the thermal properties of hadrons, finite temperature effects could be introduced by considering black hole solutions. The temperature $T$ is then identified with the Hawking temperature of the black hole. A simple choice is to consider the AdS-Schwarzchild black hole whose metric reads
\begin{equation}
    ds^2=\frac{1}{z^2}\left (f(z)dt^2-dx^idx^i-\frac{1}{f(z)}dz^2\right ),
\end{equation}
with 
\begin{eqnarray}
f(z)=1-\frac{z^4}{z_h^4}\,.
\end{eqnarray}
Here, $z_h$ is the horizon radius of the black hole, which is related to the temperature $T$ by $z_h=1/(\pi T)$.

As emphasized in Refs.\cite{Chelabi:2015gpc,Chelabi:2015cwn,Chen:2018msc}, the above model with $\Phi(z)=\mu_g^2 z^2, m_5^2=-3$, is not able to describe spontaneous chiral symmetry breaking. On the contrary, it gives a vanishing chiral condensate in the chiral limit (where all the quarks are massless). To introduce spontaneous chiral symmetry breaking, one has to modify either the dilaton profile \cite{Chelabi:2015gpc,Chelabi:2015cwn,Chen:2018msc} or the $5$D mass of the bulk scalar field $X$~\cite{FANG201686}. According to the study in Ref.\cite{Chen:2018msc}, the key point of these modifications is to guarantee that the mass squared of the scalar meson vanishes at a critical temperature $T_c$. Above such a critical value, the scalar mass squared becomes negative and a second order phase transition occurs in the chiral limit (for more details, we refer to Ref.~\cite{Chen:2018msc}).   
Additionally, a correct description of the meson spectra and of the low temperature chiral condensate requires higher order corrections in the scalar potential. A quartic term is the simplest correction that can be added. Taking all these considerations into account, we will use the following extension of the original soft-wall model potential,
\begin{equation}
    V(|X|)=m_5^2(z)\,|X|^2+\lambda \,|X|^4,
\end{equation}
where the $z$ dependent 5D mass $m_5^2(z)$ takes the following form
\begin{equation}\label{mm}
    m_5^2=-3-\mu_c^2 z^2.
\end{equation}
Here, two additional parameters, $\lambda$ and $\mu_c$, have been introduced. Since the dilaton will not be modified and its profile will remain as in Eq.\eqref{eq-dilaton}, the negative corrections in $m_5^2(z)$ could also be thought as originating from the interaction between $\phi$ and $X$ (e.g., a bulk term of the form $\phi|X|^2$). However, here we will treat it as a simple phenomenological model without caring about the microscopic origin of this term.  

So far, we have three free parameters, $\mu_c,\mu_g, \lambda$, in the model. By comparing the model predictions with the experimental data for the light mesons, the best fitting of these parameters could be obtained as ~\cite{FANG201686}
\begin{eqnarray}\label{inputparameters}
\mu_g=440\, \rm{MeV}, \ \ \ \ \mu_c=1450 \,\rm{MeV}, \ \ \ \ \  \lambda =80\,.
\end{eqnarray}
This choice reveals the presence of massless pions in the spectrum in the chiral limit, confirming their Goldstone nature within this model \cite{Cao:2021tcr,Cao:2020ryx}. Also, when taking the physical and finite values for the quark masses, it gives a pion mass of around $140 $ MeV and a reasonable value of chiral condensate, about $0.015\rm{GeV}^3=(246MeV)^3$. In this work, we would like to extend the study of this model to finite temperature and to focus on the dissipative properties of the (pseudo-) Goldstone modes, i.e. the pions. 

\subsection{The Pseudo-Goldstone modes in the soft-wall model}\label{sec:correlators}
As mentioned above, the 5D matrix-valued scalar $X^{\alpha\beta}$ is dual to the 4D operator $\bar{\psi}^\alpha \psi^\beta$. The spontaneous breaking of chiral symmetry is due to the non-zero expectation value for the operator $\bar{\psi} \psi$. Thus, in this case, we do expect a non-zero background field $X$. On the contrary, the meson states are treated as perturbations above the vacuum. Therefore, following Ref.\cite{karch_linear_2006}, we take a simple ansatz for $X$ given by
 \begin{equation}
 X=\frac{\Sigma}{2}\,e^{i 2\varphi}\,.
 \end{equation}
Here, $\Sigma$ describes the chiral condensate and it is considered as a background field, while  $\varphi\equiv\varphi^a\tau^a$ describes the perturbations of the phase, which is dual to the pseudo-Goldstone mode (i.e. the pion meson). For convenience, other kinds of perturbations which are irrelevant for our analysis will be neglected.
 
Inserting this ansatz into the action in Eq.~\eqref{bulkaction}, one can easily derive the equation of motion (EOM) for $\Sigma$ as
 \begin{equation}\label{EOM:chi}
     \Sigma''+\left(3A'+\frac{f'}{f}-\Phi '\right)\Sigma'- \frac{e^{2A}}{f}\left(m_5^2+\frac{\lambda \Sigma^2}{2}\right)\Sigma=0,
 \end{equation}
 where $'$ indicates the derivative with respect to $z$. From this equation, one can obtain the asymptotic solution near the UV boundary ($z=0$) as
 \begin{eqnarray}
     \Sigma(z)&=&m_q \zeta z+\frac{1}{4}m_q\zeta [(m_q\zeta)^2\lambda+4\mu_g^2-2\mu_c^2]z^2\ln(z)+\nonumber\\
     & &\frac{\bar{\sigma}}{\zeta} z^3+\mathcal{O}\left(z^3\right).
 \end{eqnarray}
 From the dual field theory perspective, the two integration constants $m_q$ and $\bar{\sigma}$ correspond to the quark mass and the chiral condensate respectively, i.e. 
\begin{equation}
    m_q=m_u=m_d,\qquad \bar{\sigma}=\langle \bar{\psi}\psi\rangle.
\end{equation}
The normalization constant $\zeta=\sqrt{N_c}/(2\pi)$ is introduced, in order to match the holographic results of two-point function of the scalar operator to the 4D field theory calculation (for more details, see Ref.~\cite{Cherman:2008eh}).

At the horizon $z=z_h$, $\Sigma(z)$ has an asymptotic solution given by
\begin{equation}
    \Sigma(z)=c_0+\frac{c_0(c_0^2\lambda-2z_h^2\mu_c^2-6)}{8z_h}(z_h-z)+\mathcal{O}[(z_h-z)^2],
\end{equation}
with $c_0$ an arbitrary integration constant. Taking concrete values for the temperature $T$ (or equivalently the horizon radius $z_h$) and the quark mass $m_q$, and imposing regularity conditions for $\Sigma$ at the horizon, one can numerically solve Eq.~\eqref{EOM:chi} and extract the chiral condensate $\bar{\sigma}$.  For more details about the numerical algorithm, we refer the Reader to Ref.~\cite{Cao:2020ryx}. Following this scheme, the chiral condensate is self-consistently extracted dynamically, instead of being externally fixed as in the hard-wall model \cite{erlich_qcd_2005}. In the next sections, we will apply these techniques and show the corresponding numerical results for the condensate and other quantities. 
 
After constructing the background solution with finite chiral condensate, we are interested in the dynamics of the fluctuations on top of it. We take the four dimensional wave-vector to be $q=(\omega,0,0,k)$. Using these notations, the fluctuations of the axial-vector field divide into the transverse ones, $a_1$ and $a_2$, and the longitudinal ones, $a_0$ and $a_3$ in the $a_z=0$ gauge. In the 4D momentum space, we can derive the linearised EOMs for coupled fields $a_0$, $a_3$ and $\varphi$,
\begin{subequations}\label{EOM:coupledpion}
  \begin{eqnarray}
& &{a_0}''+ \left(A'-\Phi '\right){a_0}'-\left(\frac{ k^2+e^{2A}g_5^2 \Sigma ^2}{f}\right)a_0-\nonumber\\
& &\left(\frac{\omega k a_0+i\omega e^{2A}g_5^2\Sigma^2}{f}\right)\varphi=0\,,\\
& &{a_3}''+ \left(A'+\frac{f'}{f}-\Phi '\right){a_3}'+\left(\frac{ \omega^2-e^{2A}g_5^2 \Sigma ^2f}{f^2}\right)a_3+\nonumber\\
& &\left(\frac{\omega k a_3+i k  e^{2A}g_5^2\Sigma^2f}{f^2}\right)\varphi=0\,,
 \end{eqnarray}
 and
 \begin{eqnarray}
     & &\varphi''+(3A'+\frac{f'}{f}-\Phi'+\frac{2\Sigma'}{\Sigma})\varphi'+\nonumber\\
     & &\left(\frac{\omega^2-k^2f}{f^2}\right)\varphi-i\left(\frac{\omega a_0+ kf a_3}{f^2}\right)=0.
 \end{eqnarray}
  \end{subequations}
 By using Eqs.~\eqref{EOM:coupledpion}(a), (b) and (c), the equation for the bulk field $\varphi$ can be reduced to a non-dynamical first-order differential equation which reads:

 \begin{eqnarray}\label{EOM:varphi}
 & & i k f a_3'+ i \omega a_0'-e^{2 A} \Sigma^2 g_5^2 f{\varphi}' =0.
 \end{eqnarray}

To determine the two point Green's functions, we consider the on-shell action for the longitudinal axial-vector and pseudo-scalar channels which is given by
\begin{eqnarray}\label{eq:onshellactionforaxialvector}
S_{\rm{on}}&=&-\frac{1}{2g_5^2}\int d q^4 \bigg\{e^{A(z)-\Phi(z)} \big[{a_0}(-q,z){a_0}'(q,z)-\nonumber\\
& &{a_3}(-q,z)f(z){a_3}'(q,z)\big]+e^{3 A-\Phi}g_5^2 f(z)  \varphi(-q,z) \nonumber\\ 
& &  \Sigma(z)^2\varphi'(q,z)\bigg\}_{z=\epsilon},
\end{eqnarray}
with $\epsilon$ an infinitesimally small constant representing our UV cutoff.

When the quark mass is finite, i.e. $m_q\neq 0$, the $\rm{SU}(2)$ symmetry is explicitly broken and the second order chiral phase transition turns to be a crossover. The leading order of the asymptotically solution of $\Sigma(z)$ is $m_q\zeta z$. As a result, at the boundary $z=0$, the asymptotic solutions to Eqs.~\eqref{EOM:coupledpion}  behave as
 \begin{subequations}\label{eq:boundaryvarphi}
 \begin{eqnarray}
     a_0(z)&=&a_{t0}+a_{tl}z^2\ln(z)+a_{t2}z^2+\mathcal{O}(z^3),\\
     a_3(z)&=&a_{x0}+a_{xl}z^2\ln(z)+a_{x2}z^2+\mathcal{O}(z^3),\\
     \varphi(z)&=&\varphi_{0}+\varphi_{l}z^2\ln(z)+\varphi_2 z^2+\mathcal{O}(z^3),
 \end{eqnarray}
 \end{subequations}
with
 \begin{eqnarray}
     a_{tl}&=&-\frac{1}{2}\left[\omega(a_{t0} k+a_{x0}\omega)-(m_q\zeta)^2g_5^2(  a_{x0}-i  k\varphi_0)\right],\nonumber\\ 
     a_{xl}&=&\frac{1}{2}\left[k(a_{t0}k+a_{x0}\omega)+(m_q\zeta)^2g_5^2(  a_{t0}+i  \omega\varphi_0)\right],\nonumber\\
     \varphi_l&=&\frac{1}{2}[i(a_{t0}\omega+a_{x0}k)-(\omega^2-k^2)\varphi_0],\nonumber
 \end{eqnarray}
 in which $a_{t0}$, $a_{x0}$, $a_{t2}$, $a_{x2}$, $\varphi_0$ and $\varphi_2$ are integration constants. Since the second order differential equation for $\varphi$ can be reduced to a first order equation, one can express $\varphi_2$ in terms of the other integration constants
 \begin{equation}
     \varphi_2=\frac{i(a_{t2}\omega+a_{x2}k)}{g_5^2(m_q\zeta)^2}\,.
 \end{equation}
Considering these asymptotic solutions, the on-shell action reduces to a boundary term,
 \begin{eqnarray}\label{Action:onshellmq}
S_{\rm{on}}&=&-\int dq^4\left\{\frac{a_{t0}(-q)a_{t2}(q)}{g_5^2}-\frac{a_{x0}(-q)a_{x2}}{g_5^2}+\right.\nonumber\\
& &\left.{\varphi_0(-q)\varphi_2(q)(m_q\zeta)^2}\right\}
\end{eqnarray}
where higher-order corrections are not shown.
According to the holographic dictionary, we take $a_{t0}(q)$ and $a_{x0}(q)$ as the 4D field theory sources for the axial vector current $A_\mu$ and $2\varphi_0 m_q\zeta$ as the source for the 4D pion operator $\pi$, respectively. Consequently, the subleading terms in the expansions $\frac{a_{t2}}{g_5^2}$,    $\frac{a_{x2}}{g_5^2}$ and $(1/2)\varphi_2 m_q\zeta$ represent the expectation values for the corresponding field theory operators.

Following the standard prescriptions, we derive the retarded Green's functions by taking the second derivative of the on-shell action with respect to the corresponding sources. From Eq.~\eqref{Action:onshellmq}, the correlator for the dimensionless pion field (i.e. the phase) is then given by \footnote{To verify the validity of this procedure, in appendix~\ref{appendix}, we compare our results with those in Refs.~\cite{Son:2002ci,Son:2001ff}.}
\begin{eqnarray}\label{Eq:correlatorpi}
G_{\varphi\varphi}(q)&=&\frac{\delta S_{\rm{on}}}{2m_q^2 \bar{\sigma}^2\delta \varphi_0(-q)\delta \varphi_0(q)}\nonumber\\
&=&\frac{1}{(2m_q\bar{\sigma})^2}\frac{2\varphi_2(q) (m_q\zeta)^2}{\varphi_0(q)}\nonumber\\
&=&\frac{1}{(2m_q\bar{\sigma})^2}\frac{2i[a_{t2}(q)\omega+a_{x2}(q)k]}{g_5^2\varphi_0(q)}.
\end{eqnarray}

In order to compute the retarded correlator, the appropriate boundary conditions at the horizon must be imposed \cite{Son:2002sd}. For that purpose, we consider ``incoming'' boundary conditions at the horizon, $z=z_h$. We then obtain the solutions close to the horizon which are given by
\begin{subequations}\label{eq:horizonvarphi}
\begin{eqnarray}
& &a_0(z)=(z_h-z)^{-\frac{i\omega}{4\pi T}}\bigg \{-\frac{4(a_{b0}k z_h^2+4 i c_0^2\pi^2\varphi_{b0})}{(4i+z_h\omega)z_h^2}(z_h-z)\nonumber\\
 & &\ \ \ \ \ \  +\mathcal{O}[(z_h-z)^2]\bigg\}-i\varphi_{b1}\omega,\\
& &a_3(z)=(z_h-z)^{-\frac{i\omega}{4\pi T}}\bigg\{a_{b0}+\bigg\{\frac{4\pi^2 c_0^2\varphi_{b0}k}{4i+\omega z_h}+a_{b0}z_h\omega\nonumber\\
& &\ \ \ \ \ \ \bigg[ \frac{z_h^2k^2}{4i+\omega z_h}+\frac{4c_0^2\pi^2}{z_h\omega}+\frac{i(8z_h^2\mu_g^2+3iz_h\omega-2)}{4}\bigg]\bigg\}\bigg/\nonumber\\
& &\ \ \ \ \ \ \bigg[2z_h(2-i\omega z_h)\bigg](z_h-z)+\mathcal{O}[(z_h-z)^2]\bigg\}+i\varphi_{b1} k,\nonumber\\
\\
& &\varphi(z)=(z_h-z)^{-\frac{i\omega}{4\pi T}}\bigg\{\varphi_{b0}+\bigg\{\frac{2(c_0^2\pi^2\varphi_{b0}\omega-a_{b0}z_h k)}{(2i+z_h\omega)(4-i z_h\omega)}+\nonumber\\
& &\ \ \ \varphi_{b0}\bigg[\frac{4iz_h k^2-c_0^2\lambda\omega+z_h(2z_h\mu_c^2-8z_h\mu_g^2-3i\omega)\omega}{8(2i+z_h\omega)}\bigg]\bigg\}\nonumber\\
& &\ \ \ \ \ \ (z_h-z)+\mathcal{O}[(z_h-z)^2]
\bigg\} +\varphi_{b1},
\end{eqnarray}
\end{subequations}
where $a_{b0}$, $\varphi_{b0}$ and $\varphi_{b1}$ are integration constants. Combing the boundary and horizon expansions, Eqs.~\eqref{eq:boundaryvarphi} and ~\eqref{eq:horizonvarphi}, one can numerically find unique solutions for bulk EOMs in Eqs.~\eqref{EOM:coupledpion}.

From the holographic model, one can extract further information about the dynamics of pions. First, the pion decay constant is defined as $f_\pi q_\mu=\langle 0|A_\mu|\pi(q)\rangle$ at zero temperature, in which $| 0\rangle $ is the vacuum state and $|\pi\rangle$ the one-pion state. For the zero temperature case, there are some studies in the holographic AdS/QCD models~\cite{erlich_qcd_2005, MartinContreras:2021yfz}. At finite temperature, Lorentz symmetry is broken. Therefore, the pion decay constant at finite temperature splits into a time component $f_t$  and a space one $f_s$. Following the analysis in Ref.~\cite{erlich_qcd_2005}, the generalized pion decay constants are given by
\begin{eqnarray}\label{definition:fpi}
    f_t^2=-\frac{2 a_{t2}(0)}{a_{t0}(0)g_5^2},\ \ \ \ 
    f_s^2=-\frac{2a_{x2}(0)}{a_{x0}(0)g_5^2}.
\end{eqnarray}

By looking at the solution around the boundary, one can find that in  Ref.~\cite{erlich_qcd_2005} the term ($F[q,z]/m_q$) appears. This term is clearly divergent in the chiral limit, $m_q$ approaching zero, and therefore the results of Ref.~\cite{erlich_qcd_2005} must be revisited for zero quark mass.

In order to do that, we observe that, in the chiral limit, the expansions of the solutions at the horizon remain unchanged. Since the leading term of the expansion for $\Sigma$ becomes $(\bar{\sigma}/\zeta) z^3$ setting $m_q=0$, the asymptotic solution of $\varphi$ turns to be
 \begin{eqnarray}\label{eq:boundaryvarphi1}
     \varphi(z)&=&z^{-2}\left\{\bar{\varphi}_{0}+\frac{1}{2}(q^2-2\mu_g^2+\mu_c^2) \bar{\varphi}_0 z^2\ln(z)+\right.\nonumber\\
     & & \left.\bar{\varphi}_2 z^2+\mathcal{O}(z^3)\right\},
 \end{eqnarray}
 with $\bar{\varphi}_0=-\frac{i(a_{t2}\omega+a_{x2}k)}{g_5^2(\bar{\sigma}/\zeta)^2}.$ Thus, the on-shell action becomes
  \begin{eqnarray}\label{Action:onshellmq}
S'_{\rm{on}}&=&-\int dq^4\bigg\{\frac{a_{t0}(-q)a_{t2}(q)}{g_5^2}-\frac{a_{x0}(-q)a_{x2}}{g_5^2}-\nonumber\\
& &{\bar{\varphi}_0(-q)\bar{\varphi}_2(q)(\bar{\sigma}/\zeta)^2}\bigg\}.
\end{eqnarray}
According to the holographic dictionary, the quantity $2\bar{\varphi}_0(q)\bar{\sigma}/\zeta$ has to be identified as the source for pion operator and $\bar{\sigma}/(2/\zeta)\bar{\varphi}_2(q)$ as its expectation value. Following the same steps as before, the correlator for the phase field $\varphi$ is given by\footnote{In appendix~\ref{appendix}, we have verified that our holographic results in the chiral limit are consistent with the thermal chiral effective field theory (EFT) analysis of  Ref.~\cite{Son:2002ci,Son:2001ff}.}
\begin{eqnarray}\label{Eq:correlatorpi}
G_{\varphi\varphi}(q)&=&\frac{\delta S'_{\rm{on}}}{4(\bar{\sigma}/\zeta)^4 \delta \bar{\varphi}_0(-q)\delta \bar{\varphi}_0(q)}\nonumber\\
&=&-\frac{1}{(\bar{\sigma}/\zeta)^2}\frac{\bar{\varphi}_2(q)}{2\bar{\varphi}_0(q)}\nonumber\\
&=&\frac{\bar{\varphi}_2(q)g_5^2}{2i[a_{t2}(q)\omega+a_{x2}(q)k]}.
\end{eqnarray}

\section{Results}
\label{sec:results}
Based on the above discussion, we are now in the position to explore all the features of our holographic model. In this section, we will present the main results of our work.

\subsection{The chiral condensate and phase transition}
\begin{figure}[!h]
  \centering
  \includegraphics[width=3.22in]{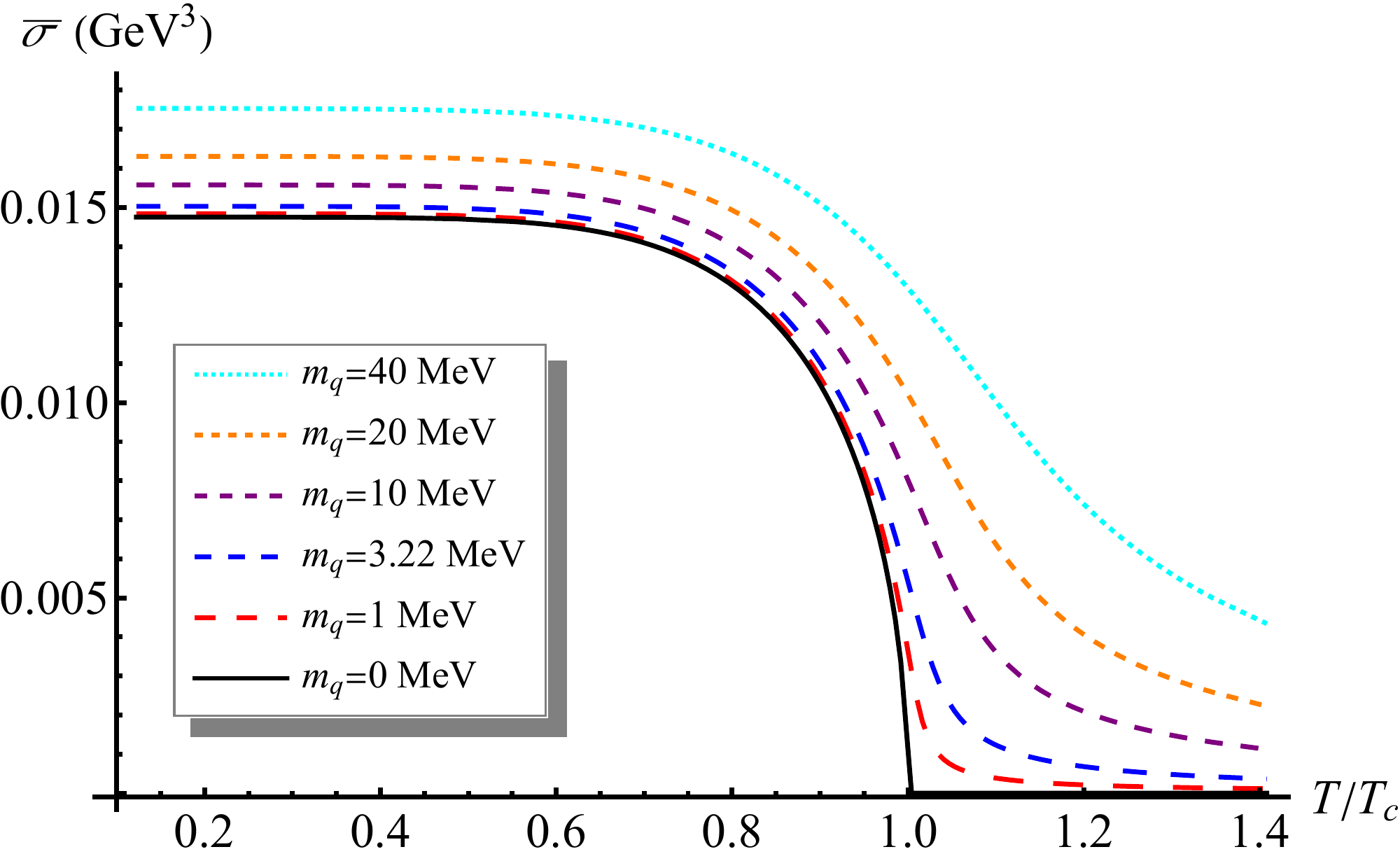}
  \caption{The chiral condensate $\bar{\sigma}\equiv \langle \bar{\psi}\psi\rangle$ as a function of the $T_c$ normalized temperature $T/T_c$, for different values of the current quark mass $m_q$. The black solid line represents the result in the chiral limit, which implies a second order phase transition at $T_c=163\rm{MeV}$. The red, blue, purple, orange dashed and cyan dotted lines represent the results for $m_q=1, 3.22, 10, 20, 40 \rm{MeV}$ respectively, and show a smooth crossover.}\label{fig1}
\end{figure}

The Goldstone modes are tightly related with the spontaneous chiral symmetry breaking, whose order parameter is given by the chiral condensate $\bar \sigma \equiv \langle \bar \psi \psi \rangle$. In Fig.\ref{fig1}, we show the behavior of the chiral condensate $\bar \sigma$ in function of the reduced temperature $T/T_c$ for different values of the quark mass $m_q$. The result in the chiral limit, $m_q=0$, corresponds to the solid black line therein. In that limit, the chiral condensate $\bar{\sigma}$ is finite at low temperature and reaches a constant value $\approx0.015 \rm{GeV}^3$. At the critical temperature, $T_c\approx 0.163 \rm{MeV}$, the chiral condensate vanishes following a square root law, $\propto \sqrt{T_c-T}$, typical of mean field second order phase transitions. Above the critical temperature, we have the chiral symmetry restored phase with zero condensate, usually also labelled as the \textit{normal phase}.

Then, we assume a finite and increasing quark mass, $m_q=1,3.22,10,20,40 \,\rm{MeV}$\footnote{Here, when $m_q=3.22\,\rm{MeV}$ (blue line), the pion mass in the vacuum is around the physical value $139.5\,\rm{MeV}$. Therefore, we will consider this value as the ``physical quark mass''.}, and re-compute the chiral condensate in those cases. The results are shown by the red, blue, purple, orange, cyan lines in Fig.\ref{fig1} respectively. There, we can see that when the quark mass becomes finite, the chiral condensate at low temperature becomes larger than its value in the chiral limit. This might be due to the contribution from the explicit symmetry breaking. Moreover, the condensate $\bar{\sigma}$ does not vanish at any temperature. A finite tail appears in the high temperature region, which could be considered as the pure contribution from the quark masses. Near $T_c$, we could see that the transition from the broken phase to the symmetry restored phase is no longer sharp. Instead, it becomes a continuous crossover. This is reasonable, since the finite quark mass breaks the exact symmetry explicitly rendering the notion of spontaneous symmetry breaking imprecise.\footnote{Notice how that is not necessarily always the case. See for example the discussions in \cite{Ammon:2021pyz,Andrade:2020hpu}.}

\subsection{Pions in the chiral limit}

\begin{figure}[ht!]
    \centering
    \begin{overpic}[width=.8\linewidth]{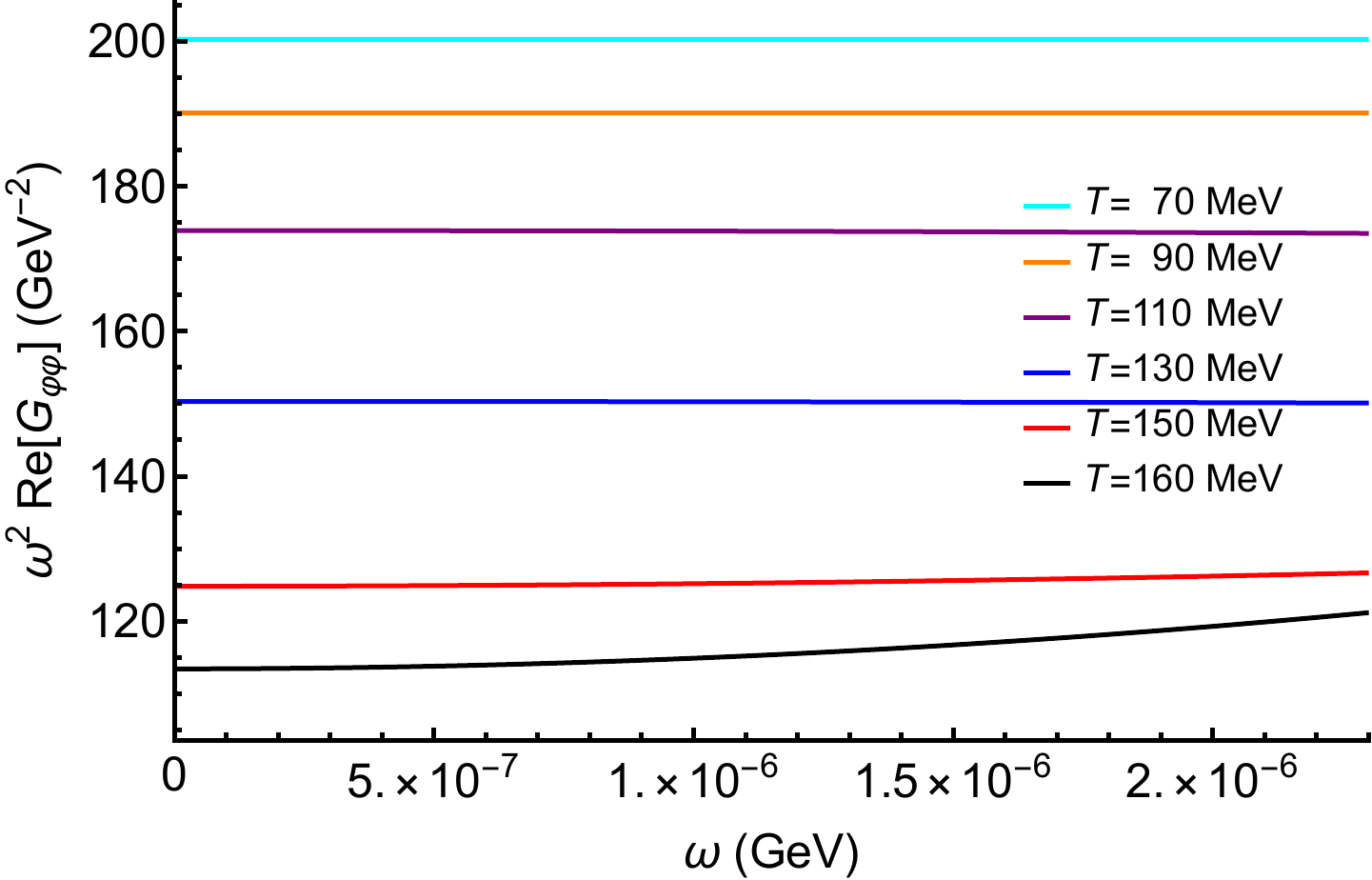}
  \put(85,63){\bf{(a)}}
    \end{overpic}
   
   \vspace{0.2cm}
   
    \begin{overpic}[width=.8\linewidth]{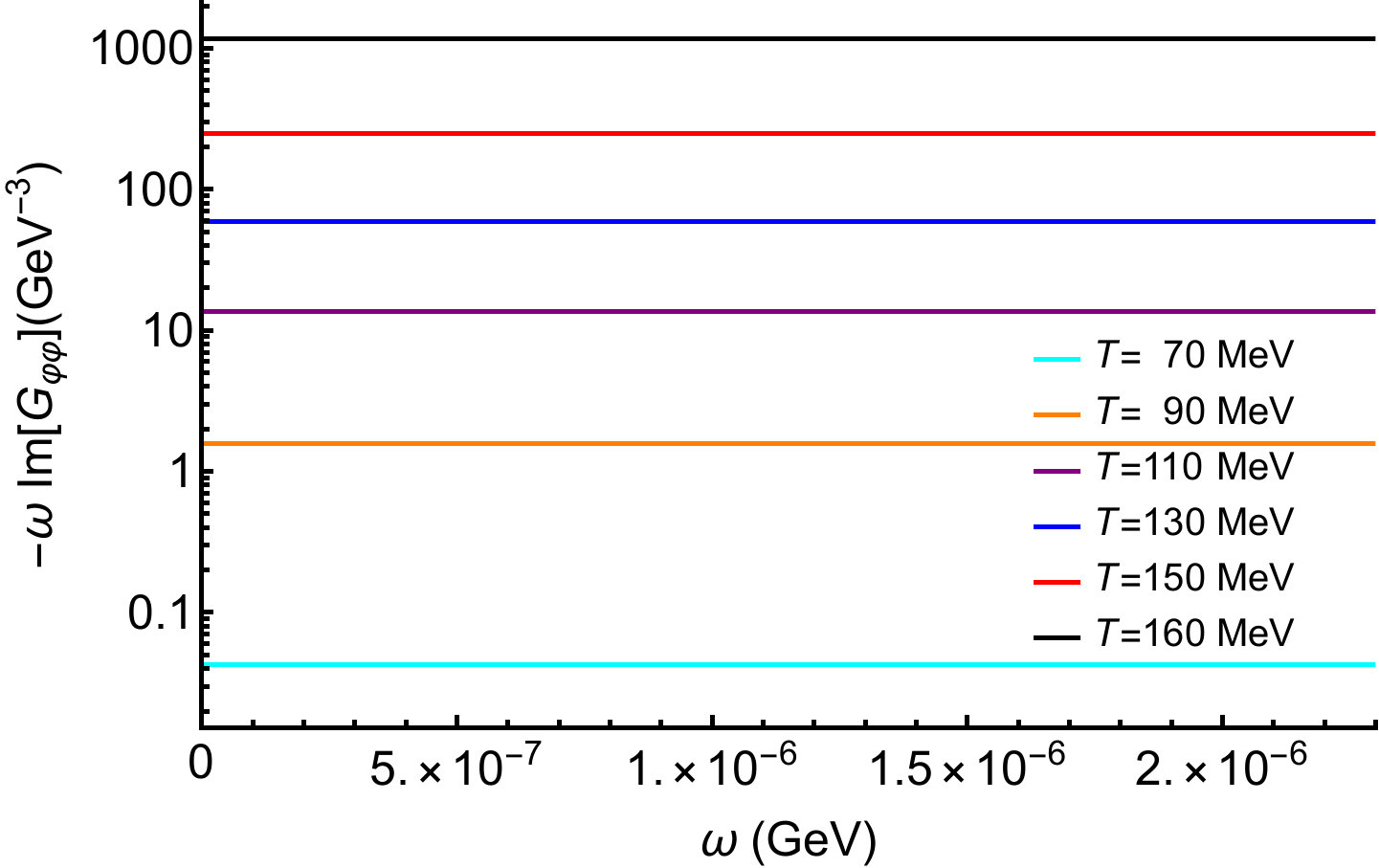}
        \put(85,62){\bf{(b)}}
    \end{overpic}
    
   \vspace{0.2cm}
   
    \begin{overpic}[width=.75\linewidth]{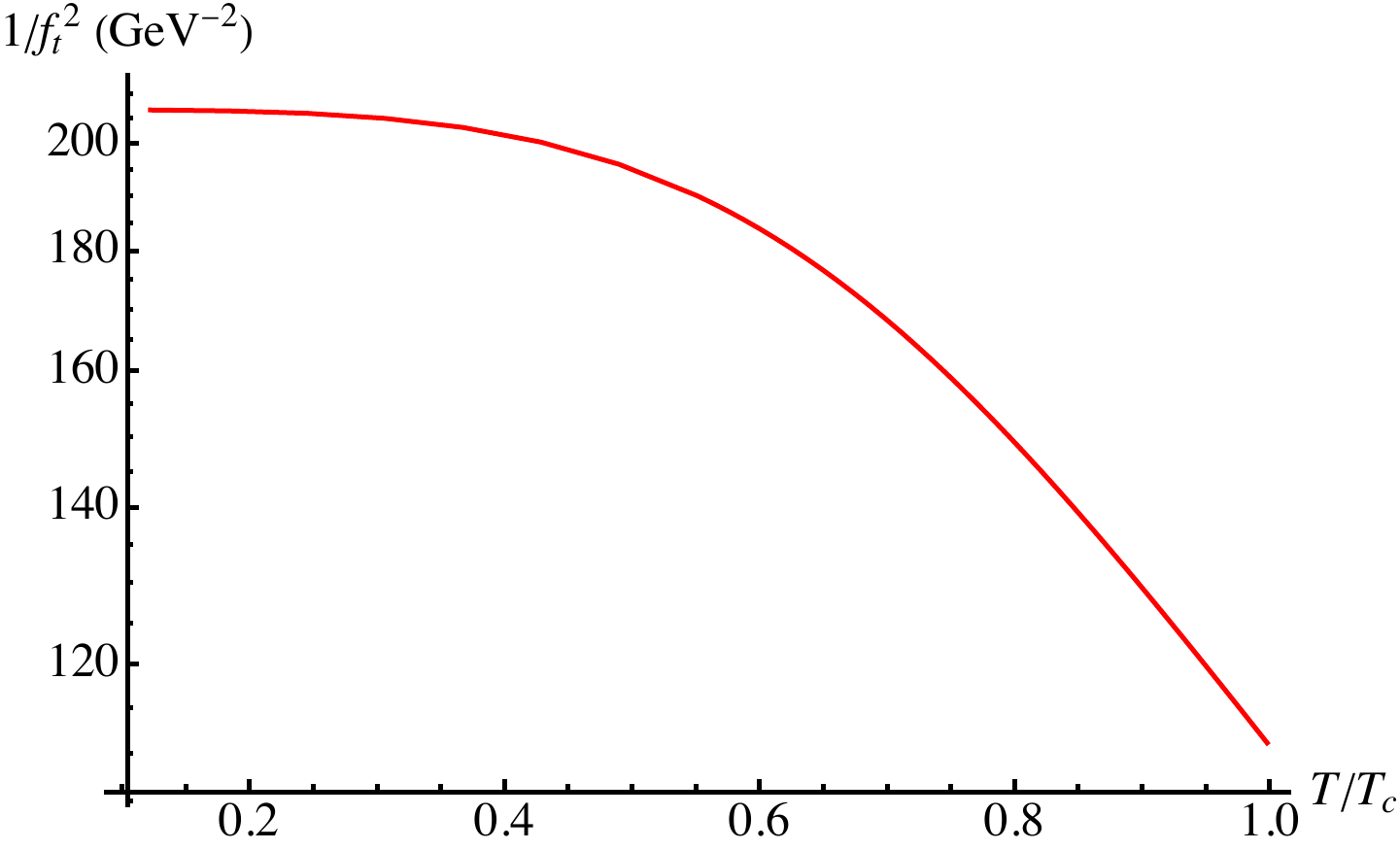}
        \put(90,50){\bf{(c)}}
    \end{overpic}
    
   \vspace{0.2cm}
   
    \begin{overpic}[width=.75\linewidth]{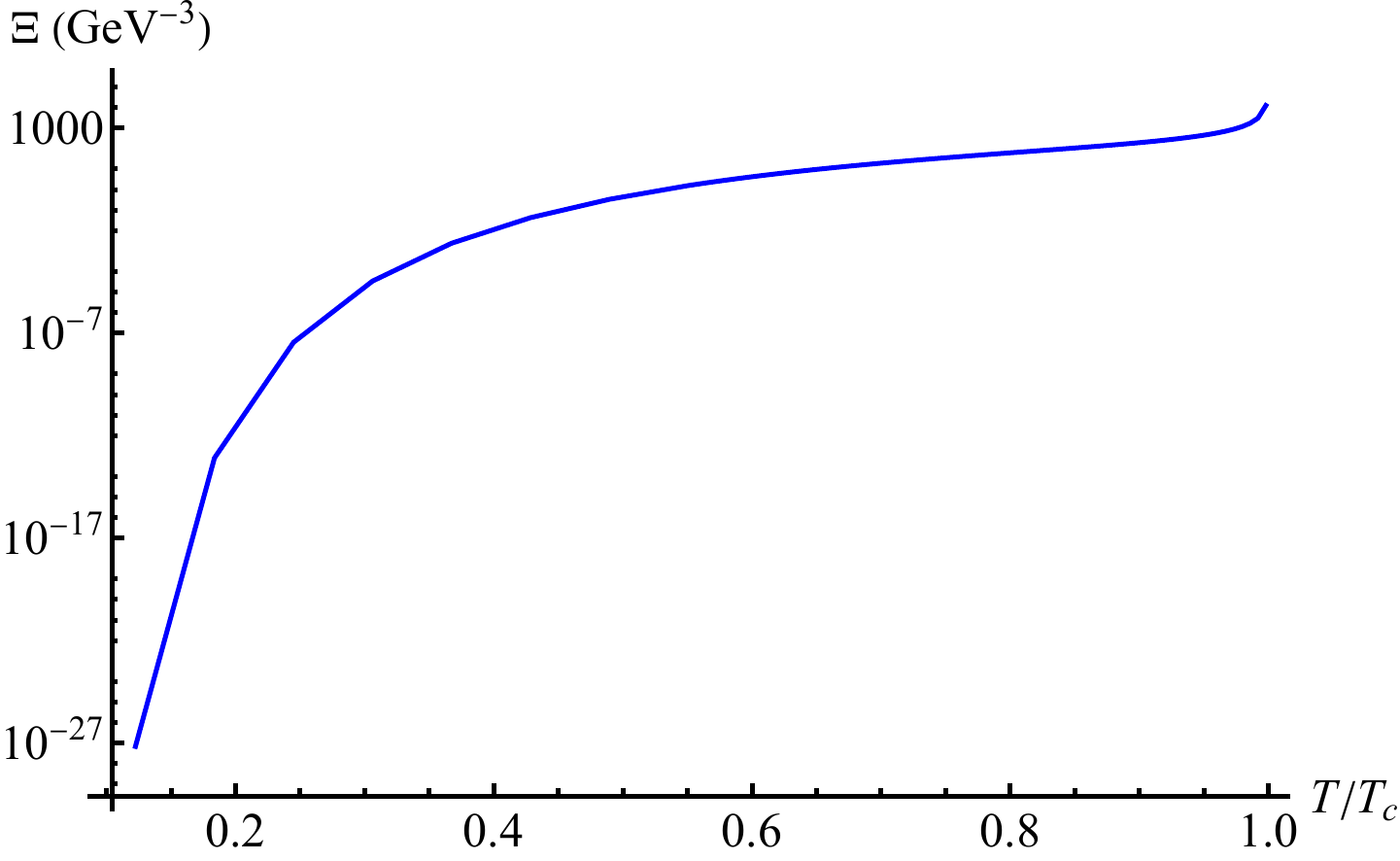}
        \put(90,50){\bf{(d)}}
    \end{overpic}
    \caption{The low frequency behavior of the Goldstone correlator. \textbf{(a)} $\omega^2{\rm{Re}[G_{\varphi\varphi}]}$ and \textbf{(b)} $-\omega{\rm{Im}}[G_{\varphi\varphi}]$ as a function of $\omega$ at $k=0$. The thermodynamic susceptibilities extracted from the low-frequency limit of the dynamical Goldstone correlator, Eq.\eqref{green1}. \textbf{(c)} The dimensionless axial charge susceptibility $1/f_t^2$ and \textbf{(d)} the Goldstone diffusivity $\Xi$ as a function of the reduced temperature $T/T_c$.}
    \label{fig:2}
\end{figure}

After extracting the information about the chiral phase transition from the order parameter, we continue to investigate the dynamical and thermal properties of the low-energy (quasi-)particles. In this section, we will focus on the case with massless quarks. We start by studying the properties of the Goldstone modes, the pions.

From Eq.~\eqref{eq:Goldstonediffusivity}, the pion diffusivity can be obtained from the ratio of the dissipative coefficient $\Xi$ and the Goldstone susceptibility $\chi_{\varphi\varphi}$. These two quantities can be extracted from the temporal correlations in Eq.~\eqref{stat}.  

We take $m_q=0$ and compute the dynamical correlation function of the Goldstone modes $G_{\varphi\varphi}(\omega, k=0)$ numerically. The results are shown in panels (a) and (b) of Fig.\ref{fig:2}, where we have taken $T=70, 90, 110, 130, 150, 160 \rm{MeV}$ (cyan, orange, purple, blue, red, black lines respectively). In order to show the scaling behavior near $\omega=0$, we have rescaled the real part and the imaginary part of $G_{\varphi\varphi}$ with additional factors of $\omega^2$ and $\omega$ respectively. From the figures, one can easily seen that both the rescaled real and imaginary parts are almost constant when $\omega$ is small. Thus, we have  
\begin{equation}
    \mathrm{Re}G_{\varphi\varphi}(\omega,k=0) \propto \omega^{-2}\,,\qquad \mathrm{Im}G_{\varphi\varphi}(\omega,k=0) \propto \omega^{-1}\,,
\end{equation}
which is exactly of the same form as Eq.\eqref{green1}. 

Then, $f_t^2$ and $\Xi$ could be obtained numerically from the values of the reduced real and imaginary parts in the limit of small frequency. The numerical results are given in Fig.~\ref{fig:2}(c) and (d) respectively. From Fig.\ref{fig:2}(c), we could see that, at low temperature,  $1/f_t^2(T=20 \rm{MeV})=1/0.0048\ \ (\rm{GeV})^{-2}$, which is very close to the square of the zero temperature value, $1/f_t^2\approx 1/0.0049\  {\rm{(GeV)}}^{-2}$~\cite{FANG201686}. Then, it decreases with increasing the temperature and reaches a finite value around  $1/f_t^2(T=163\rm{MeV})=1/0.0090\ (\rm{GeV})^{-2}$ near $T_c$. This implies that $f_t$ is finite at $T_c$, which is consistent with the analysis in Ref.~\cite{Son:2002ci, Son:2001ff}. From Fig.\ref{fig:2}(d), $\Xi$ vanishes at low temperature, as expected for any dissipative coefficient.  Along with the increasing of temperature, $\Xi$ increases and becomes of order $10^4 (\rm{GeV^{-3})}$ near $T_c$. 

\begin{figure}
    \centering
    \begin{overpic}[width=.85\linewidth]{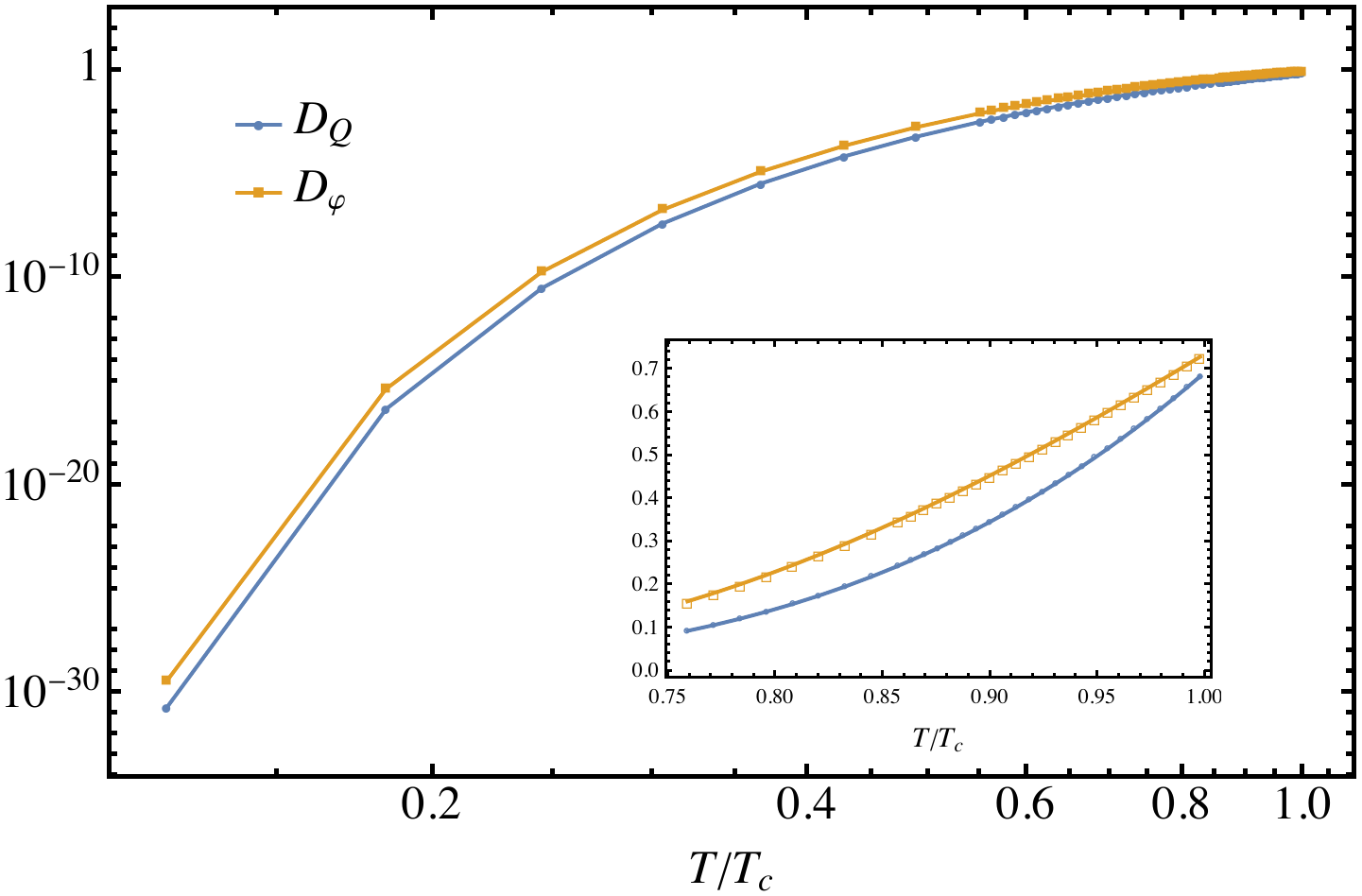}
    \end{overpic}
    \caption{The reduced temperature dependence of the dimensionless diffusivities $D_{\varphi}$ and $D_Q$. The inset is a zoom for $T/T_c$ in the range $[0.75,1]$.}
    \label{fig:3}
\end{figure}

\begin{figure}[ht!]
    \centering
    \begin{overpic}[width=.8\linewidth]{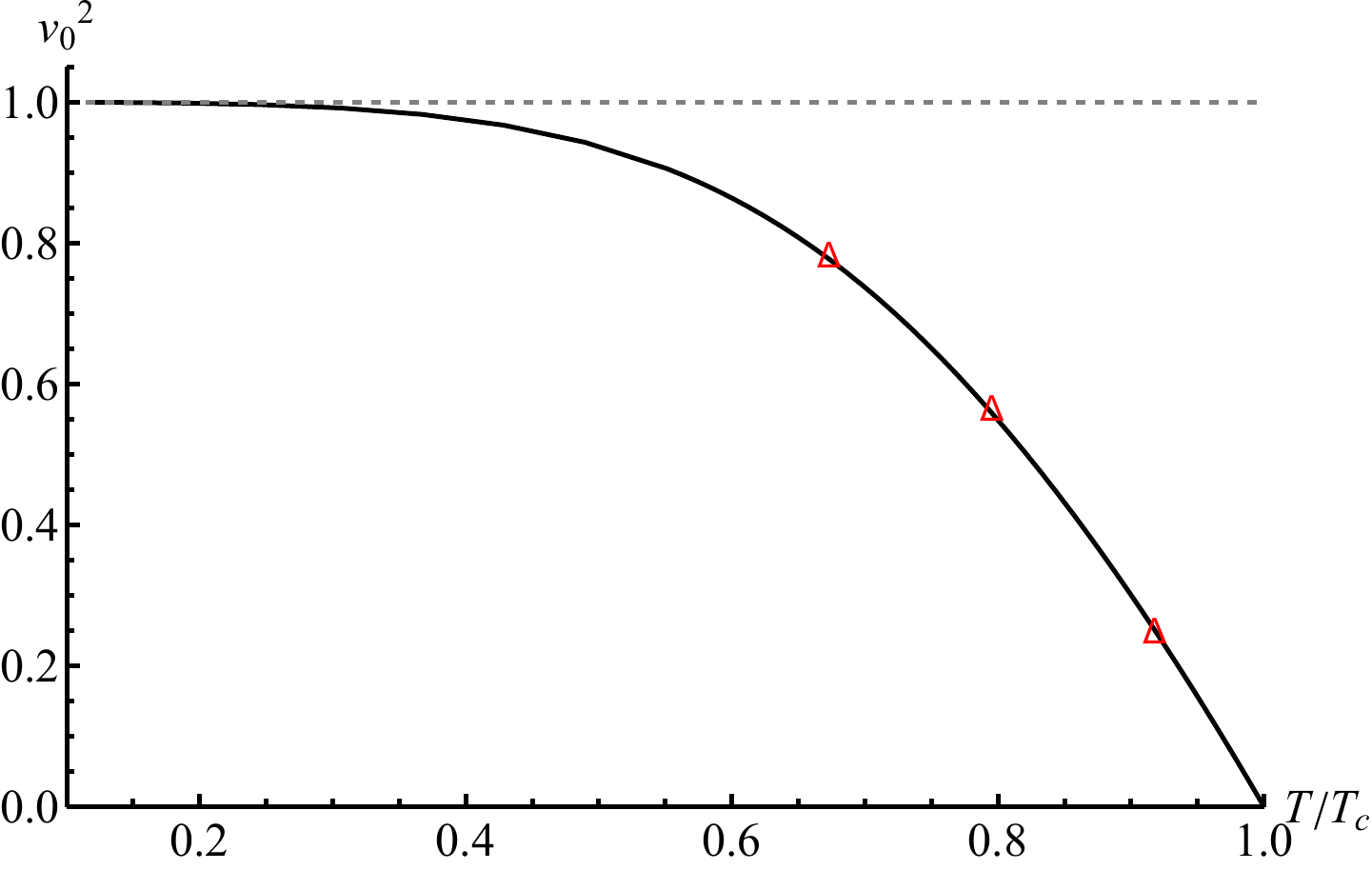}
        \put(85,57){\bf{(a)}}
    \end{overpic}
    
    \vspace{0.4cm}
   
    \begin{overpic}[width=.85\linewidth]{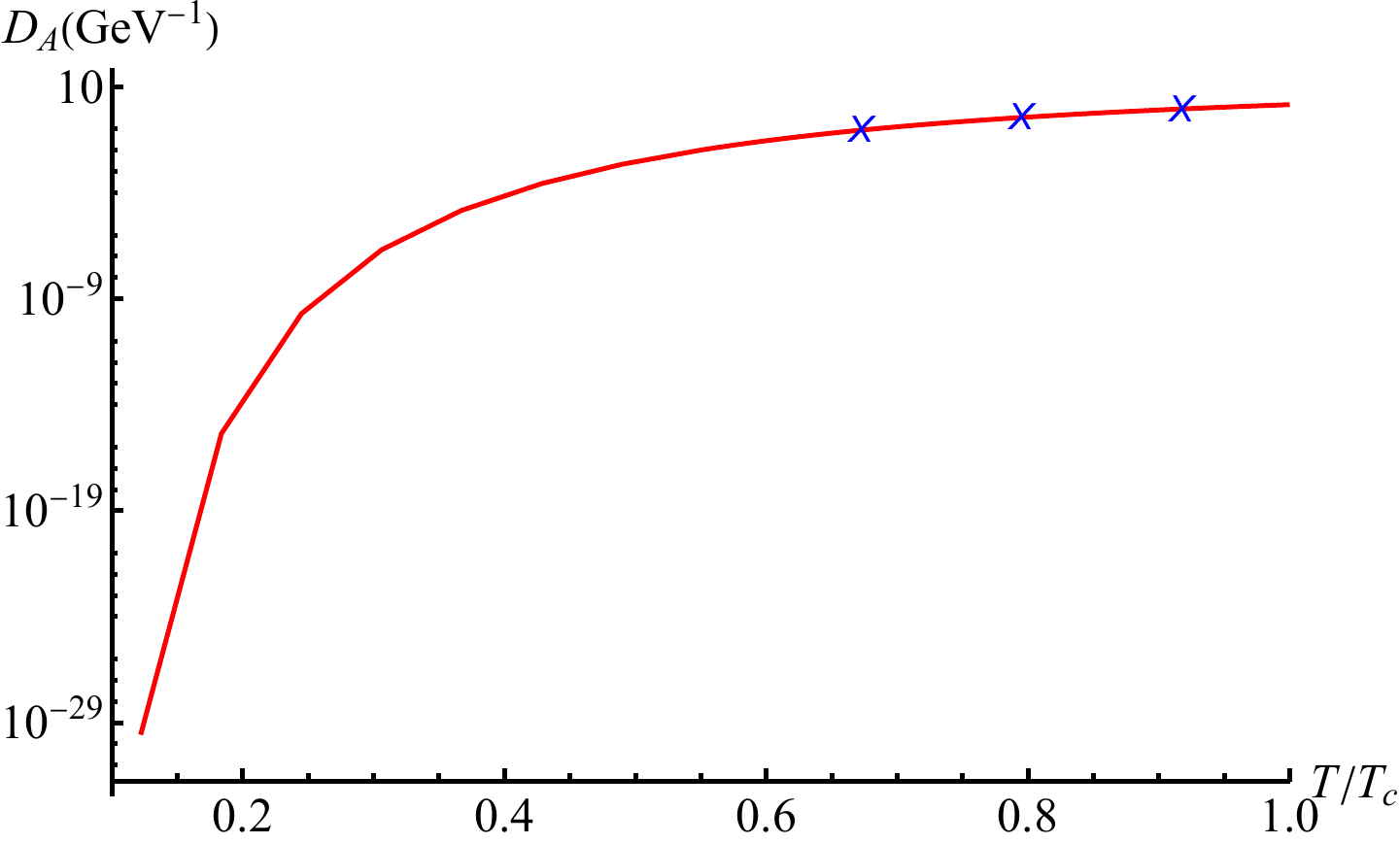}
        \put(85,55){\bf{(b)}}
    \end{overpic}
   
    \caption{\textbf{(a)} The pion velocity $v_0^2$ in the chiral limit in function of the reduced temperature $T/T_c$. The horizontal gray dashed line is as a guide to the eye. \textbf{(b)} The attenuation constant $D_A$ in function of the reduced temperature $T/T_c$. The solid line are the values extracted using the quantities appearing in the various correlators. The symbols are the values extracted by fitting the numerical dispersion relations in Fig.\ref{fig:3p}.}
    \label{fig:4}
\end{figure}

\begin{figure}[ht!]
    \centering
    \begin{overpic}[width=.8\linewidth]{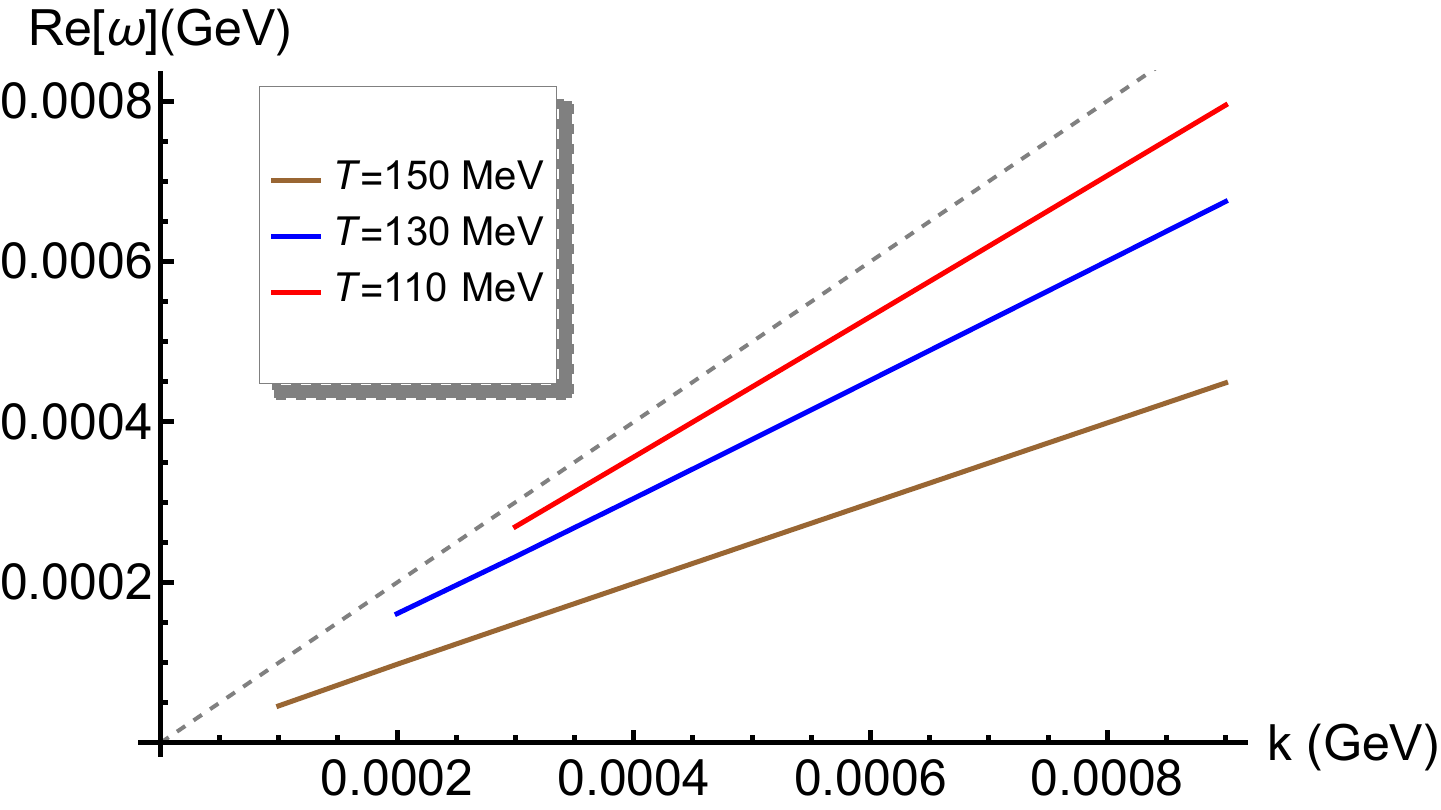}
        \put(65,50){\bf{(a)}}
    \end{overpic}
    
    \vspace{0.4cm}
   
    \begin{overpic}[width=.85\linewidth]{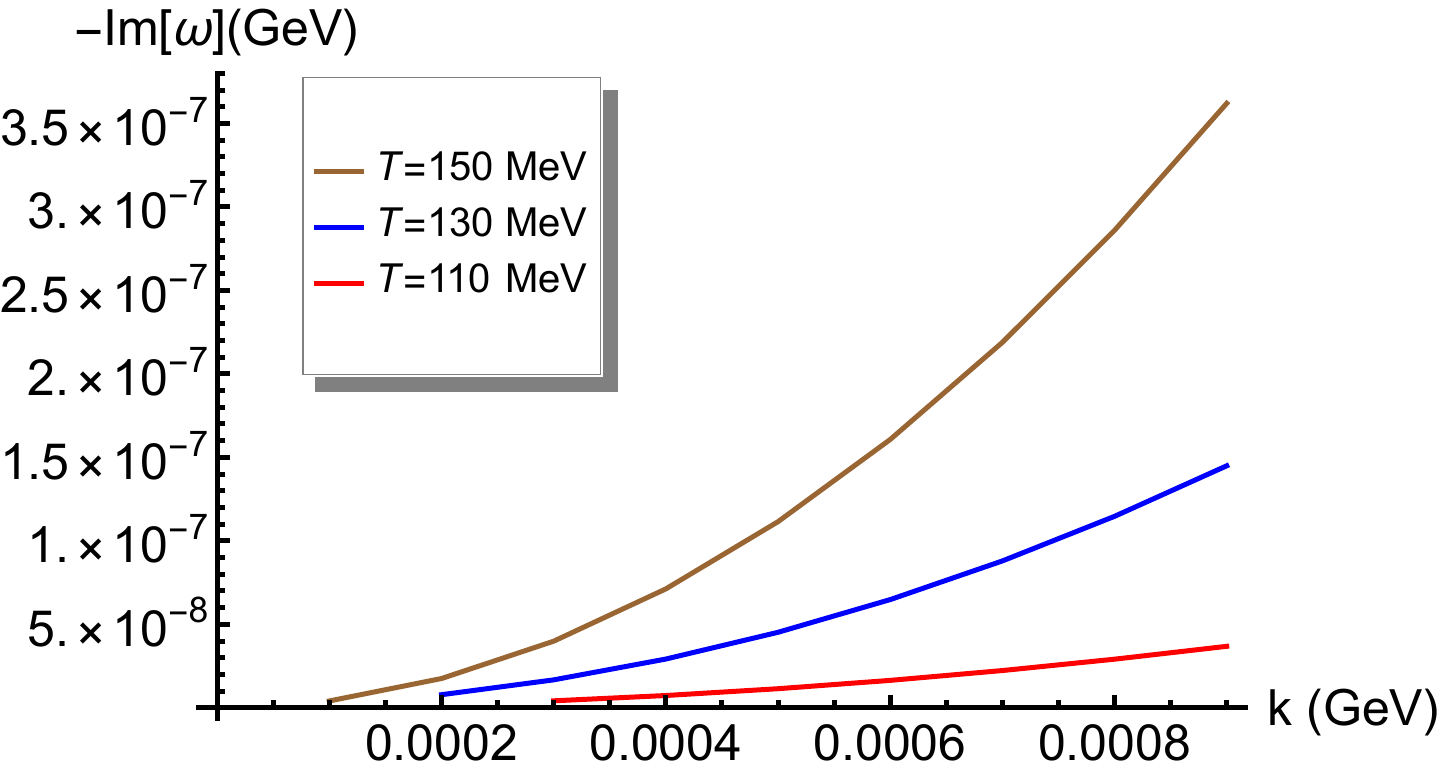}
        \put(65,50){\bf{(b)}}
    \end{overpic}
    \caption{\textbf{(a)} The real and \textbf{(b)} imaginary part of the dispersion relation in the chiral limit. The gray dashed line guides the eyes towards the light-like dispersion, ${\rm{Re}}[\omega]=k$.}
    \label{fig:3p}
\end{figure}

\begin{figure}[ht!]
    \centering
        \begin{overpic}[width=.85\linewidth]{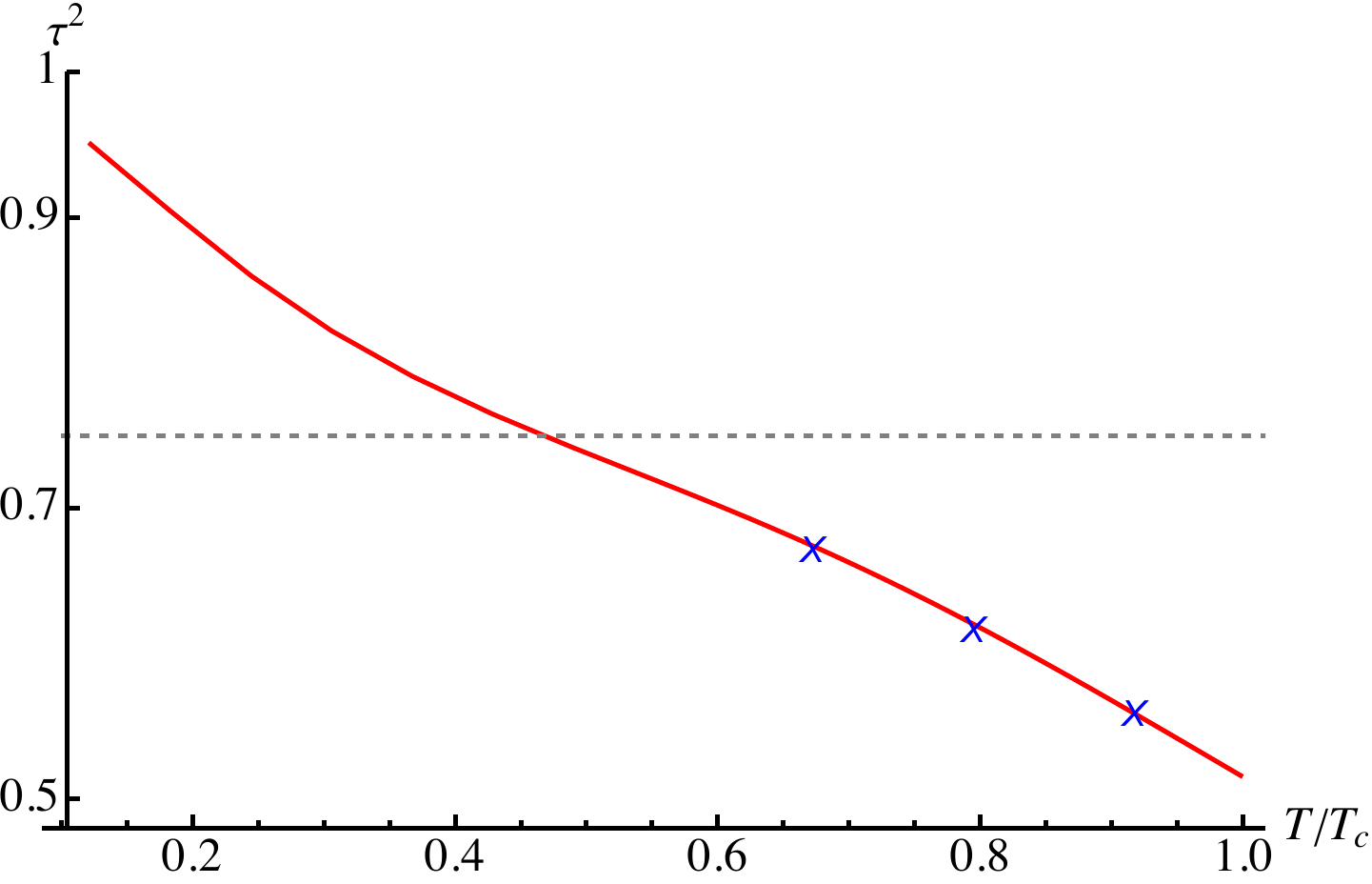}
    \end{overpic}
    \caption{The value of the $\mathfrak{r}^2$ parameter defined in Eq.\eqref{parameterr}. The symbols are the fitting results from the dispersion relations shown in Fig.\ref{fig:3p}. The dashed line indicates the chiral perturbation theory value $\mathfrak{r}^2=3/4$ \cite{Torres-Rincon:2022ssx}.}
    \label{fig:4p}
\end{figure}
After obtaining numerically the susceptibility $f_t$ and the dissipative coefficient $\Xi$, it is straightforward to get the Goldstone diffusivity $D_{\varphi}$ from their ratio. In Fig.~\ref{fig:3}, we show the Goldstone diffusivity $D_\varphi$ together with the charge diffusivity $D_Q$ extracted from the ratio between the conductivity $\sigma_Q$ in Eq.~\eqref{eq:sigmaQ} and the corresponding susceptibility $f_t$. The two diffusivities display a similar behavior as a function of the reduced temperature $T/T_c$. More precisely, they vanish at low temperature and they approach a constant value close to the critical point $T \approx T_c$, \{$D_{\varphi}(T=163\rm{MeV})=0.7269 (\rm{Gev}^{-1})$ , $D_{Q}(T=163\rm{MeV})=0.6815  (\rm{Gev}^{-1})$\}. \footnote{In general, the sound attenuation constant $D_A$ is expected to scale as $D_A\propto (1-T/T_c)^{\nu(d-4)/2}$ \cite{Son:2002ci} where $\nu$ is the critical exponent determining the class of universality of the critical point. The holographic model we used in this paper belongs to the mean field universality class, i.e. ${\nu(d-4)/2}=0$. Therefore we consistently find that $D_A \rightarrow \text{const.}$ for $T \rightarrow T_c$.}

We then move to the discussion of the pion velocity and attenuation constant, $v_0^2, D_A$ (where the subfix $0$ indicates the chiral limit). According to the definition of the sound velocity in Eq.~\eqref{eq:soundvelocity1}, we extract the results and show them in Fig.\ref{fig:4}(a). The pion velocity approaches the speed of light at low temperature and, as expected, it goes to zero at the critical temperature as:
\begin{equation}
    v_0^2 \propto (T_c-T)\,.
\end{equation}
The attenuation constant, $D_A=D_\varphi+D_Q$, is shown in the panel (b) of Fig.\ref{fig:4}. It goes from a constant value at the critical point $T_c$ to zero in the limit of $T \rightarrow 0$.

To confirm our computations, we numerically obtain the dispersion relation of the pions in the chiral limit. The results are shown in Fig.\ref{fig:3p} for three different temperatures. We find that the dispersion relation is consistently described by Eq.\eqref{didi} in the small wave-vector limit. By fitting the dispersion relations, we can then find the value of the attenuation constant $D_A$ and the sound speed $v_0$. The results are shown with colored symbols in Fig.\ref{fig:4} and are in perfect agreement with the predictions from the correlators (solid lines therein).

Finally, in Fig.\ref{fig:4p}, we show the $\mathfrak{r}$ parameter defined in Eq.\eqref{parameterr}. It is predicted to be equal to $3/4$ in the limit of zero temperature using chiral perturbation theory (see Ref.~\cite{Torres-Rincon:2022ssx}. In our holographic model, we find that $\mathfrak{r}^2$ approaches $1$ in the zero temperature limit. It monotonously decreases by increasing temperature and reaches a constant value $\approx 0.5$ in the limit $T \rightarrow T_c$. We will comment about this outcome in the conclusions.

\subsection{Pions as pseudo-Goldstone modes}

\begin{figure}
\centering
    \begin{overpic}[width=0.8\linewidth]{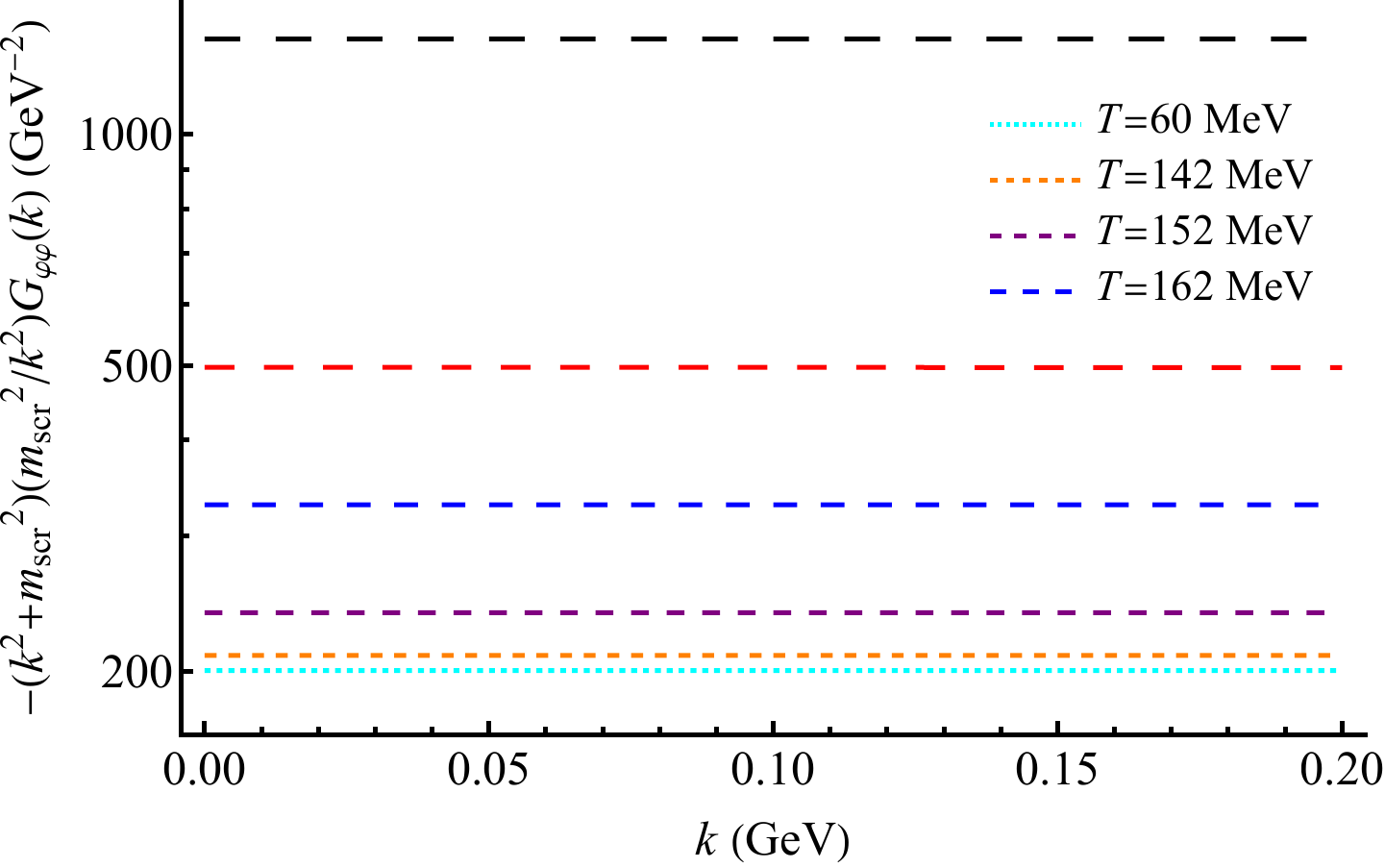}
        \put(80,62){\bf{(a)}}
    \end{overpic}
    
    \vspace{0.2cm}

    \begin{overpic}[width=0.8\linewidth]{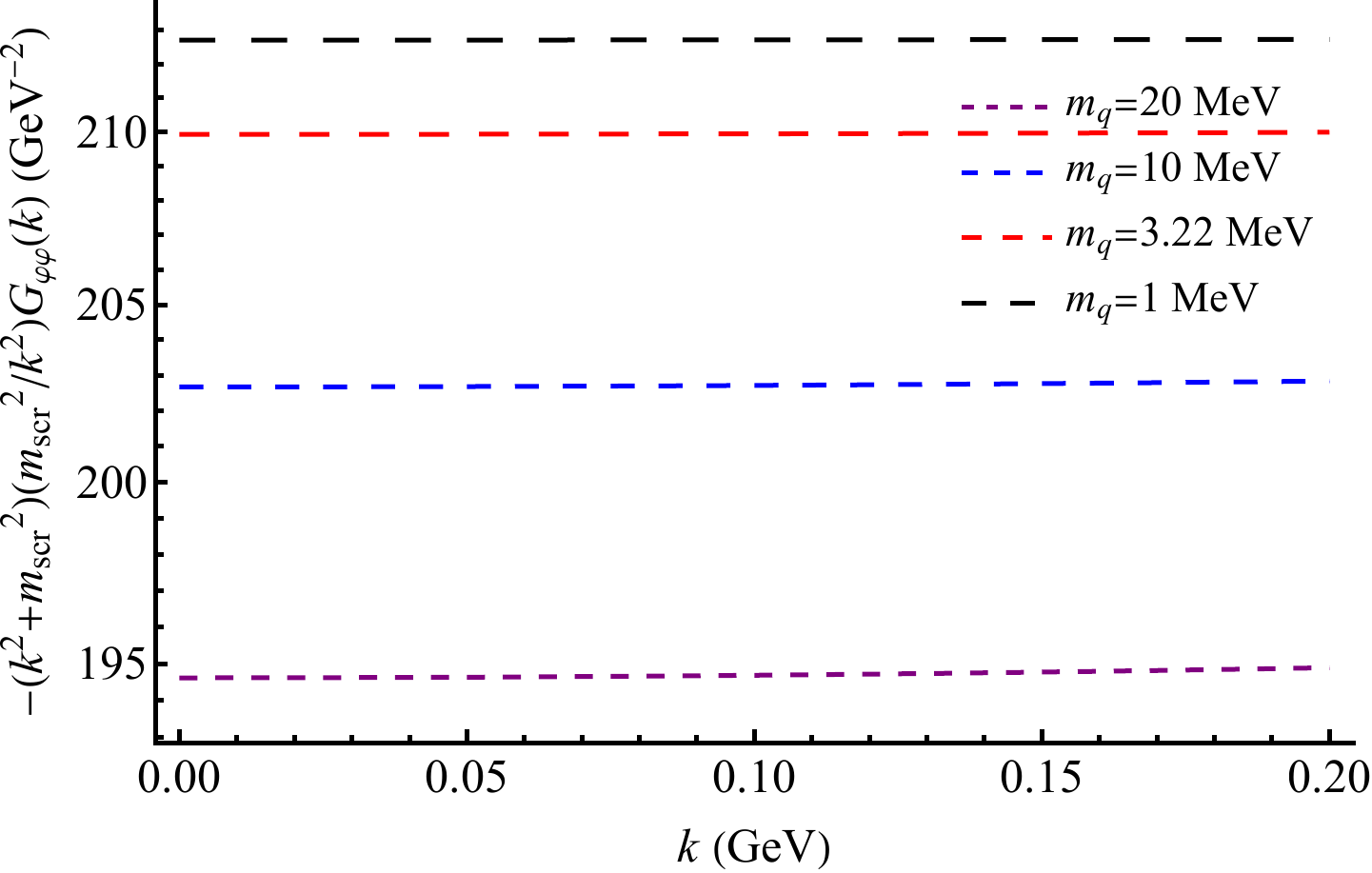}
    \put(80,62){\bf{(b)}}
    \end{overpic}
    
    \vspace{0.2cm}

    \begin{overpic}[width=0.8\linewidth]{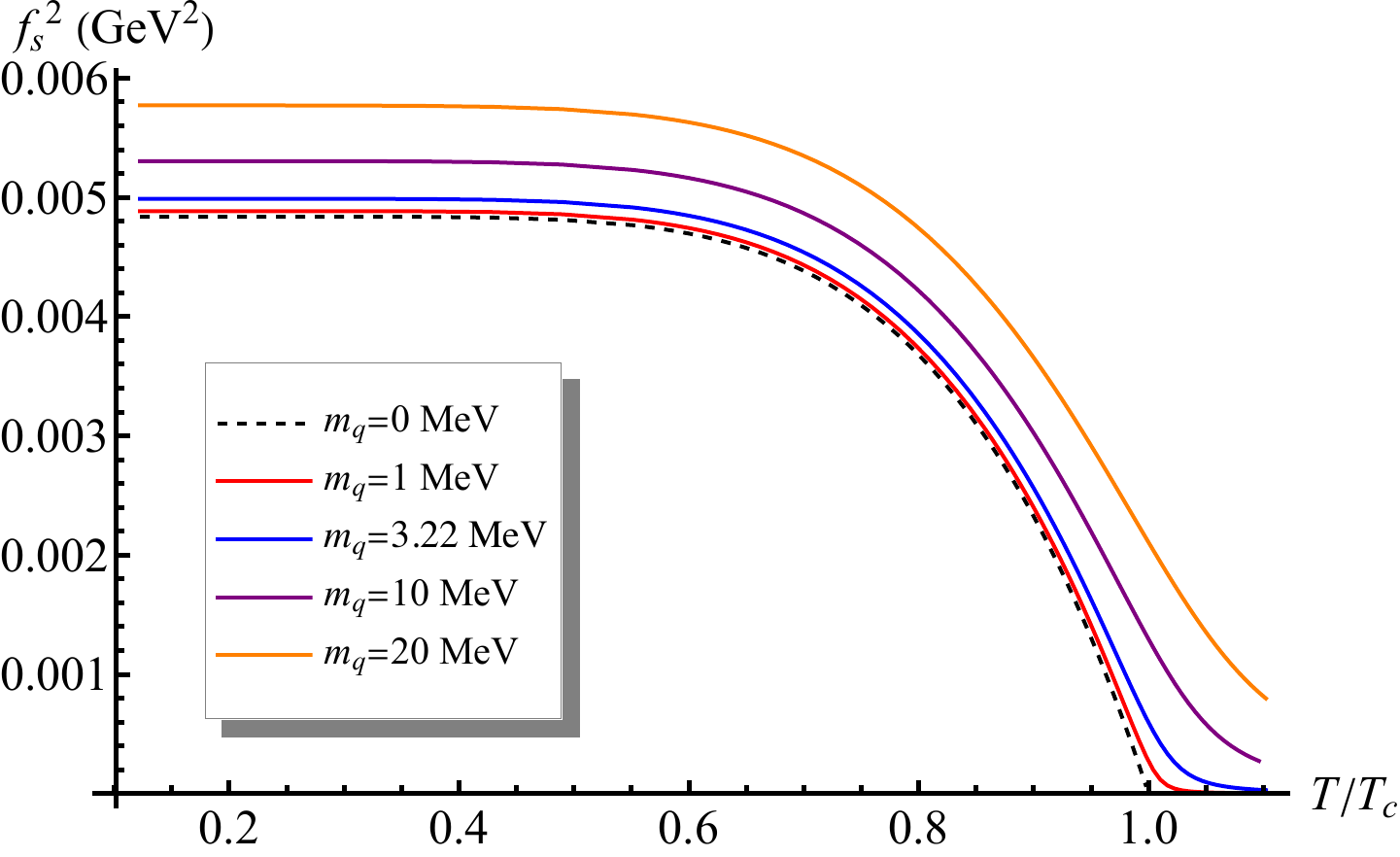}
    \put(80,55){\bf{(c)}}
    \end{overpic}
    
    \vspace{0.2cm}

    \begin{overpic}[width=0.8\linewidth]{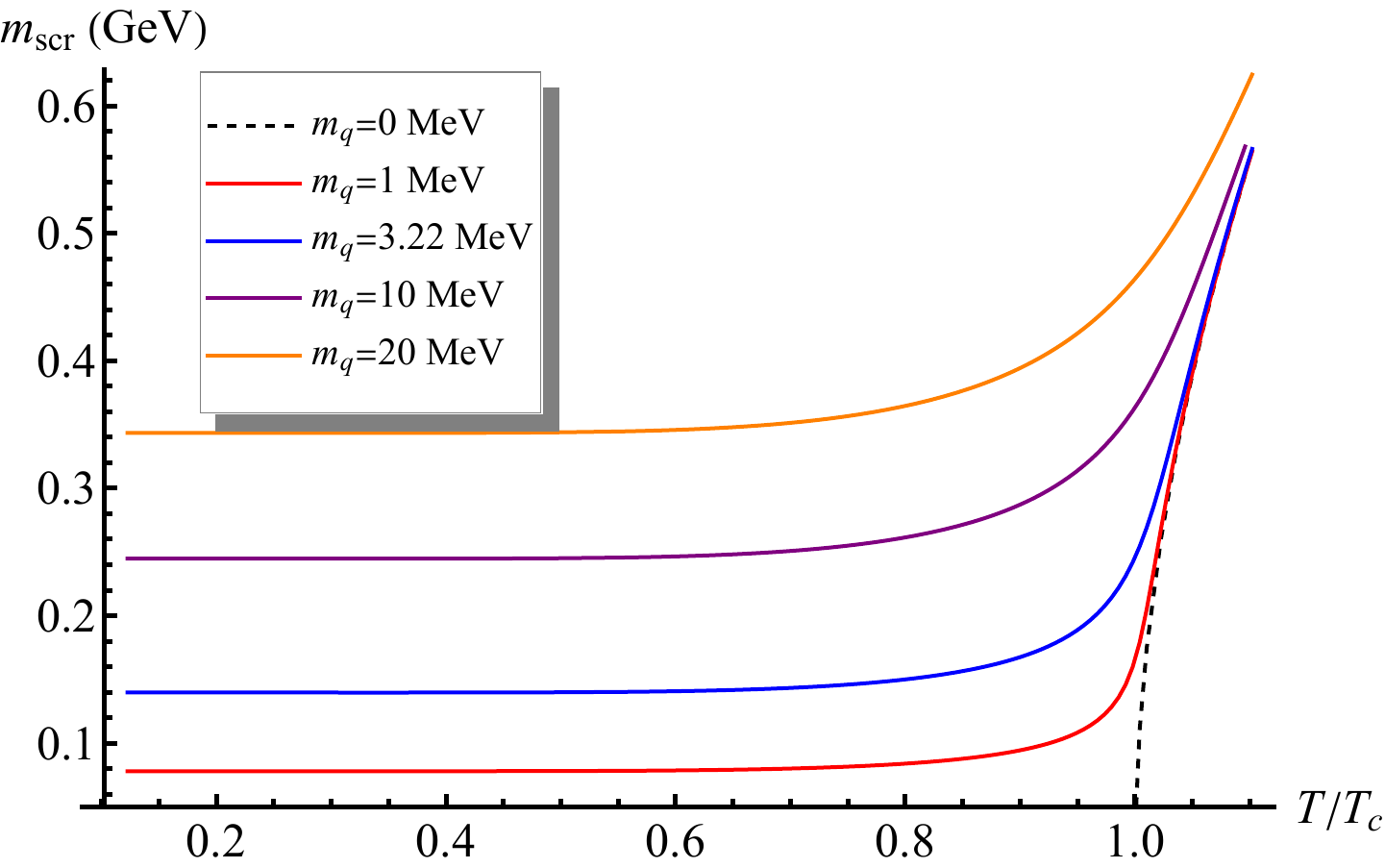}
    \put(80,55){\bf{(d)}}
    \end{overpic}
    \caption{The combination $-(k^2+m_{\rm{scr}}^2)(m_{\rm{scr}}^2/k^2) G_{\varphi\varphi}(k)$: \textbf{(a)} at different temperatures and fixed $m_q=3.22$ MeV and \textbf{(b)}  at different quark masses and $T=100$ MeV. \textbf{(c)} The pion decay constant. \textbf{(d)} The temperature dependence of screening masses for different quark masses.}\label{fig:5}
\end{figure}

\begin{figure}[ht!]
    \centering
    \includegraphics[width=0.85\linewidth]{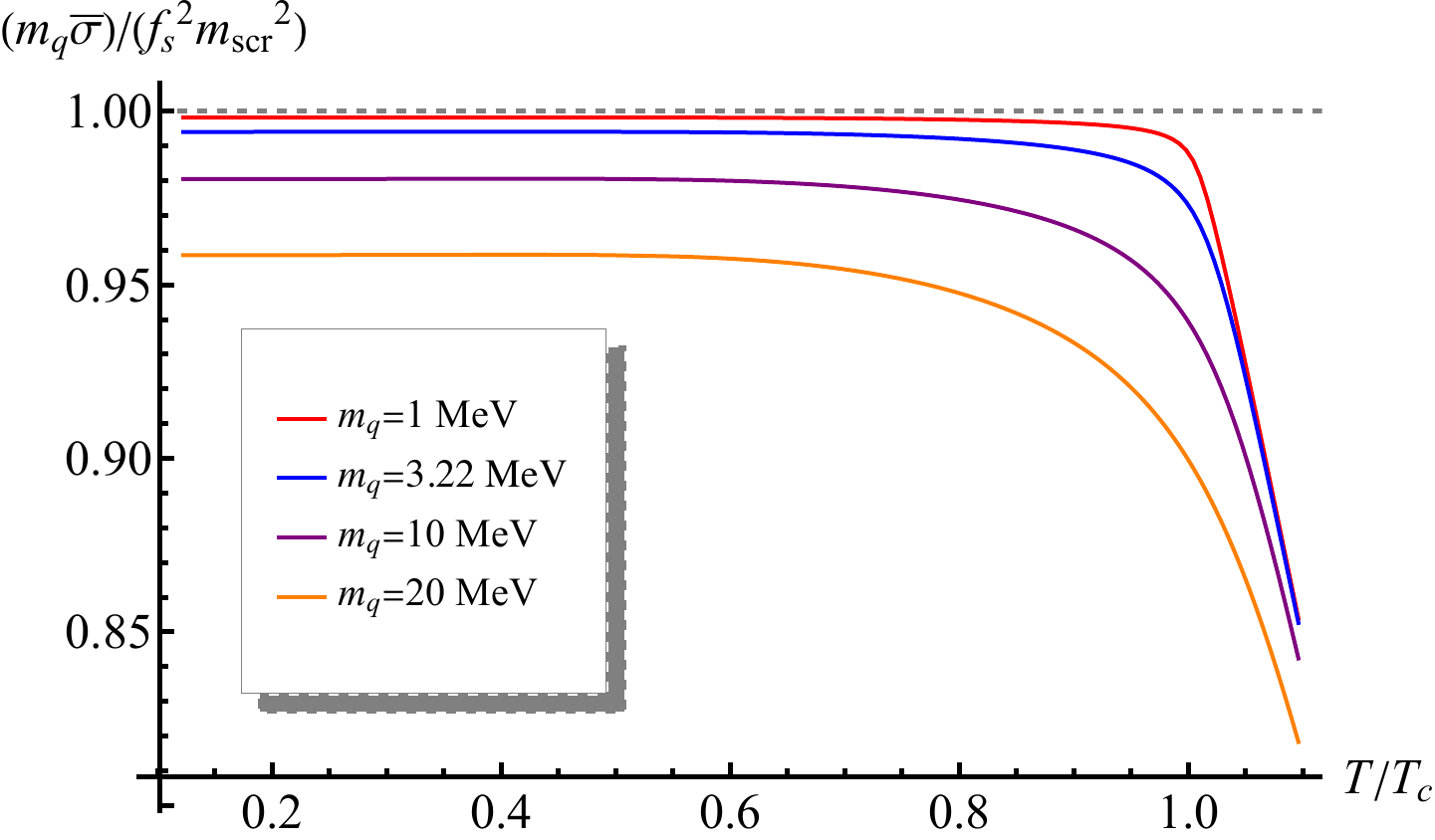}
    \caption{The validity of the GMOR relation at finite quark masses. The horizon gray dashed line as a guide to the eye.}
    \label{fig:6}
\end{figure}

So far, we have focused on the chiral phase transition and the dynamics of the pions in the chiral limit where the masses of the quarks are assumed to be zero. However, in real world, chiral symmetry appears to be softly broken since the quarks are not massless. We therefore extend here the discussion in presence of a small quark mass.

We start by considering the behavior of the pion correlator in the static limit, $\omega=0$. In panels (a) and (b) of Fig.~\ref{fig:5}, we plot the combination $-(k^2+m_{\rm{scr}}^2)(m_{\rm{scr}}^2/k^2) G_{\varphi\varphi}(\omega=0,k)$ as a function of the momentum $k$ at different temperatures and different quark masses. At low momentum, all the curves approach a constant value which is well described by the leading order approximation of the static Green's function, Eq.~\eqref{stat}. From such a limit we can define the pion decay constant as $f_s^2=1/\chi_{\varphi\varphi}$. As evident in Eq.~\eqref{definition:fpi}, we can calculate the finite pion decay constant using the solution of the axial-vector field equation at the singularity pole $q=0$. As shown in Fig.~\ref{fig:5}(c), the pion decay constant decreases with temperature and goes to zero near $T_c$ in the limit of zero quark mass, $m_q=0$. In the chiral limit, the pion decay constant behaves as $f_\pi^2\propto |T-T_c|$ near the critical point.\footnote{Our numerical fit gives a power scaling $|T-T_c|^{1.2}$. The deviation from the expected linear power is just a result of numerical accuracy and of a too small number of points close to $T_c$.} Away from the chiral limit, the pion decay constant does not goes anymore to zero at the bare critical temperature $T_c$ and it has a finite tail reminiscent of the behavior of the chiral condensate in Fig.\ref{fig1}. Following the same analogy, the value of the pion decay constant grows with the quark mass $m_q$. At last, in Fig.~\ref{fig:5}(d)~\footnote{The data for the pion screening mass in the chiral limit is obtained from Ref.~\cite{Cao:2021tcr}.}, we plot the pion screening mass as functions of temperature for different quark masses. The screening mass is related to the inverse of the correlation length ($\zeta$) and increases monotonically with temperature. In the chiral limit $m_q=0$, the screening mass equals to zero in the chiral broken phase and behaves as $m_{scr}^{-1}=\zeta\propto|T-T_c|^{-\nu}$ with $\nu=0.5$ in the chiral restored phase~\cite{Cao:2021tcr}. This is yet another manifestation of the mean-field nature of the critical point in the chiral limit.

\begin{figure*}[bht!]
\centering
\vspace{0.5cm}
  \begin{overpic}[width=0.3\linewidth]{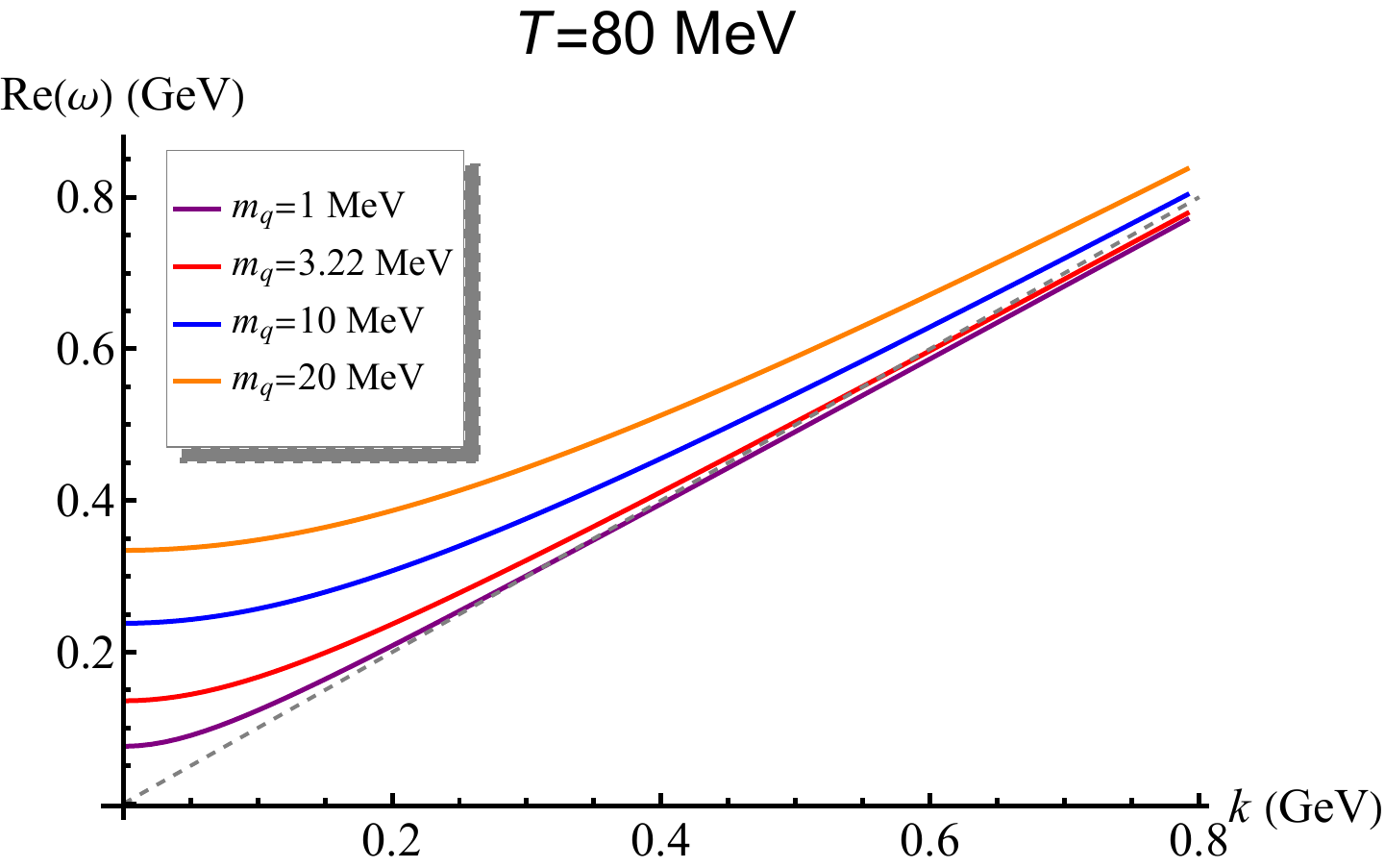}
   \put(80,10){\bf{(a)}}
   \end{overpic}
      \begin{overpic}[width=0.3\linewidth]{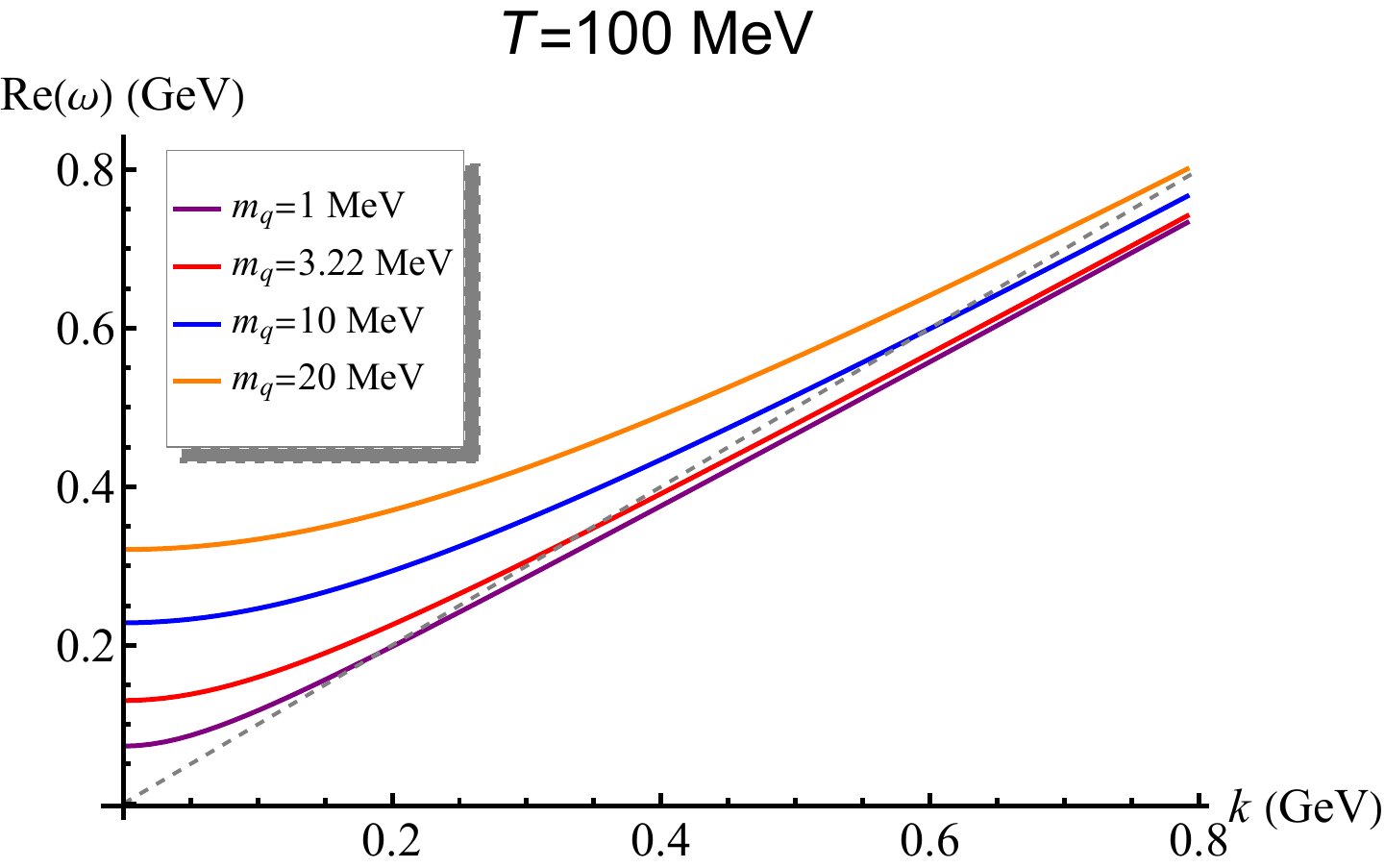}
   \put(80,10){\bf{(b)}}
   \end{overpic}
\begin{overpic}[width=0.3\linewidth]{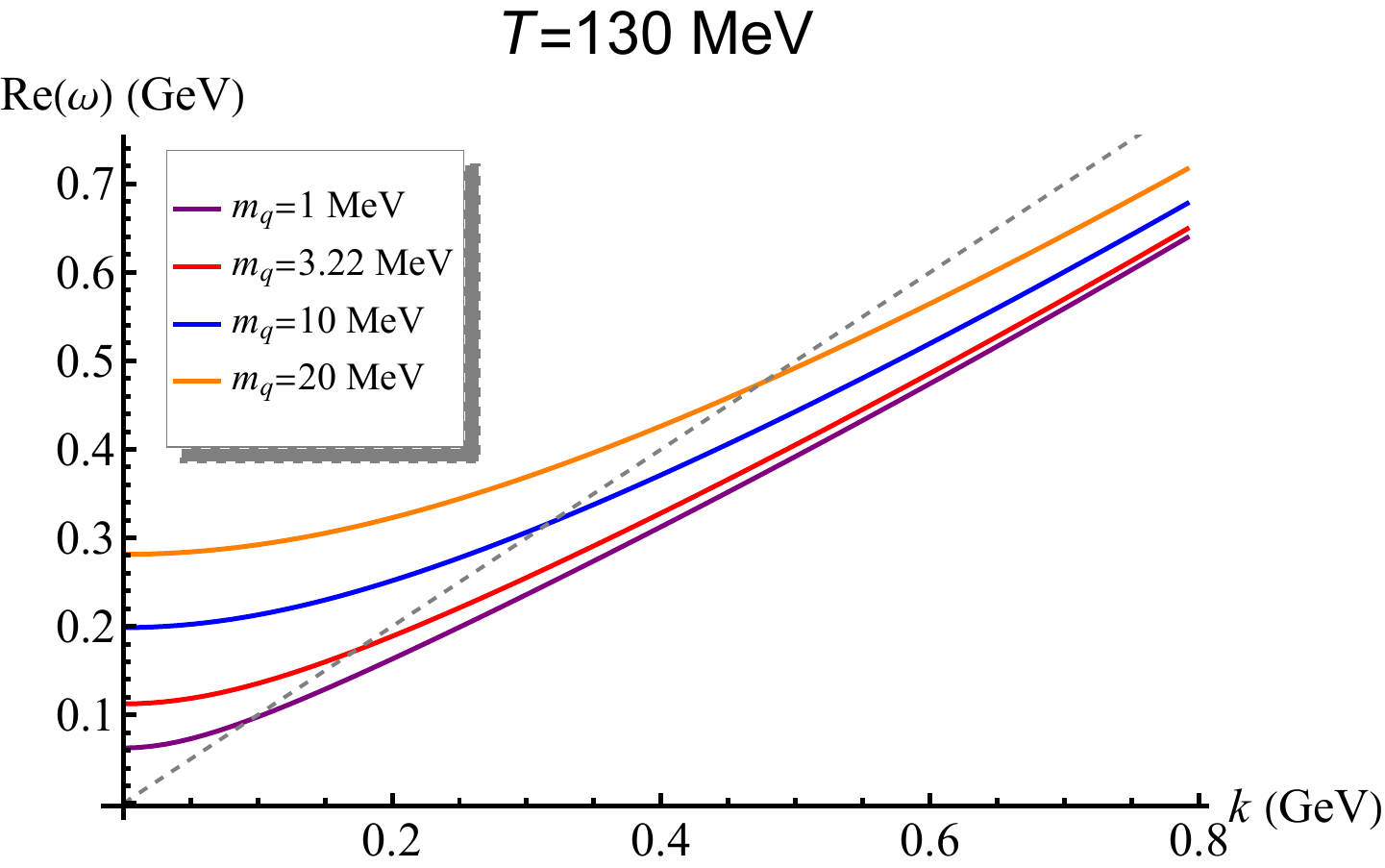}
   \put(80,10){\bf{(c)}}
   \end{overpic}
\vspace{0.3cm}

  \begin{overpic}[width=0.3\linewidth]{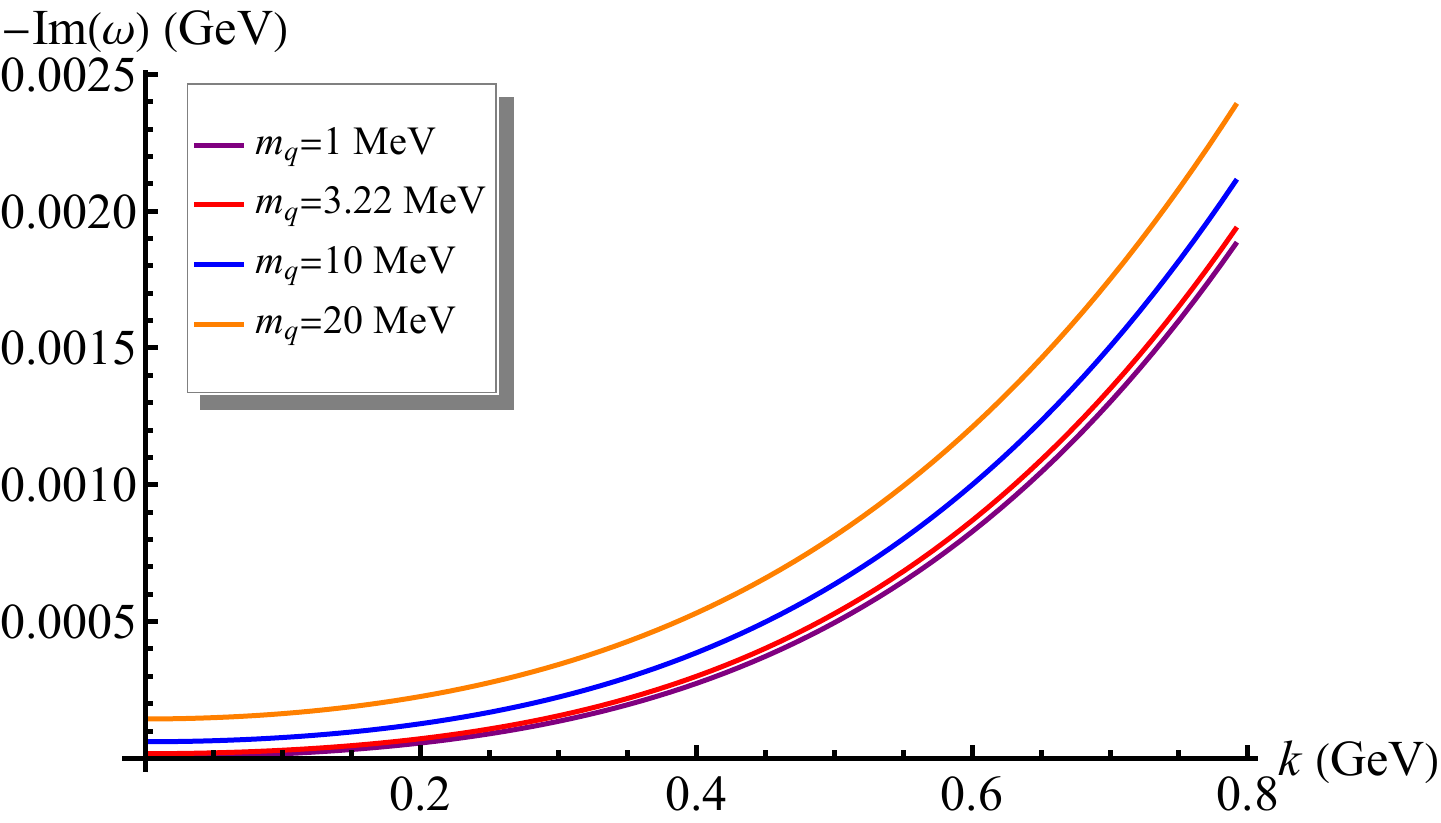}
   \put(80,10){\bf{(d)}}
   \end{overpic}
      \begin{overpic}[width=0.3\linewidth]{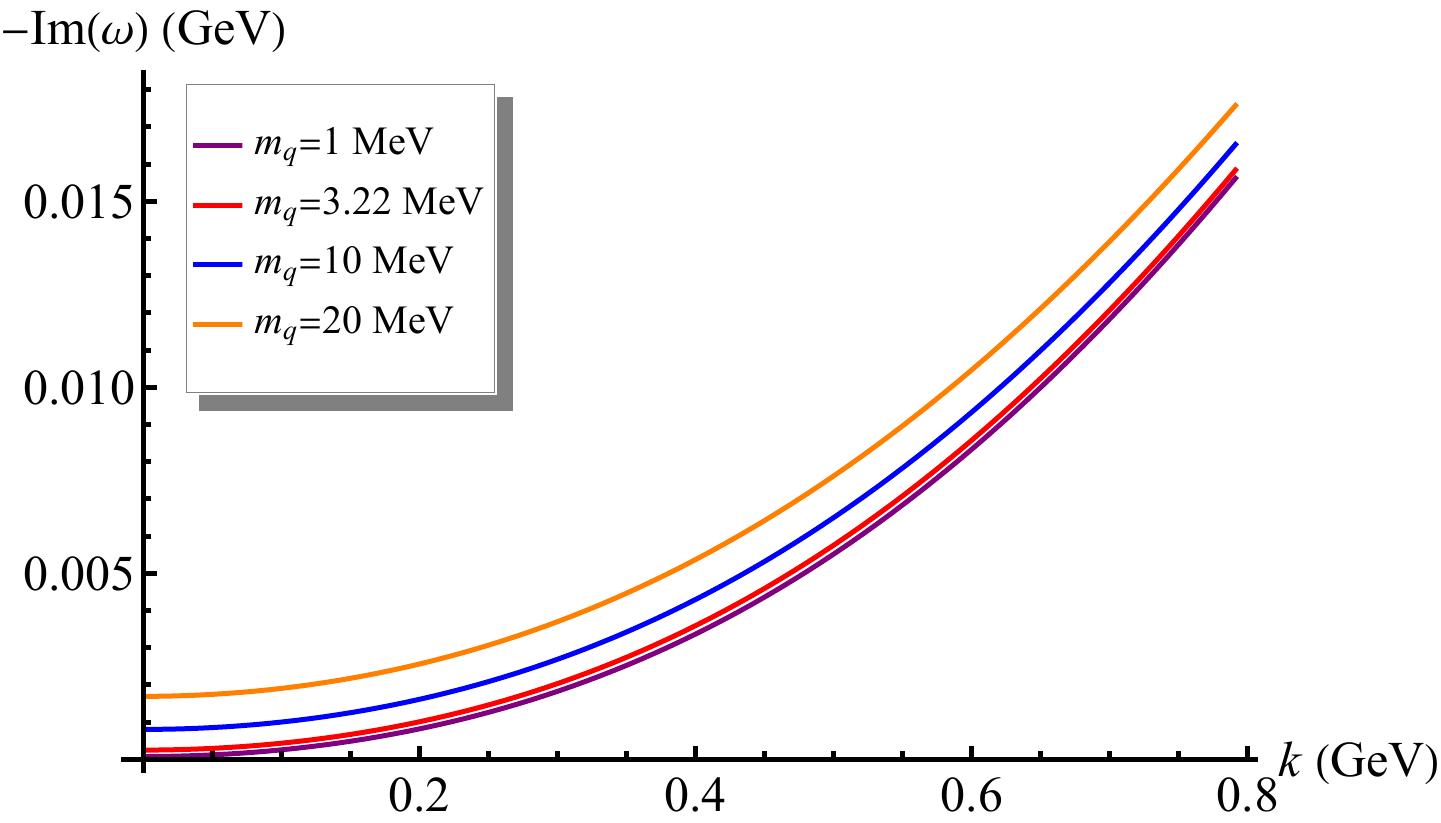}
   \put(80,10){\bf{(e)}}
   \end{overpic}
\begin{overpic}[width=0.3\linewidth]{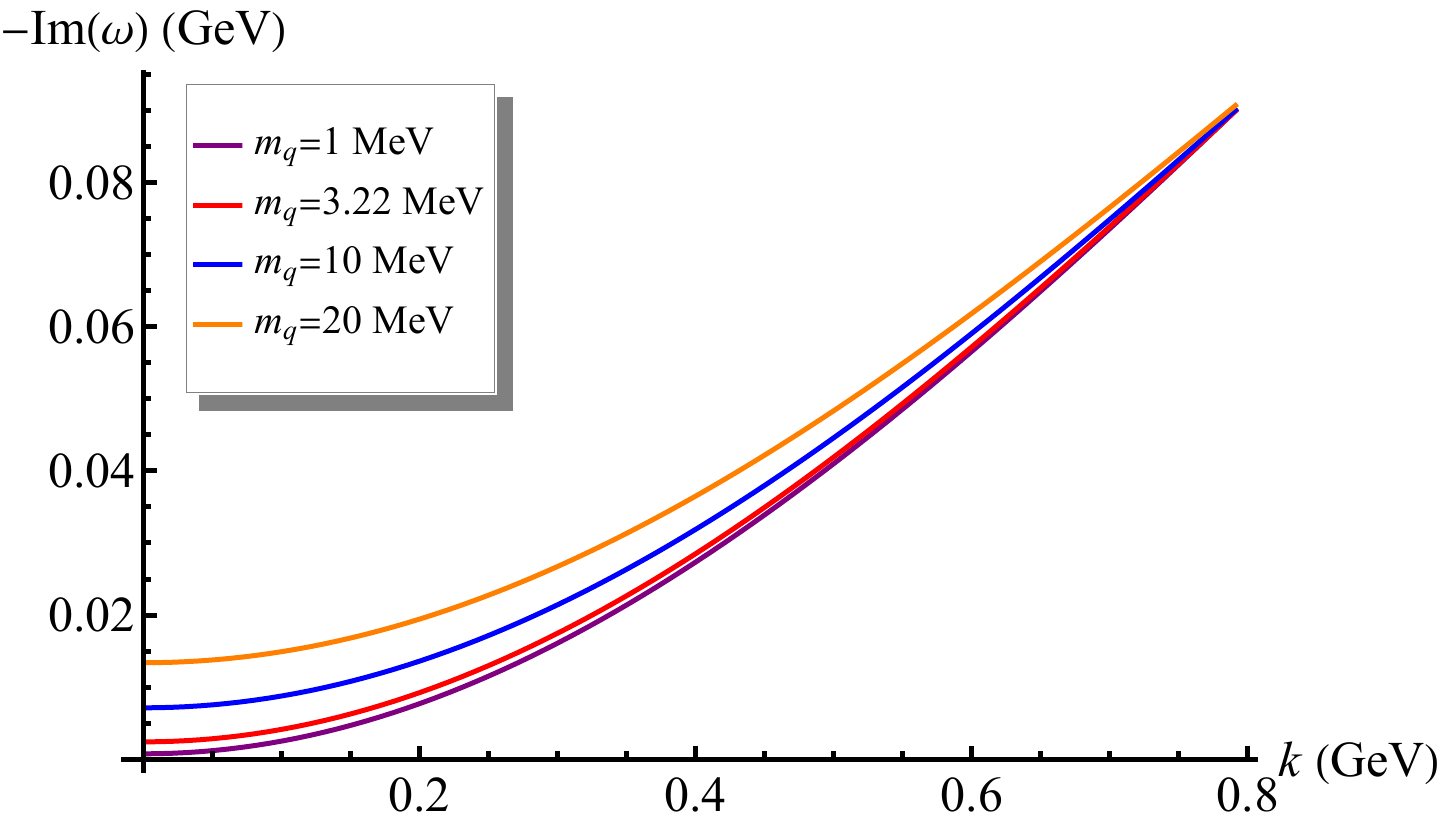}
   \put(80,10){\bf{(f)}}
   \end{overpic}
    \caption{Real part (${\rm{Re}}[\omega]$) and imaginary part ($-{\rm{Im}}[\omega]$) of the dispersion relation of the pions at \textbf{(a)} and \textbf{(d)} $T=80$ MeV; \textbf{(b)} and \textbf{(e)} $T=100$ MeV; \textbf{(c)} and \textbf{(f)} $T=130$ MeV. The gray dashed lines guide the eyes towards the light-like dispersion, ${\rm{Re}}[\omega]=k$.} 
    \label{fig:7}
\end{figure*}

Since we already have the pion decay constant, the sigma condensate and the screening mass, we can immediately verify the validity of the famous GMOR relation
\begin{equation}
    f_s^2 m_{\rm{scr}}^2=2m_q\bar{\sigma}
\end{equation}
valid in the limit of finite but small quark masses. We plot the ratio $2m_q\bar{\sigma}/(f_s^2 m_{\rm{scr}}^2)$ as a function of $T/T_c$ in Fig.~\ref{fig:6}. In the limit of small quark masses, the ratio is very close to $1$ apart from a very small region close to the critical temperature. Close to $T_c$, the quark condensate becomes very small and we are not anymore in the limit of pseudo-spontaneous breaking of chiral symmetry since the explicit breaking term, the quark mass, becomes larger than the spontaneous scale, the condensate. Our analysis indicates that the GMOR relation works well in the chiral symmetry broken phase when the quark mass is very small.
By increasing the value of the quark mass, which parametrizes the explicit breaking scale, the GMOR ceases to be accurate. Nevertheless, we interestingly find that the GMOR value still acts as a lower bound for the pion mass:
\begin{equation}
  m_{\rm{scr}}^2f_s^2\geq 2 m_q \bar \sigma\,.  
\end{equation}
The larger the quark masses, the more the pion mass deviates from such a lower bound. It would be interesting to verify whether this outcome is a general property or it is specific to our holographic model. We are not aware of any discussion of this sort in the literature.

\begin{figure}
    \centering
    \begin{overpic}[width=0.83\linewidth]{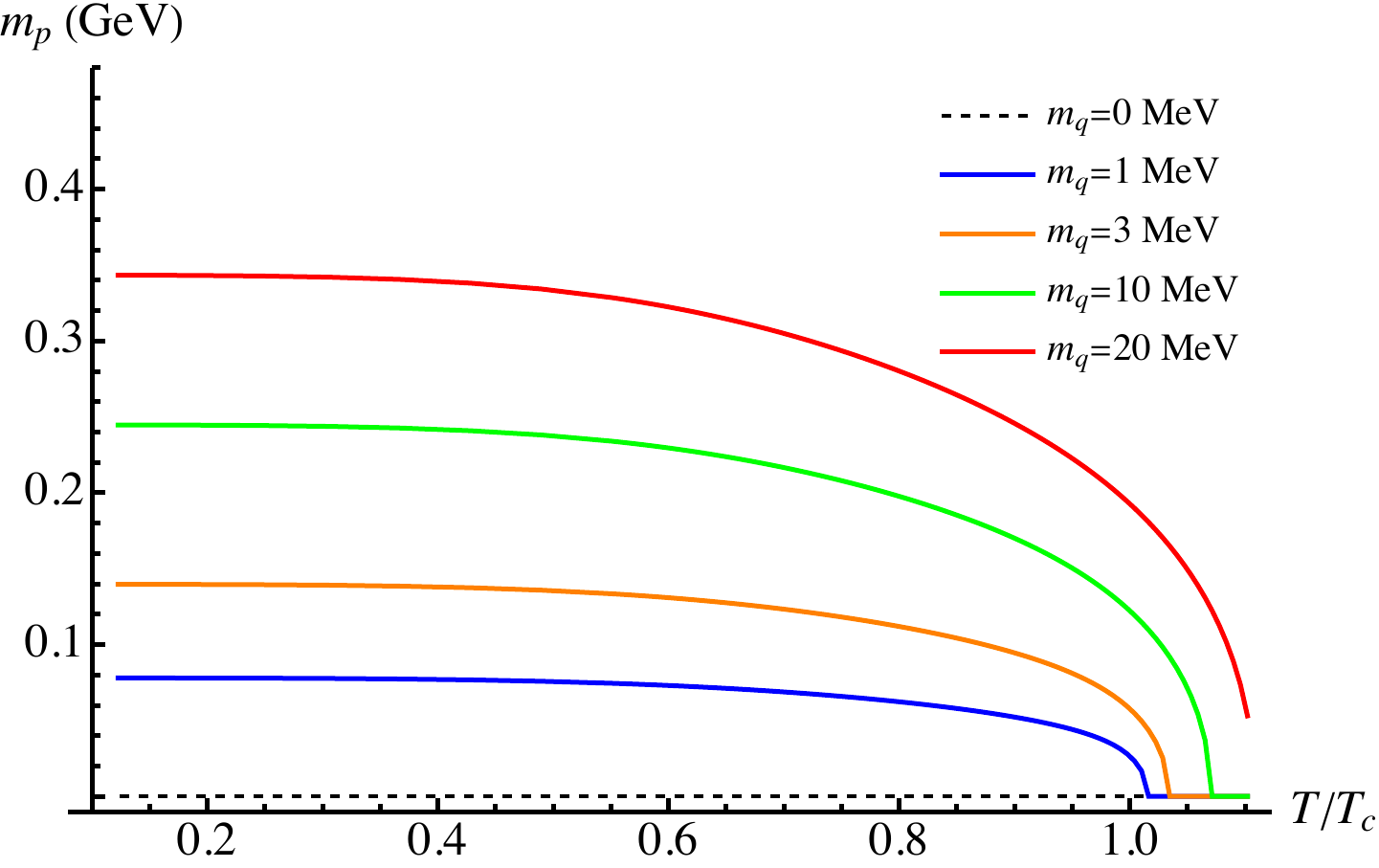}
    \put(85,58){\bf{(a)}}
    \end{overpic}
    
    \vspace{0.2cm}

    \begin{overpic}[width=0.8\linewidth]{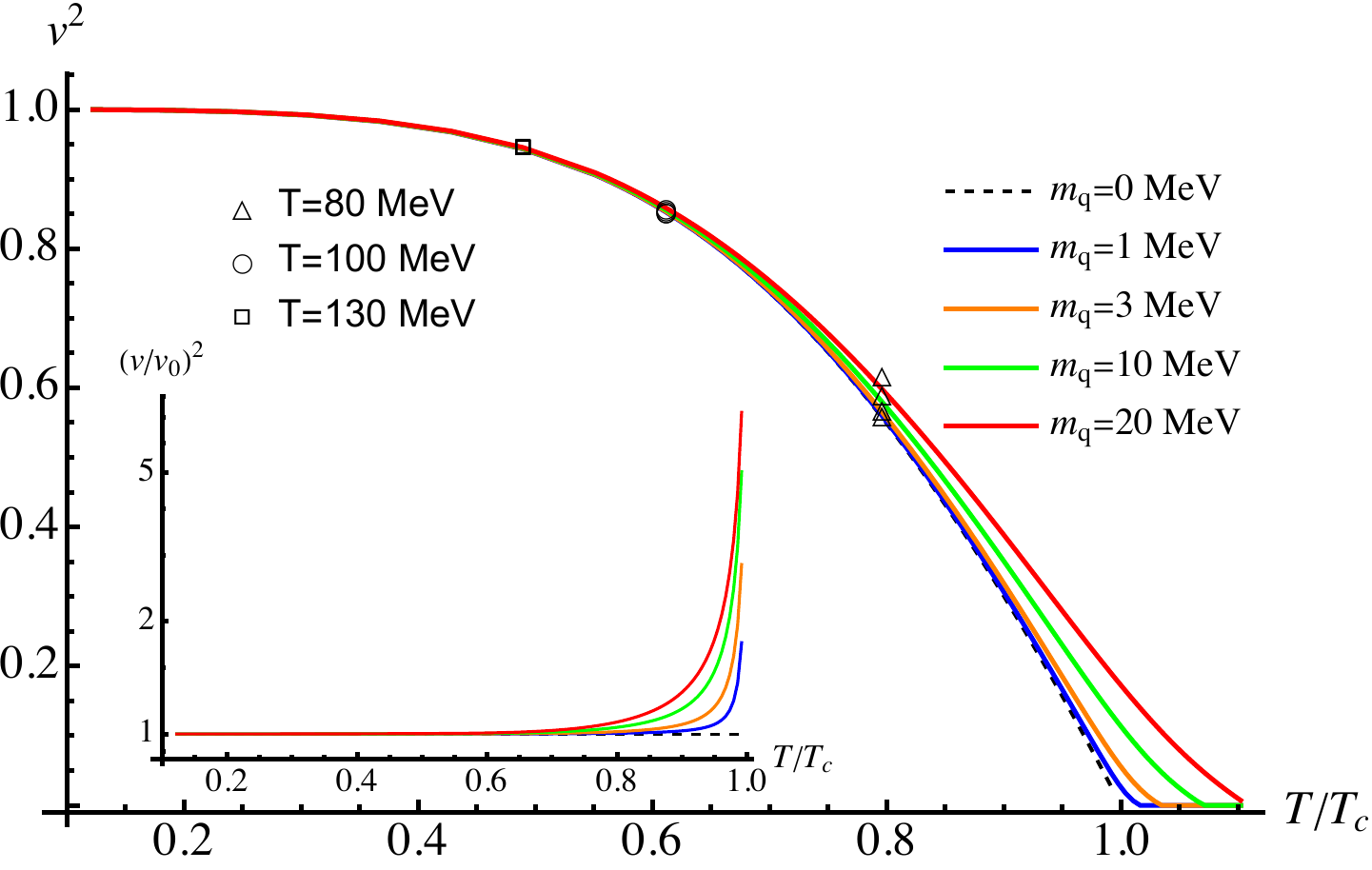}
    \put(85,56){\bf{(b)}}
    \end{overpic}
    \caption{\textbf{(a)} Temperature dependence of the pole mass $m_{\rm{p}}$ with different quark masses. \textbf{(b)} The square of the sound velocity $v$ as a function of the reduced temperature $T/T_c$. The open symbols are the values of the sound speed obtained by fitting the dispersion relations in Fig.~\ref{fig:7} with the formula ${\rm{Re}}[\omega]=\sqrt{m_p^2+v^2 k^2}$. The inset figure shows the ratio between the velocity at finite quark mass $v$ and the bare one in chiral limit $v_0$.}
    \label{fig:8}
\end{figure}

We are now ready to look at the dispersion relation of the pseudo-goldstone modes, the pions away from the chiral limit. In Fig.\ref{fig:7} we show the dispersion relation as a function of the quark mass for different values of temperature. At small wave-vector, the dispersion relation of the pions is well approximated by Eq.~\eqref{eq:dispersion} where the renormalized energy $\omega$ is given in Eq.\eqref{eq:dispersionreal} and the $k-$dependent damping $\Gamma_k$ by the expression in Eq.\eqref{nene}. A lot of information can be extracted from the dispersion relation. First, the real part of the frequency at zero wave-vector gives the pole mass, i.e. $m_{\rm{p}}={\rm{Re}}[\omega(k=0)]$. We plot the pole mass as a function of $T/T_c$ for different $m_q$ in Fig.~\ref{fig:8}(a). The pole mass increases with the quark mass as expected. Using the values for the screening mass in Fig.~\ref{fig:5}(d), we can calculate the sound velocity $v$ as described in Eq.~\eqref{eq:soundvelocity}. We plot the ratio between $v^2$ and the square of sound velocity in the chiral limit $v_0^2$  as a function of $T/T_c$ in Fig.~\ref{fig:8}(b). At low temperature, the sound velocity approaches its value in the chiral limit. More precisely, we see that the ratio is approximately one up to $T/T_c \sim 0.6$. Approaching the critical point, the ratio grows rapidly and it increases with the value of the quark mass. As shown in the inset, the velocity always approaches the speed of light in the zero temperature limit and decreases with the increasing temperature. In order to confirm that the pion velocity extracted from Eq.\eqref{eq:soundvelocity} is correct, we present also an alternative derivation. In particular, we directly fit the pion dispersion relation with the formula
\begin{equation}\label{lala}
    \mathrm{Re}[\omega(k)]=\sqrt{m_p^2+v^2 k^2}\,.
\end{equation}
Part of the data for the dispersion relations are shown in Fig.\ref{fig:7}. In panel (b) of Fig.\ref{fig:8}, we indicate the results from the fitting with open symbols for three different temperatures and different values of the quark mass. We find perfect agreement between the velocity extracted with Eq.\eqref{eq:soundvelocity} and that obtained by fitting the real part of the dispersion relation with Eq.\eqref{lala}.

\begin{figure}
    \centering
    \begin{overpic}[width=0.8\linewidth]{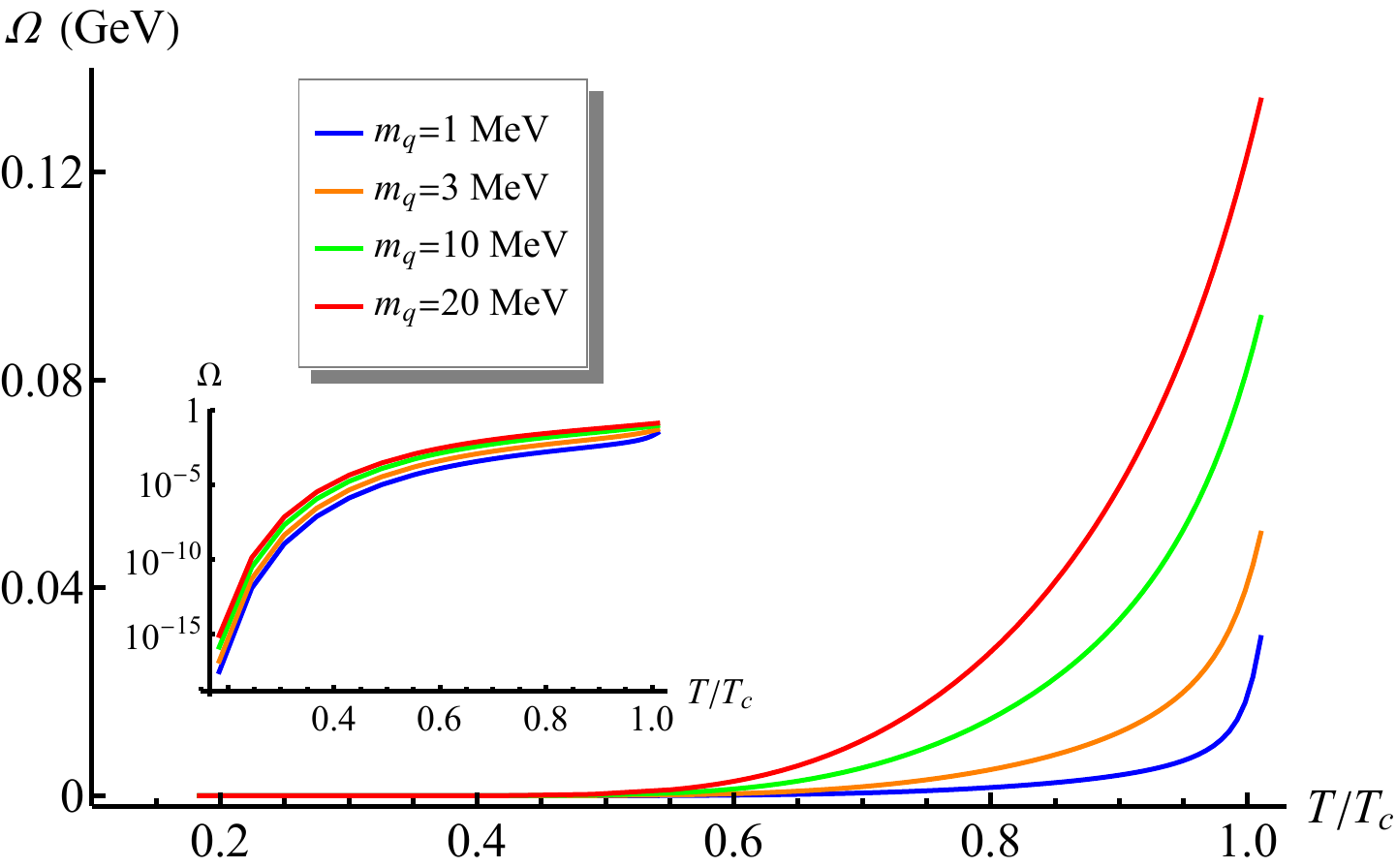}
    \put(80,50){\bf{(a)}}
    \end{overpic}
    
    \vspace{0.2cm}

    \begin{overpic}[width=0.8\linewidth]{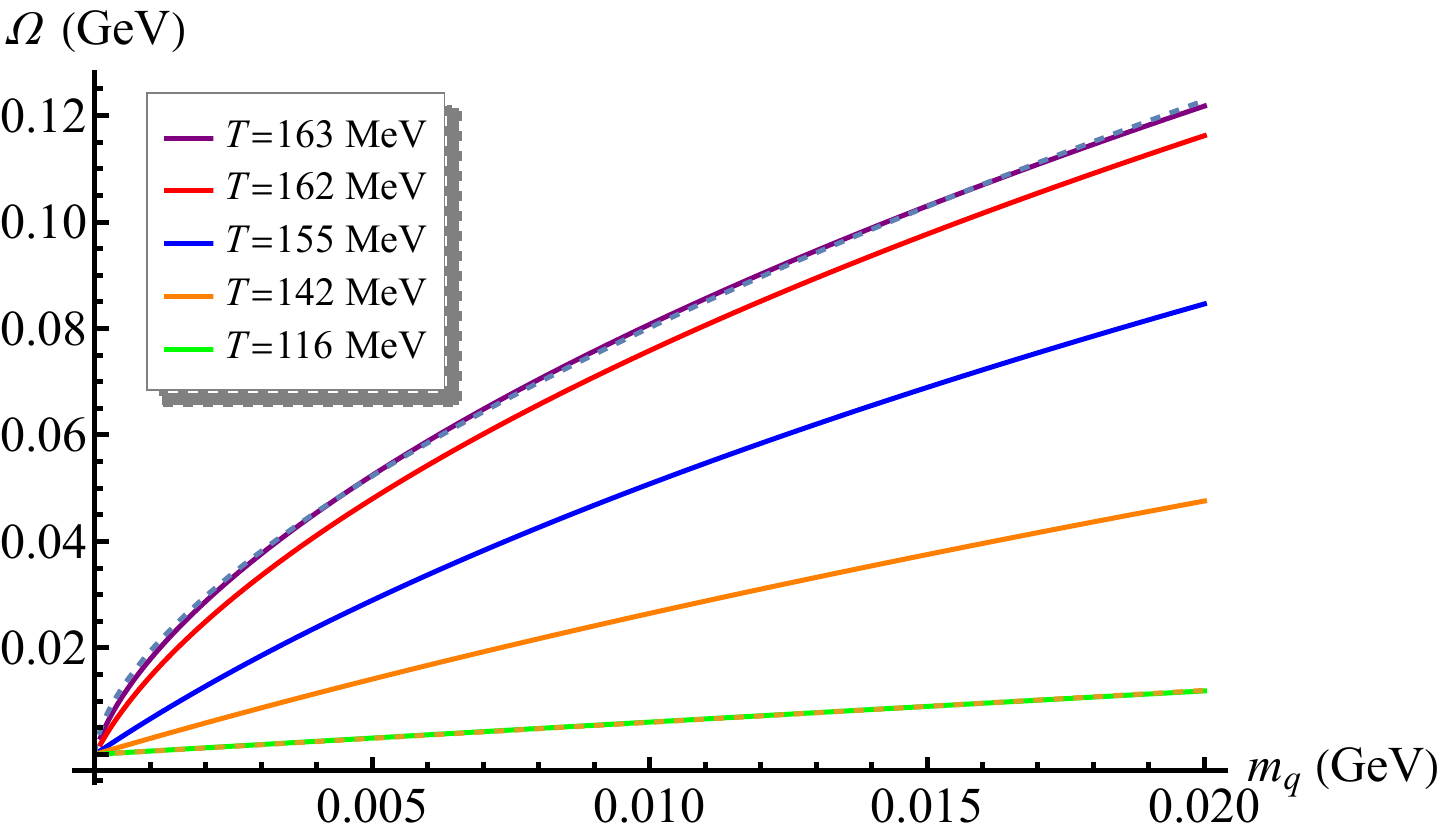}
    \put(77,50){\bf{(b)}}
    \end{overpic}
    \caption{\textbf{(a)} The phase relaxation rate $\Omega$ as a function of the reduced temperature $T/T_c$ for different quark masses $m_q$. The inset is a semi-log plot of the same data. \textbf{(b)} The same quantity as a function of $m_q$ for different $T$. The dashed lines are fitting results to a power-law scaling of the form $\Omega \propto m_q^{b}$. We have $b\approx 0.62$ at $T=163$ MeV and $b\approx 0.98$ at $T=116$ MeV.}
    \label{fig:9}
\end{figure}

Always from the dispersion relation, the imaginary part corresponds to the damping $\Gamma_k$ (or thermal width) as defined in Eq.~\eqref{dam}. Here, we are particularly interested in its value at $k=0$ which corresponds to the phase relaxation rate $\Omega$ as defined in Eq.~\eqref{eq:dampingwidth}. The numerical results for $\Omega$ are shown in Fig.~\ref{fig:9}.  Panel (a) of Fig.~\ref{fig:9} displays the phase relaxation rate as a function of $T/T_c$. The inset shows the details in the low temperature regime. Panel (b) shows the phase relaxation rate as a function of the quark mass $m_q$. Away from the critical point, for $T \ll T_c$, the phase relaxation rate is linear in the quark mass:
\begin{equation}
    \Omega \propto m_q\,.
\end{equation}
On the contrary, close to the critical point, $T\approx T_c$, it satisfies a scaling relation:
\begin{equation}
    \Omega\propto m_q^{\nu z/(\beta\delta)}
\end{equation}
with the critical exponents $\beta=1/2$ and $\delta=3$. This behavior is consistent with the analysis of the critical point in Ref.~\cite{Cao:2022mep}. This observation is also compatible with the nonlinear scalings found for the GMOR relation and for the phase relaxation rate in certain holographic models with broken translations \cite{Andrade:2020hpu,Andrade:2018gqk,Andrade:2017cnc}.

\subsection{The relaxation rate of Pions}
\begin{figure}
    \centering
    \includegraphics[width=0.9\linewidth]{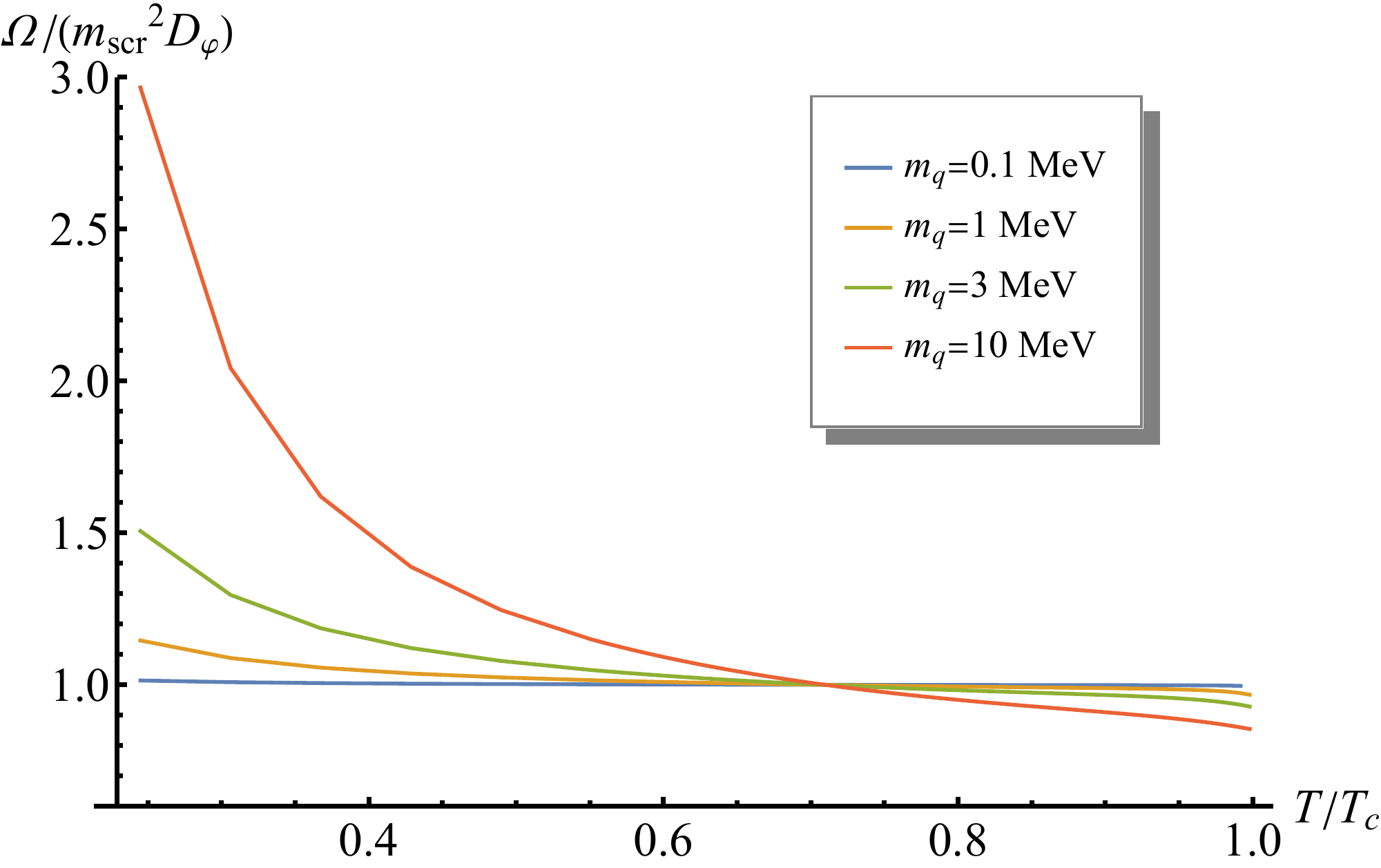}
    \caption{The dimensionless ratio $\Omega/({m_{\rm{scr}}^2} D_{\varphi})$ as a function of the dimensionless temperature $T/T_c$ four different fixed quark masses from $m_q=0.1$ MeV to $m_q=10$ MeV.}
    \label{fig:last}
\end{figure}

Finally, we are in the position to verify the universal damping relation $\Omega=D_{\varphi} m_{\rm{scr}}^2$ which is expected to be valid for any pseudo-Goldstone modes. In Fig.~\ref{fig:last}, we plot the dimensionless ratio $\Omega/(m_{\rm{scr}}^2 D_{\varphi})$ as a function of reduced temperature $T/T_c$ for different values of the quark mass. Whenever the explicit symmetry breaking parameter $m_q$ is small, this ratio is equal to $1$ in almost all the range of temperatures apart from a very small region close to the critical point. Around the critical point, once more, the condensate becomes small and therefore the pseudo-spontaneous limit ceases to be valid. Moreover, if the breaking parameter becomes large, the universal relation is gradually breaking (as expected) at any temperature. Still, for reasonable values of the mass, the universal relation still holds approximately well at intermediate temperatures. Finally, let us notice how this relation appears to be violated in both directions. This means, that, differently from the GMOR relation and the pion mass, there is no lower or upper bound for the phase relaxation rate $\Omega$ away from the chiral limit.

\section{Outlook}
\label{sec:out}
In summary, we have studied in detail the low-energy dynamics of pseudo-Goldstone modes (pions) in a holographic soft-wall QCD model with ${\rm{SU}}(2)_L \times {\rm{SU}}(2)_R$ broken symmetry at finite temperature. By matching to the hydrodynamics framework \cite{Grossi:2021gqi}, we have extracted the dissipative transport coefficients and the dispersion relation of the pNGMs. The holographic model serves as a ``microscopic'' description of the pions dynamics which could play a role complementary to the kinetic theory \cite{Torres-Rincon:2022ssx} and simulations \cite{Florio:2021jlx} approaches. As a direct test of the results in \cite{Grossi:2021gqi}, and more generally in \cite{Delacretaz:2021qqu,Armas:2021vku}, we have numerically verified the universal relation for the damping  of pseudo-Goldstone modes, Eq.\eqref{uni}, which was previously verified only in toy models with $\rm{U}(1)$ symmetry \cite{Ammon:2021pyz} or for the case of translations \cite{Amoretti:2018tzw}.\\

There are several  direct questions which need to be addressed. As a concrete example, following \cite{Grossi:2020ezz,Grossi:2021gqi}, it would be interesting to compute the effects of a finite pion mass on the QCD transport properties such as the shear viscosity or the axial conductivity and investigate further the scaling behaviors of the transport coefficients close to the critical point. A different question regards the value of the parameter $\mathfrak{r}^2$ and the observed deviations from the chiral perturbation theory value $3/4$ \cite{Torres-Rincon:2022ssx} together with their meaning and universality (e.g. in different holographic QCD models). Is it an effect of strong coupling? Is it an effect of large-N? Further investigation in this direction is needed.

Additionally, in our analysis the dynamics and role of the amplitude mode (the $\sigma$ meson) has been completely neglected. Close to the critical temperature, the effects of the amplitude mode could be dramatic (see for example \cite{Donos:2022www}).\footnote{This could be simply understood from the fact that the amplitude mode becomes massless at $T_c$.} Moreover, whenever the symmetry is also broken explicitly a clear separation between the massless Goldstone modes and the amplitude mode is missing even away from the critical point. This could produce even stronger effects. It would be interesting to enlarge the hydrodynamic framework by considering the amplitude mode and use holography as a tool in this direction as initiated in \cite{Donos:2022xfd,Donos:2022qao}.

Finally, it would be fruitful to extend our analysis to the time dependent dynamics and analyze the thermalization properties of this system near the critical point as done in the chiral limit in \cite{Cao:2022mep}.

We leave some of these questions for the near future.

\subsection*{Acknowledgments} 
We thank S.~Grieninger, A.~Donos, V.~Ziogas, A.~Soloviev and B.~Gout\'eraux for fruitful discussions on the topic of this paper. X.C. is supported by the National Natural Science Foundation of China under Grant Nos. 12275108 and the Fundamental Research Funds for the Central Universities under grant No. 21622324. M.B. acknowledges the support of the Shanghai Municipal Science and Technology Major Project (Grant No.2019SHZDZX01) and the sponsorship from the Yangyang Development Fund. M.B. would like to thank IFT Madrid, NORDITA and GIST for the warm hospitality during the completion of this work and acknowledges the support of the NORDITA distinguished visitor program and GIST visitor program.  H.L. is supported by the National Natural Science Foundation of China under Grant No. 11405074. D.L. is supported by the National Natural Science Foundation of China under Grant Nos. 12275108, 12235016, 11805084, and the Guangdong Pearl River Talents Plan under Grant No. 2017GC010480.

 \appendix
 \section{Comparison between the holographic result for $G_{\varphi\varphi}^R$ and the thermal chiral effective field theory predictions}\label{appendix} 
In this work, we focused on the dissipative properties of soft pions, which are intrinsically embodied in the pion correlators. Thus, it is very important to verify the validity of the procedure to extract the correlators derived within the holographic soft-wall model. In this Appendix, we compare the results for the holographic correlators to the correlators obtained in thermal chiral effective field theory in Refs.~\cite{Son:2002ci,Son:2001ff}.

\begin{figure*}[htb!]
    \centering
   \begin{overpic}[width=0.31\linewidth]{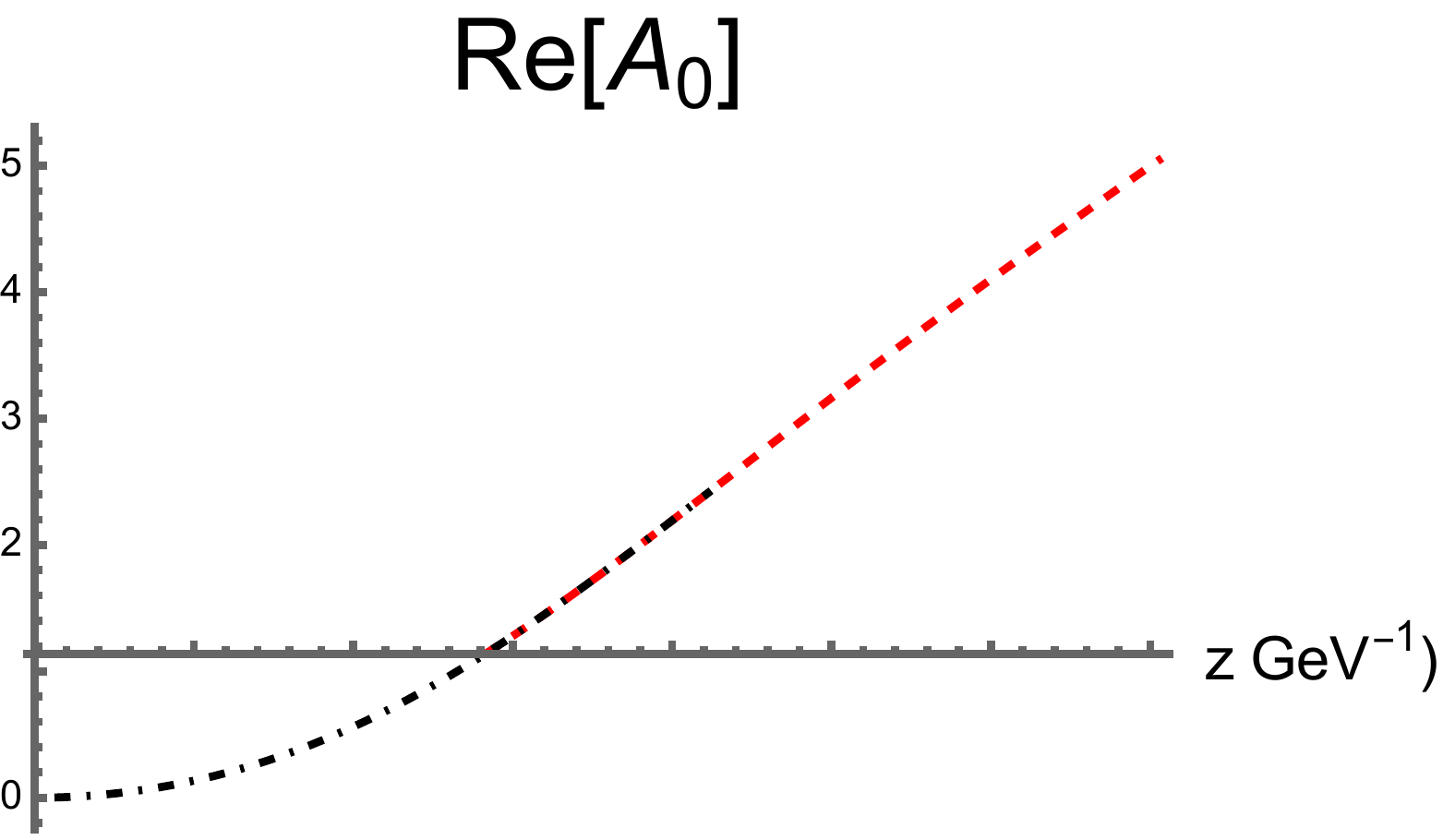}
     \end{overpic}
\vspace{0.2cm}
    \begin{overpic}[width=.31\linewidth]{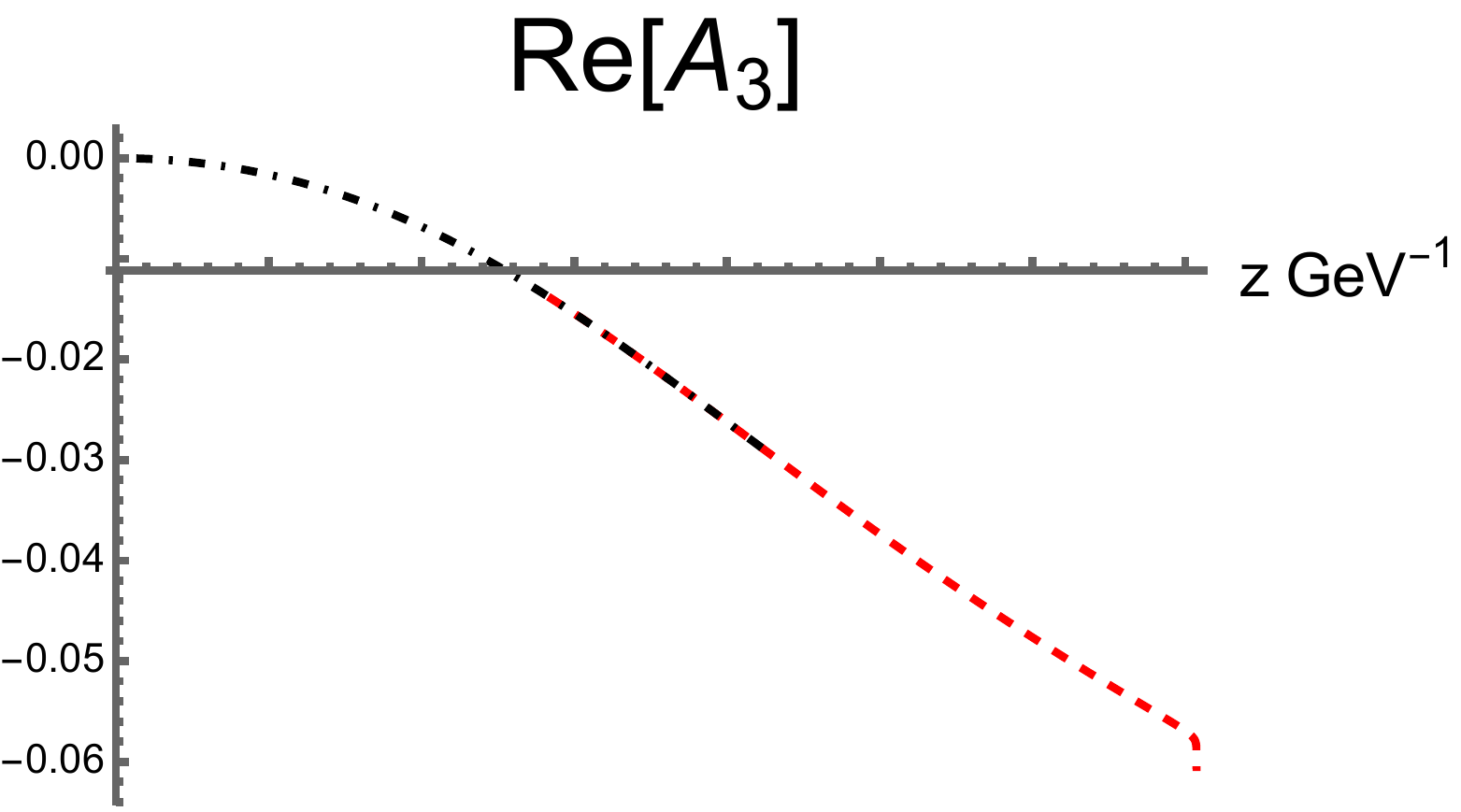}
    \end{overpic}
    \vspace{0.2cm}
    \begin{overpic}[width=.31\linewidth]{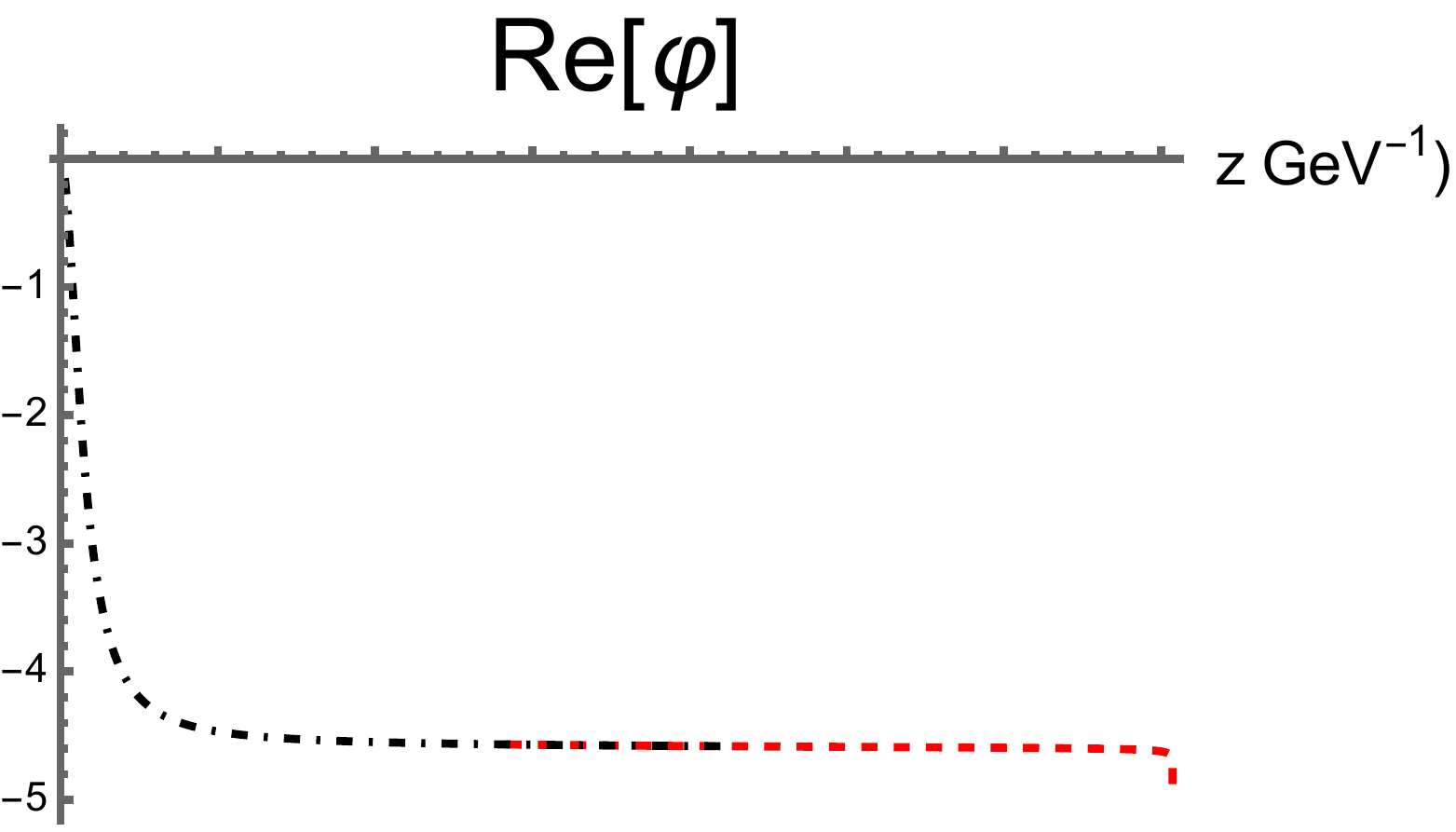}
    \end{overpic}
       \begin{overpic}[width=0.31\linewidth]{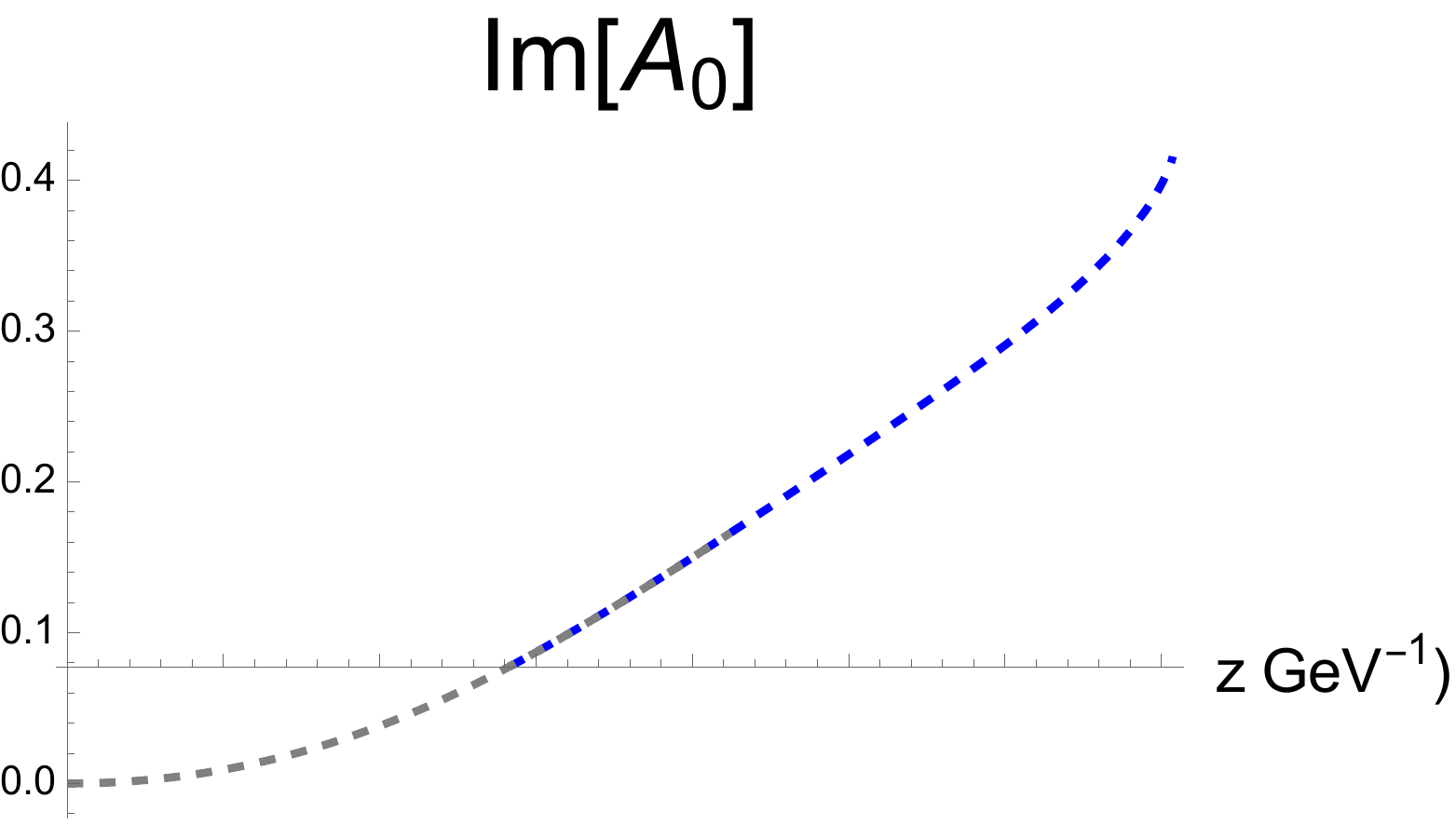}
     \end{overpic}
\vspace{0.2cm}
    \begin{overpic}[width=.31\linewidth]{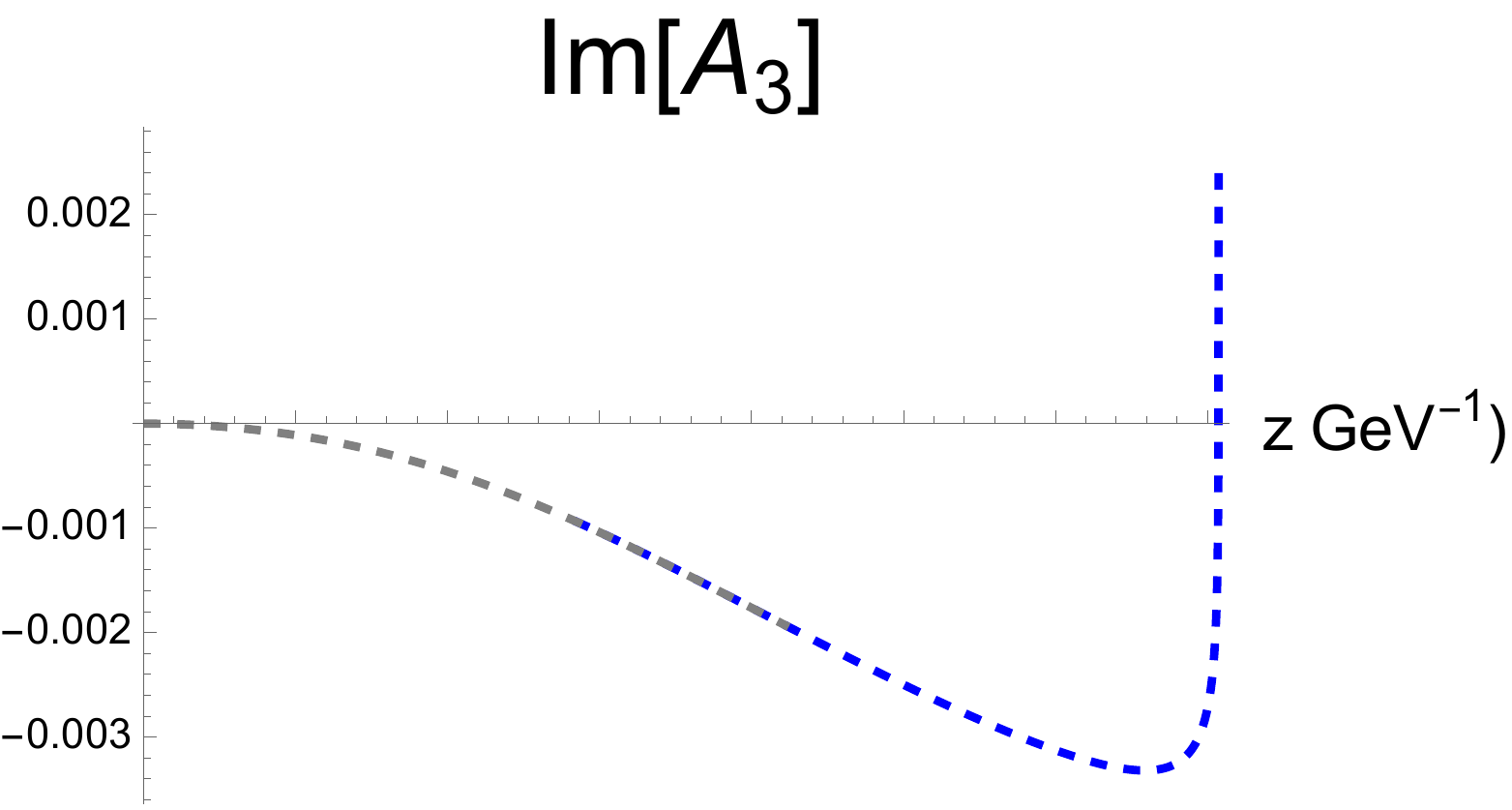}
    \end{overpic}
    \vspace{0.2cm}
    \begin{overpic}[width=.31\linewidth]{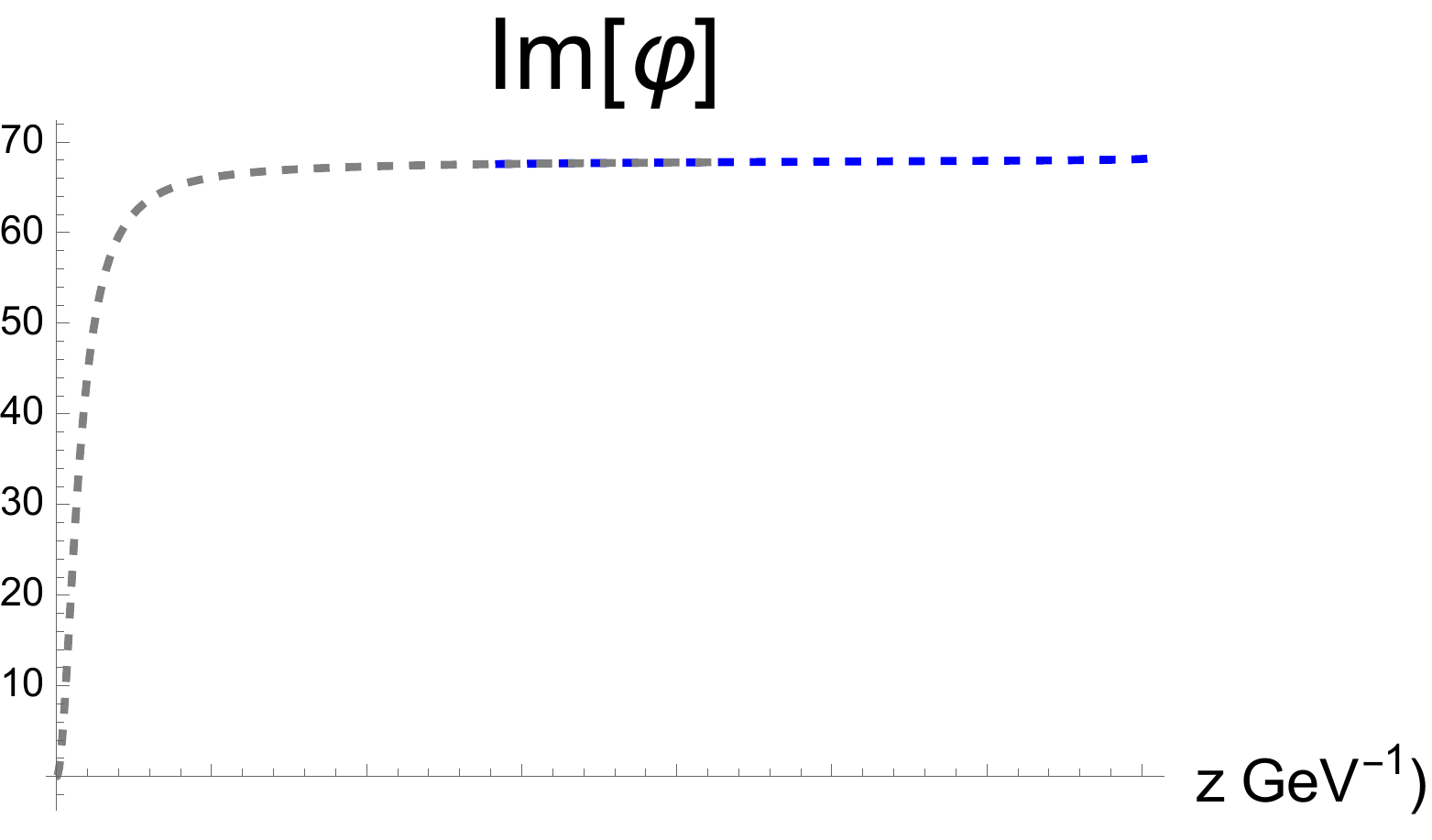}
    \end{overpic}
    \caption{The real and imaginary parts of numerical solutions of $a_0(z)$, $a_3(z)$ and $\varphi(z)$ for our benchmark example discussed in Appendix \ref{appendix}. The colored red and blue dashed lines are the solutions from the IR shooting. The black dot-dashed and gray dashed lines those obtained from the UV shooting.}
    \label{fig:a1}
\end{figure*}

\begin{figure*}
    \centering
    \begin{overpic}[width=.31\linewidth]{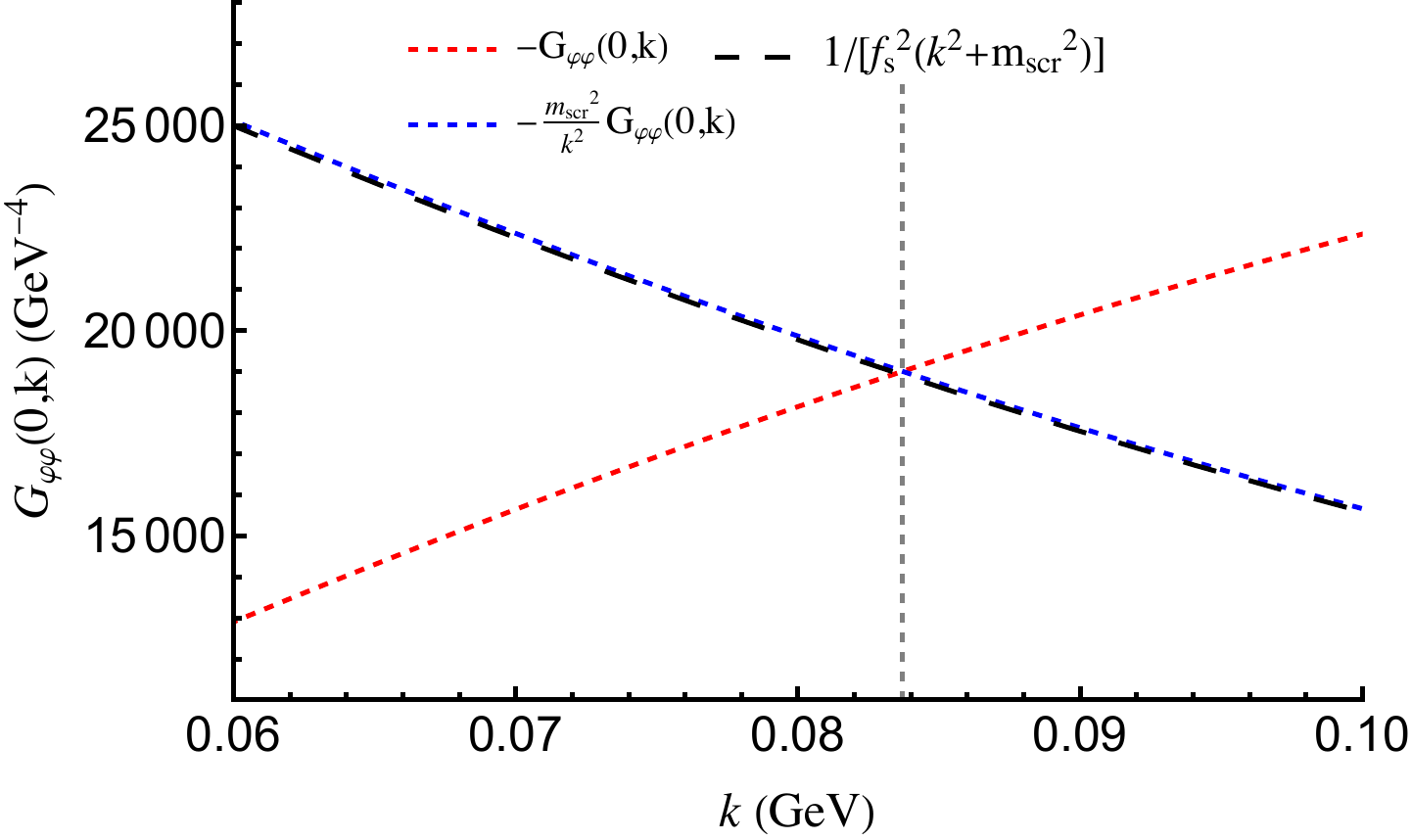}
        \put(85,55){\bf{(a)}}
    \end{overpic}
    \vspace{0.2cm}
        \begin{overpic}[width=.31\linewidth]{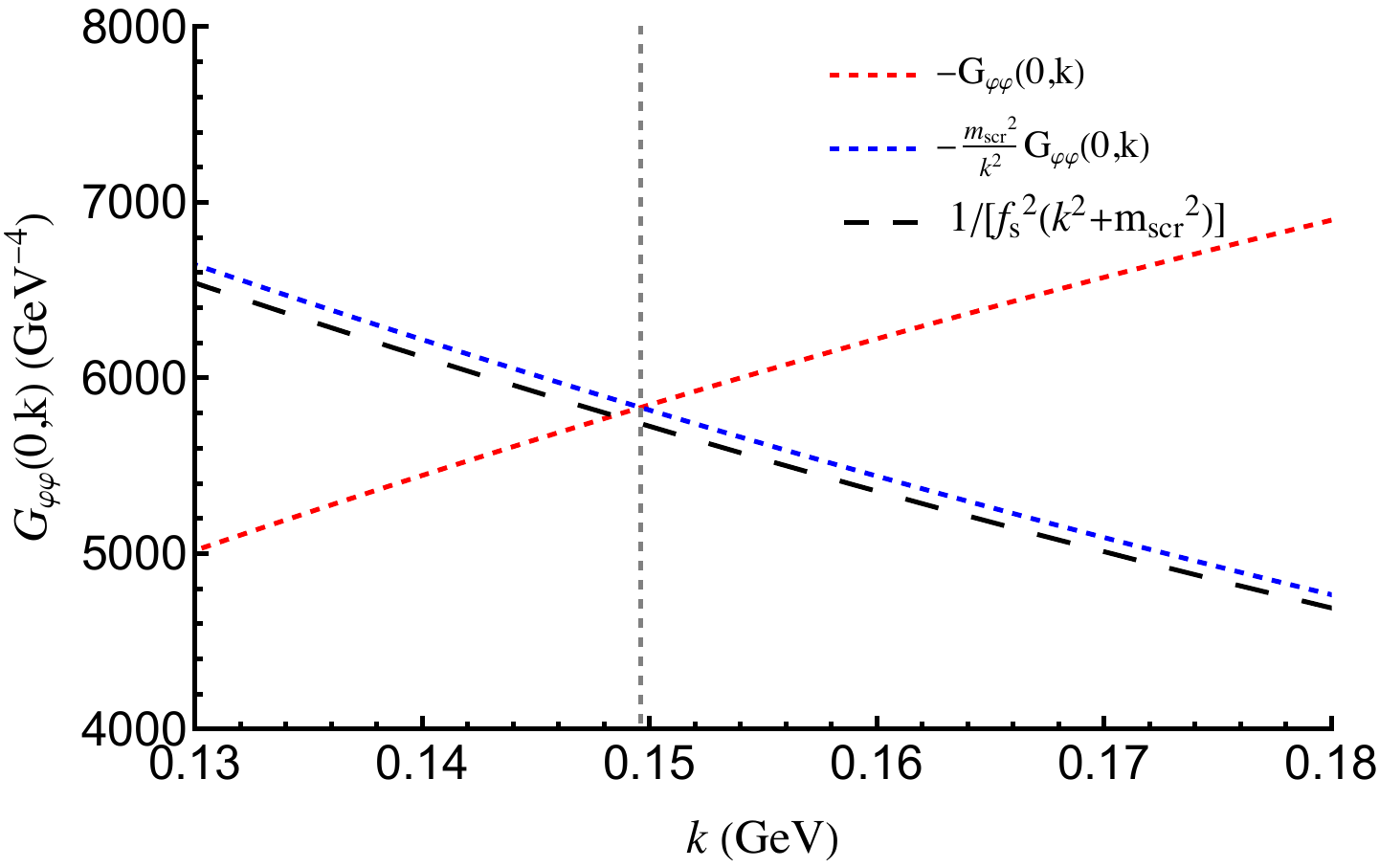}
        \put(85,55){\bf{(b)}}
    \end{overpic}
    \vspace{0.2cm}
    \begin{overpic}[width=0.31\linewidth]{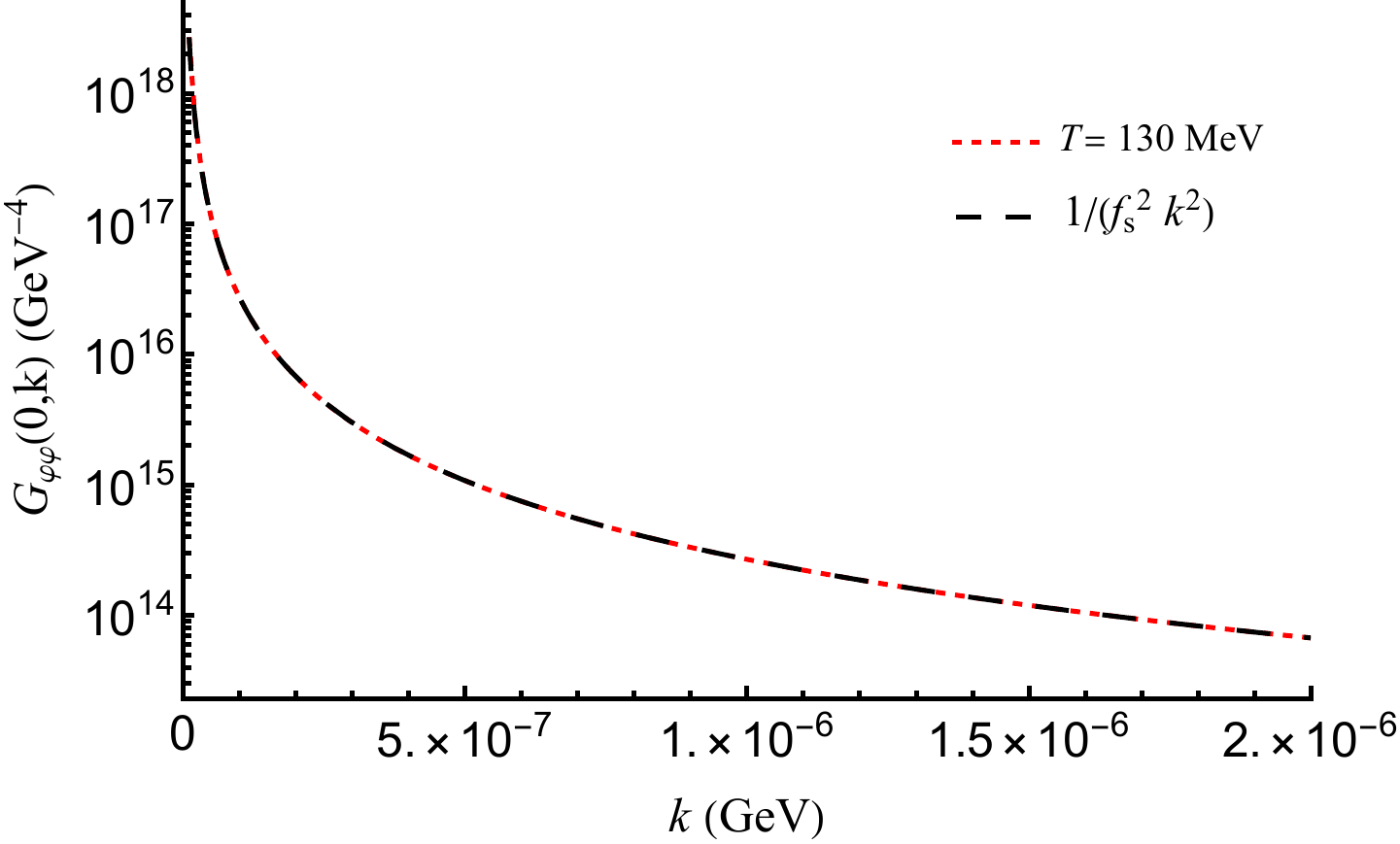}
    \put(85,55){\bf{(c)}}
    \end{overpic}
    \caption{The comparison of $G_{\varphi\varphi}(0,k)$ at $T=130 {\rm{MeV}}$ between the holographic numerical results and the results from thermal chiral effective field theory obtained in Refs.~\cite{Son:2002ci,Son:2001ff}. \textbf{(a)} $m_q=1 {\rm{MeV}}$;\textbf{(b)} $m_q=3.22 {\rm{MeV}}$; \textbf{(c)} chiral limit, $m_q=0$. The vertical gray dashed lines mark the  momentum positions $k^2=m_{\rm{scr}}^2$.}
    \label{fig:a2}
\end{figure*}

In section \ref{sec:correlators}, we have introduced the correlator of the pion operator in the soft-wall AdS/QCD model. To obtain the correlator, we calculate the three coupled second-order differential equations in Eqs~\eqref{EOM:coupledpion}. At the AdS boundary, we have three solutions ~\eqref{eq:boundaryvarphi}- \eqref{eq:boundaryvarphi1} with six undetermined integration constants, $a_{t0}$, $a_{t2}$, $a_{x0}$, $a_{x2}$, $\varphi_{0}$ (or $\bar{\varphi}_1$) and $\varphi_2$ (or $\bar{\varphi}_2$).\footnote{Considering the reduced functions in Eq.~\eqref{EOM:varphi} would cancel out one of the integration constants.} Near the horizon, we have asymptotic solutions ~\eqref{eq:horizonvarphi} with three undetermined integration constants ($a_{b0}$, $\varphi_{b0}$, $\varphi_{b1}$). Besides, the frequency $\omega$ and wave-vector $k$ appear also in the coupled equations.

Since these are linear equations, one can set one of the integration constants to a fixed value. For convenience, we fix $\varphi_{b0}=1$. We solve the equations using the so-called ``double shooting method'', in which the equations are integrated shooting both from the horizon $z=z_h$ and the UV boundary $z=0$ and then matched at an intermediate radial position, $0<z_0<z_h$.

As a concrete example of this procedure, we consider the pion correlator in Eq.~\eqref{Eq:correlatorpi} at $T=90$ MeV, $k=1$ MeV and $m_q=1$ MeV. Since we are not interested in other correlators, we set the source for the gauge field operator $A_0$ and $A_3$ to be zero, i.e. $a_{t0}=a_{x0}=0$. To obtain the pole of the correlator, one needs to search for the value of the frequency $\omega$ which satisfies the zero source boundary condition at the UV boundary, $\varphi_0=0$. To continue, we have five undetermined integration constants, $a_{t2}$, $a_{x2}$, $\varphi_2$, $a_{b0}$ and $\varphi_{b1}$, one free parameter, the frequency $\omega$, and six matching conditions. The latter are given by the continuity of the bulk functions and their first derivatives at the matching point $z_0$,
\begin{align}
    &a_0^b(z_0)=a^h_0(z_0)\,,\quad a_3^b(z_0)=a^h_3(z_0)\,,\nonumber\\
    &\varphi_0^b(z_0)=\varphi^h_0(z_0)\,,\quad \partial_z a_0^b(z_0)=\partial_z a^h_0(z_0)\,,\nonumber\\
    &\partial_z a_3^b(z_0)=\partial_za^h_3(z_0)\,,\quad \partial_z\varphi_0^b(z_0)=\partial_z\varphi^h_0(z_0)
\end{align}
in which the indices $b$ and $h$ label respectively the numerical solutions solved with boundary conditions from the UV, $z=\epsilon$, or with horizon conditions from the horizon $z=(1-\epsilon)z_h$. In our numerics, we take $\epsilon=10^{-6}$.  Besides, we numerically obtain the chiral condensate $\bar{\sigma}=0.0149$ $\rm{GeV}^3$ and $c_0=0.756149$ GeV through the EOM for $\Sigma(z)$ in Eq.~\eqref{EOM:chi} using a similar method. We plot the numerical results form the various bulk functions $A_0(z)$, $A_3(z)$ and $\varphi(z)$ in Fig.~\ref{fig:a1}. The two sets of solutions for the functions $A_{0}(z)$, $A_{3}(z)$, $\varphi(z)$ and their first derivatives are continuous at the matching point $z_0$. In this case, the choice of parameters which satisfies all the matching conditions is given by:
\begin{align*}
    & a_{t2}=0.527998+ 0.035872 i\  \rm{GeV}^{2}\,,\\
    &a_{x2}=-0.006417-0.000432 {\rm{GeV}^{2}}\,,\\
    & \omega=0.074525-0.000023 i\ {\rm{GeV}}\,,\\
    &\varphi_2=-886.85+13114.6\ {\rm{GeV}^2}\,,\\
    &a_{b0}=0.010709+0.000357 i\  {\rm{GeV}}\,,\\
    &\varphi_{b1}=-5.600+67.91 i\  {\rm{GeV}^2}\,.
\end{align*}
If we are not only interested in the poles of the correlator but in its full structure, the procedure is slightly different. We now keep the value of the frequency $\omega$ as an external parameter and solve on the contrary the matching conditions in terms of the integration constant $\varphi_0$. 

To verify the validity of the correlator in Eq.~\eqref{Eq:correlatorpi}, we compare $G_{\varphi\varphi}(q)$ at $T=130$ MeV as a benchmark example. The expected form for such a correlator is given by \cite{Son:2002ci,Son:2001ff} 
\begin{equation}\label{eq:eftcorrelator}
G_{\varphi\varphi}(0,k)=-\frac{1}{f_s^2(k^2+m_{\rm{scr}}^2)}\,.
\end{equation}
In Fig.~\ref{fig:a2}(a) and (b), we choose $m_q=1$ MeV and $m_q=3.22$ MeV and we plot the static correlator $G_{\varphi\varphi}(0,k)$ as a function of the wave-vector $k$. We expand the correlator around $k^2=m_{\rm{scr}}^2$ and only retain the leading term $(m_{\rm{scr}}^2/k^2)G_{\varphi\varphi}$. The dashed lines are the results from Eq.~\eqref{eq:eftcorrelator} with the pion decay constants obtained through Eq.~\eqref{definition:fpi}, $f_s^2=0.003772$ and $0.003892\ {\rm{GeV}^2}$, respectively. In Fig.~\ref{fig:a2}(c), we follow the same procedure in the chiral limit, $m_q=0$. We compare once more the holographic results with those from Eq.~\eqref{eq:eftcorrelator}  in which the pion decay constant $f_s^2=0.003718\ {\rm{GeV}^2}$ is obtained using Eq.~\eqref{definition:fpi}.
When the quark mass is small, we find that the holographic results are in perfect agreement with the predictions from thermal chiral effective field theory confirming the validity of our computations.

\bibliographystyle{apsrev4-2}
\bibliography{refs}
\end{document}